\documentclass[preprint]{aastex}
\usepackage[caption=false]{subfig}

\usepackage{graphicx}
\usepackage{multirow}
\usepackage{xcolor}
\usepackage{ulem}

\usepackage{ifpdf}  
\newif\ifmod
\modfalse
\newif\ifdel
\delfalse
\newcommand{\methanol}{\mbox{CH$_3$OH }}

\newcommand{\water}{\mbox{H$_2$O }}
\newcommand{\ghz}{\mbox{\,GHz}}

\newcommand{\kms}{\mbox{\,km\,s$^{-1}$}}

\newcommand{\htwo}{\mbox{H$_2$}}

\newcommand{\rew}[1]{\ifmod\textcolor{black}{\textbf{#1}}\else{#1}\fi}
\newcommand{\mod}[1]{\ifmod\textcolor{magenta}{\textbf{#1}}\else{#1}\fi}
\newcommand{\del}[1]{\ifdel\textcolor{blue}{\sout{#1}}\else\fi}

\begin{document}
\shortauthors{Kang et al.}

\title{Simultaneous observation of water and class I methanol masers toward class II methanol maser sources}

\author{Hyunwoo Kang\altaffilmark{1,2}, Kee-Tae Kim\altaffilmark{1}, Do-Young Byun\altaffilmark{1}, Seokho Lee\altaffilmark{3} and Yong-Sun Park\altaffilmark{2}}
\altaffiltext{1}{Korea Astronomy and Space Science Institute, 776 Daedeokdae-ro, Yuseong-gu, Daejeon 305-348, Korea; orionkhw@kasi.re.kr, ktkim@kasi.re.kr}
\altaffiltext{2}{Astronomy Program, Department of Physics and Astronomy, Seoul National University, Seoul 151-742, Korea}
\altaffiltext{3}{\rew{School of Space Research,} Kyung Hee University, \rew{Yongin-si, Gyeonggi}-do 449-701, Korea}

\begin{abstract}
We present a simultaneous single-dish survey of 22\ghz~water maser and 44\ghz~and 95\ghz~class I methanol masers toward 77 6.7\ghz~class II methanol maser sources, 
which were selected from the Arecibo methanol maser Galactic plane survey (AMGPS) catalog.
Water maser emission is detected in 39 (51\,\%) sources, of which  {15} are new detections. Methanol maser emission at 44\ghz~and 95\ghz~is found in 25 (32\,\%) and 19 (25\,\%) sources, of which 21 and 13 sources 
are newly detected, respectively. 
We find 4 high-velocity ($>$30~\kms) water maser sources, including 3 dominant blue- or redshifted outflows.
The 95\ghz~masers always appear with the 44\ghz~maser emission. They are strongly correlated with 44\ghz~masers in velocity, flux density, and luminosity, while they are not correlated with either water or 6.7\ghz~class II methanol masers.  The average peak flux density ratio of 95\ghz~to 44\ghz~masers is close to unity, which is two times higher than previous estimates. 
The flux densities of class I methanol masers are more closely correlated with the associated BGPS core mass than those of water or class II methanol masers. 
\rew{Using the large velocity gradient (LVG) model and assuming unsaturated class I methanol maser emission, we derive the fractional abundance of methanol to be in a range of $4.2\times10^{-8}$ to $2.3\times10^{-6}$, with a median value of $3.3\pm2.7\times10^{-7}$.}
\end{abstract}


\section{Introduction}
Masers are widely recognized as indirect tools for understanding the early stages of massive star formation. {Water} masers 
are  {frequently} detected in star-forming regions. They are pumped by collisions with neutral particles and appear in accretion disks or outflows 
\citep{elitzur89,torrelles97,mosca06}. Furthermore, water masers are detected near young stellar objects (YSOs) regardless of mass, and 
more massive YSOs tend to have stronger maser emission \citep{genzel77,furuya03,bae11}. Water masers often exhibit {significant variability in velocity and flux density \citep{rodri80,bae11,choi12}, and sometimes high-velocity features near outflows \citep{gwinn92, gwinn94}.} \cite{beuther02} investigated the \rew{spatial relationship between water and 6.7\ghz~}methanol masers. 
Both maser species are closely associated with the early stages of massive star formation, but no spatial correlation was observed.

\rew{Many methanol maser transitions show maser activity. These maser} transitions are empirically divided into two groups: 
class I and class II \citep{menten91}. Class I methanol masers such as the $4_{-1}-3_{0}~E$, $7_{0}-6_{1}~A^{+}$, $5_{-1}-4_{0}~E$, and 
$8_{0}-7_{1}~A^{+}$ lines at 36, 44, 84, and 95\ghz, respectively, are \rew{normally detected offset from HII regions,} infrared sources, 
OH, and \water masers \citep{plambeck90, kurtz04}. However, class II methanol masers such as the $5_{1}-6_{0}~A^{+}$, $2_{0}-3_{-1}~E$, 
$9_{2}-10_{1}~A^{+}$, and $J_{0}-J_{-1}~E$ lines at 6.7, 12, 23, and 157\ghz, respectively, are found close to other star formation indicators 
\citep{walsh98}. Using a theoretical model, \cite{cragg92} inferred that class I methanol masers are pumped by collisions, while 
class II masers are excited by radiative pumping.

{Methanol masers are primarily found in massive star-forming regions \citep{minier03, breen13}, although weak class I methanol masers are detected in several low- and intermediate-mass star forming regions as well \citep{kalenskii10, bae11}.}
Class I methanol masers {usually} appear in the interface 
{between outflows and the surrounding regions} \citep{plambeck88, plambeck90, kurtz04, cyg09, voronkov06, voronkov14}.
\cite{valtts00} reported that the peak flux density of \rew{44\ghz~masers is on average about three times higher than that of the associated 95\ghz~class I methanol maser.} Class II methanol masers, on the other hand, are detected \rew{at an early age evolutionary phase of} massive star formation. 
{There is ongoing debate about} whether class II masers trace {accretion} disks or outflows \citep{nor93, pes04, walsh98, minier00, dodson04}.
 
\cite{slysh94} {surveyed} 44\ghz~class I methanol maser {emission} with 5 to 10 Jy detection limits (3$\sigma$) toward about 250 HII regions and water and 6.7\ghz~class II methanol maser sources. Of the 55 detected 44\ghz~maser sources, 64\,\%  are related to 6.7\ghz~masers. They suggested that class I and class II \rew{masers} have anti-correlations of velocity and flux density. However, their observations were based on single-dish data with no accurate positions and made \rew{towards a source sample with significant selection biases.} {\citet{ellingsen05} searched for similar anti-correlations between 95\ghz~class I and 6.7\ghz~masers, but did not find any anti-correlation, either in velocity or flux density.}

In this study, 
we report a simultaneous survey of 22\ghz~water and 44\ghz~and 95\ghz~class I methanol masers toward 6.7\ghz~class II methanol masers, which were selected from the Arecibo methanol maser Galactic plane survey (AMGPS) catalog \citep{pand07}.
The AMGPS covered the inner Galactic region of $35.2\arcdeg \lesssim l \lesssim 53.7\arcdeg$ and $|b|\lesssim 0.41\arcdeg$ with a detection limit of 0.27 Jy (3$\sigma$).  
The properties of the AMGPS sources such as distance, luminosity, 
and kinetic temperature have been subsequently studied by \citet{pand09, pand11, pand12}.
The AMGPS sources may represent various \rew{evolutionary} stages of massive star formation, because they were identified by a {very sensitive unbiased survey.} 
We statistically investigate the detection rates of the observed maser {species} and correlations between them. We are particularly interested in the relationship between class I and II methanol masers. Since water and class I methanol masers are both pumped by collision mechanisms, one may expect a significant correlation between them.  
\cite{chen12} recently suggested selection criteria for 
the Bolocam Galactic Plane Survey \rew{(BGPS) clumps likely to have an associated 95\ghz~class I methanol maser.} We also {examine} the validity of these criteria using our data sets.

Section~2 gives the source selection and observation methodology. Section~3 presents the detection statistics, new detections, and short descriptions of some interesting sources. Section~4 describes velocity characteristics, flux densities and luminosities, comparison with the BGPS clump parameters, and methanol column densities and abundances. Section~5 summarizes the main results.

\section{Source Selection and Observations}

\subsection{Source Selection}

We selected 77 sources from the AMGPS catalog of 88 6.7\ghz\ methanol 
maser sources. The catalog originally consisted of 86 sources \citep{pand07}, 
and later 2 new sources (G37.04-0.04 and G49.42+0.32) were added via MERLIN observations 
\citep{pand11}. 
Considering the beam size of the telescope at 95\ghz, 33$\arcsec$ (see \S~2.2),
we excluded 9 sources in two very crowded regions: W49N (G43.15+0.02, G43.16+0.02, G43.17+0.01, G43.17-0.00, and G43.18-0.01) and W51 (G49.47-0.37, G49.48-0.40, G49.49-0.37, and G49.49-0.39).
We also rejected G49.42+0.32 due to its proximity to G49.41+0.33 \citep{pand11},
and G41.87-0.10 due to its uncertain coordinates \citep{pand07}.
The systemic velocities of the individual sources were taken from the previous molecular line observations \citep{pand09, pand12}. 
Table \ref{Table:list} lists the 77 sources in our sample together with their equatorial coordinates and systemic velocities.

\subsection{Observations}
Using the Korean VLBI Network (KVN) 21\,m telescopes, we performed simultaneous 
surveys of the \water\ $6_{16}-5_{23}$ (22.23508\ghz), \methanol\ $7_{0}-6_{1} A^{+}$ 
(44.06943\ghz), and \methanol\ $8_{0}-7_{1} A^{+}$ (95.169463\ghz) masers toward the 
77 class II methanol maser sources. The KVN consists of three 21\,m telescopes
\citep{kim11, lee11}. 
Each telescope is equipped with a multi-frequency receiving system with 22\ghz, 44\ghz, 86\ghz, and 129\ghz~bands \citep{han08}. 
{Two telescopes at the Yonsei and Ulsan sites\footnote[1]{See http://kvn.kasi.re.kr/index.html} were used in single-dish mode.}
Table \ref{Table:units} presents the aperture efficiencies, half-power beam widths (HPBWs), and conversion factors of the telescopes.
{ All spectra were obtained by position switching.
For calibration,
we adopted the standard chopper-wheel method, and the line intensity was observed on the $T^*_{\rm A}$ scale.}
The pointing and focus were checked every 2$-$3 hours. The system temperature (T$_{sys}$) ranges were 70$-$90\,K at 22\ghz, 100$-$150\,K at 44\ghz, and 150$-$250\,K at 95\ghz. 
{ The aperture efficiencies of the used telescopes remain constant within 15\% for elevations \rew{of} $\ge$20$\arcdeg$ at the observing frequencies\footnote[2]{See Gain Curve at http://kvn.kasi.re.kr/status$\underbar{ }$report$\underbar{ }$2013} \citep{lee11}.
The bandwidths were set to 32\,MHz with 4096 channels. }
The KVN receiving systems support dual-polarization mode, and the computational backend can process up to four inputs simultaneously \citep{lee11}. 
To approximately equalize root-mean-square ($rms$) noise levels in different bands, \rew{we observed a single polarization for the 22\ghz~and 44\ghz~bands and dual polarization for the 95\ghz~band.} 
{The resultant noise levels were about 1.5~Jy (3$\sigma$) at a velocity resolution of 0.2~\kms\ for all three bands.}
\rew{The KVN telescopes are} shaped Cassegrain and hence have fairly high first side lobe levels of about 5\,\% \citep{kim11, lee11}.  { We made 7$\times$7 grid maps with half-beam spacing for all detected 22 and 44~\ghz\ maser sources in order to assess if the maser emission is contaminated by any nearby strong source. 
Here 95~\ghz\ maser sources were excluded because they all correspond to 44~\ghz\ maser sources (see \S~3.1).}
\rew{The observation took a total of 18 days to complete and were undertaken between} 2011 Nov -- 2012 Mar: {15 days for the survey and 3 days for the grid mapping.}

 We also observed 4 sources in the $^{13}$CO~(J=1$-$0) (110.201353\ghz) line and one of them in the HCO$^+$~(J=1$-$0) (89.188523\ghz) line using the Taeduk Radio Astronomy \rew{Observatory} (TRAO) 14~m telescope on 2014 March 17. \rew{The aim of these} observations \rew{was} to investigate the origin of the water maser features with large velocity offsets (see \S~\ref{section:hv_feature}).
The system temperatures were 560~K at 89\ghz~and 760~K at 110\ghz. The main-beam efficiencies of the telescope are 50\% and 49\%  at 89\ghz~and 110\ghz, respectively.
All the data were reduced and analyzed with the GILDAS/CLASS package and IDL. The bisector method was adopted for the linear regression 
calculations \citep{isobe90}.

\section{Results}
\subsection{Detection Rates}
{ Table \ref{Table:list} summarizes the observational results, and Figure \ref{fig:results} displays the statistics of detected maser sources. We detected at least one maser species in {46} of the 77 sources.
Water masers were detected in  {39 (51\,\%)} sources. 44\ghz~and 95\ghz~methanol masers were detected in 25 (32\,\%) and 19 (25\,\%) sources, respectively. 
Interestingly, \rew{95\ghz~masers always accompany 44\ghz~maser emission and often (84\,\%) appear with water maser emission. 44\ghz~masers also mostly (72\,\%) accompany water maser emission.}

The 6.7\ghz~maser sources in our sample have a wide range of peak flux densities, 0.1--55.6\,Jy, and isotropic maser luminosities, $\rm10^{-9}-10^{-5}\,L_{\sun}$ \citep{pand07, pand09}. We investigate the detection rates \rew{of the three observed maser transitions along with the flux density and luminosity of the 6.7\ghz~}maser emission (Figure \ref{fig:rate_cutoff}). 
All three rates tend to increase with the flux density and luminosity, although the trend is not strong.
\cite{ellingsen05} also {surveyed} 95\ghz~maser \rew{emissions} with a flux limit of $\sim$4\,Jy (3$\sigma$) toward  {sixty six} 6.7\ghz\ methanol maser sources, {which had been} found by previous blind surveys\rew{, and detected emission in 25 sources.} The detection rate (38\%) is significantly higher than our value (25\,\%) in spite of considerably poorer sensitivity. This is presumably because their sample contains more bright sources than our sample. \rew{The fraction of sources with} flux densities \rew{of} $\ge$10~Jy \rew{is} about 60\% for their sample but about 20\% for ours.
\rew{We examine the detection rate of 44 GHz class I methanol maser emission for sources with strong} water maser emission, and find that the rate is 89\% (8/9) for the sources with peak flux densities of water maser emission $>$ 20 Jy.  Meanwhile the rate is much lower (33\%, 10/30) for the sources with flux densities $<$20Jy. We also find a similar trend for the detection rate of water maser emission. The rate is 100\% (4/4) for the sources with flux densities of 44 GHz methanol maser emission \mod{$>$10 Jy} while it is 67\% for the others.}

\subsection{New Maser Sources}
{ We identify 15 of the 39 water maser detections as new discoveries using the VizieR service\footnote{See http://vizier.u-strasbg.fr/viz-bin/VizieR} \citep{ochsenbein00} by comparing our results with previous surveys:} \cite{palagi93,codella94,han98,forster99,valdettaro01,codella04,szymczak05,urquhart11}.
{ \citet{szymczak05} also searched for water maser emission toward 4 sources in our sample (G36.70+0.09, G37.02-0.03, G37.04-0.04, G38.03-0.30), 
\rew{but did not detect emission in any of these sources.}
They used the Effelsberg 100\,m telescope (HPBW $\simeq$ 40$\arcsec$) with a flux limit of 1.5\,Jy (3$\sigma$). 
The observed positions are offset from our 6.7\ghz\ maser positions by 27$\arcsec$--87$\arcsec$. 
Thus their \rew{nondetection} may be due both to slightly different observed positions and relatively small beam size.}
\rew{We can apply the same explanation for two other sources} 
(G40.28-0.22 and G42.03+0.19) in which no water maser emission was detected by \citet{urquhart11} using the GBT 100\,m telescope with a flux limit of 0.4\,Jy (3$\sigma$).

\rew{Meanwhile, there have been many fewer surveys of} class I methanol masers. We find that 21 of the 25 detected 44\ghz~methanol maser sources are new detections \citep{kurtz04,cyg09,bayandina12}.
Thirteen of the 19 detected 95\ghz~methanol maser sources appear to be new detections {from  \rew{a} comparison with the catalogues of \cite{chen11,chen12}, who also used single-dish telescopes. Among the new sources, G37.55+0.19 and G45.49+0.13 have been observed by \cite{chen11} with the Mopra 22\,m telescope (HPBW $\simeq$ 36$\arcsec$) at a detection limit of 1.6\,Jy (3$\sigma$). Their positions are different from ours by 33$\arcsec$ --41$\arcsec$. 

\subsection{Notes on Selected Sources\label{section:sources}}

{\subsubsection{G35.03+0.35, G43.04-0.46, and G45.07+0.13}
\rew{These three are the only already-known sources both with water} and 44\ghz~class~I methanol masers in our sample. \rew{We confirm that the water masers vary substantially in both the velocity and flux density of their emission, while the 44\ghz\ masers do not.} \citet{forster99} observed water maser emission of 18.5\,Jy at 68.5\kms\ in G35.03+0.35. This is very different from our result, {62.5}$\pm${0.9}\,Jy at {44.7}\kms.
\cite{cyg09} detected 44\ghz~maser emission at 52.8\kms~using the Very Large Array (VLA). The peak velocity \rew{is consistent with our value}, {52.8}\kms.
For G43.04-0.46,  \cite{valdettaro01} observed water maser emission of 5.1\,Jy at 57.55\kms, \rew{which is very different from our measurement,} {5.6}$\pm${0.7}\,Jy at {45.7}\kms. \cite{kurtz04} observed 44\ghz~maser emission of 3.8\,Jy at  58.0\kms\ with the VLA. We also detected the maser emission of {8.7}$\pm${1.1}\,Jy at the same velocity. 
There have been several reports of water maser detection in G45.07+0.13. \rew{The peak flux densities and velocities} are 50.7\,Jy at 60.1\kms, 21.5\,Jy at 60.6\kms, 41\,Jy at 60.1\kms, and {23.2}$\pm${0.6}\,Jy at {60.1}\kms~by \cite{palagi93}, \cite{forster99}, \cite{valdettaro01}, and this study, respectively. \cite{kurtz04} showed 44\ghz~maser emission of 1.1\,Jy at 59.3\kms\ and we also detected a peak flux density of {1.7}$\pm${0.5}\,Jy at {59.2}\kms.}

\subsubsection{G36.92+0.48, G38.26-0.08, G41.58+0.04, G43.80-0.13, G44.64-0.52, and G45.07+0.13}

The systemic velocities of these sources were not determined by the NH$_{3}$ lines \citep{pand12} but by the $^{13}$CO ($J$ = 2$-$1) 
or CS ($J$ = 5$-$4) lines \citep{pand09}. G43.80-0.13 and G45.07+0.13 show strong NH$_{3}$ lines, but the line profiles are 
complicated \citep{pand12}. These two sources {have} all three maser {species}, and G43.80-0.13 is the strongest water maser source in our sample (Table 3). 
The other four sources have no detectable NH$_{3}$ line {emission}. Only G38.26-0.08 among them shows water maser emission.

\subsubsection{G37.02-0.03 and G37.04-0.04}

 {These two sources are separated by 47\arcsec.7 and overlap within the {HPBW of the telescope at 22\ghz}. They exhibit two common water maser lines at {80.7} and {81.6}~\kms.
Therefore, we count them as one detection.}
A more detailed investigation using interferometric observations is required
to clarify the origin of water maser emission in this area.
The BGPS counterpart, G037.042-00.034, is closer to G37.04-0.04 than G37.02-0.03.

\subsubsection{G38.66+0.08, G39.39-0.14, and G42.43-0.26}

These sources are associated with ultracompact (UC) HII regions \citep{pand10}. G42.43-0.26 has all three maser species, while G39.39-0.14 shows {only} class I methanol masers, and G38.66+0.08 {does not display any} maser emission.

\subsubsection{G40.28-0.22 and G48.99-0.30}

{G40.28-0.22 and G48.99-0.30 are the strongest 44\ghz~and 95\ghz~maser sources in our sample, respectively (Table 3).} \rew{G40.28-0.22 has a peak flux density of} {89.2}$\pm${1.4}\,Jy at 44\ghz~ and {72.7}$\pm${1.0}\,Jy at 95\ghz. 
This object is associated with a luminous ($L_{\rm bol}$$\simeq$2.1$\times$10$^5$~$L_{\sun}$) YSO, which might be a hypercompact HII region \citep{pand10}. Water and 44\ghz~methanol masers were newly detected, while 95\ghz~methanol maser emission has been detected by \cite{chen11} with a peak flux density of 53.2\,Jy at 72.8\kms. \cite{urquhart11} searched for water maser emission at this site, \rew{ but did not detect any.
G48.99-0.30 has a 44\ghz~peak flux density of {93.2}$\pm${1.2}\,Jy at 66.4\kms\ } and a 95\ghz~peak flux density of {54.2}$\pm${0.9}\,Jy at 66.5\kms.
\rew{Both class~I methanol maser transitions are new detections, whereas} the water maser emission has been detected by multiple surveys (see Table 1). For instance, \cite{codella94} and \cite{valdettaro01} observed water maser emission of 7\,Jy at 71.0\kms~and 8\,Jy at 70.9\kms, respectively. \cite{urquhart11} detected a peak flux density of 58.9\,Jy at 65.3\kms, {which is similar to our detection, {62.4}$\pm${0.8}\,Jy at 66.0\kms.}

\subsubsection{High velocity Features of Water Maser\label{section:hv_feature}}

We found 4 water maser sources showing very different ($>$30~\kms) velocity components from the systemic velocities. G35.59+0.06, G37.60+0.42, 
G41.08-0.13, and G41.34-0.14 have maser lines with velocity offsets of $-$60.5, +48.0, $-$77.6, and +47.5~\kms\, respectively. 
 \cite{breen10} also found 10 water maser sources with large ($>$30~\kms) relative velocity components with respect to 6.7\ghz\ methanol maser emission.  \rew{The fraction of their sources in their sample (approximately 200 sources) which show high velocity emission (5\%) is half that of this study (4/39, 10\%).} These features can be produced by high-velocity jets/outflows from the massive YSOs associated with 6.7\ghz\ methanol masers \citep{gwinn92, gwinn94} or associated with other YSOs at different \rew{distances along the same line of sight. }
In order to distinguish between the two possibilities,
we observed $^{13}$CO~(J=1$-$0) line emission toward these sources, and found no $^{13}$CO line emission related to the large velocity-offset features in  {any of them except} G41.34$-$0.14 (Figure \ref{fig:hvtrao}). We also observed G41.34$-$0.14 in the HCO$^+$~(J=1$-$0) line, which is a very reliable tracer of massive star-forming cores (e.g., \citealt{purcell06}), \rew{and detected line emission only} at the systemic velocity. 
Therefore, the large velocity-offset features are likely to be high-velocity components caused by jets/outflows, even though we cannot exclude the possibility that the feature of G41.34$-$0.14 may be associated with a low-mass YSO at a different distance in the same line of sight.  

G37.60+0.42 has a maser emission near the systemic velocity as well, while the others do not. 
G35.59+0.06 and G41.08$-$0.13 are so-called dominant blue-shifted water maser outflows. \cite{caswell08} reported \rew{three}  {such} water maser sources and suggested that they can be generated from pole-on jets.
G41.34$-$0.14 may be the red-shifted counterpart of the dominant blue-shifted water maser sources. \rew{Follow-up observations with higher resolution} are required to understand the origin of these objects.

\section{Analyses}

\subsection{Velocity Characteristics}

\rew{Figure \ref{fig:offset} shows the distribution of the peak velocity offsets from the systemic velocities for the four maser species}. The bin size is 2~\kms.
\rew{The average offsets of the 6.7\ghz,} 22\ghz, 44\ghz, and 95\ghz~masers are $-$0.0$\pm$5.0, $-$0.6$\pm$4.8, 0.2$\pm$1.1, and 0.4$\pm$1.1\kms, respectively. \rew{In these calculations, the four high-velocity water maser sources have been excluded.}
\rew{The velocity offsets of the 44\ghz~}and 95\ghz~masers are always within $\pm$4~\kms.
\rew{This is consistent with the findings of previous studies} (e.g., \citealt{fontani10, bae11}). \rew{\cite{plambeck90} and \cite{kurtz04} suggest that the class I methanol maser emission is} associated with ambient molecular cloud that is heated by outflows and is not accelerated by the outflow. \rew{Our results support this hypothesis.}
The offsets of 6.7\ghz~class II methanol masers are much more widely distributed, but are still within $\pm$10~\kms\ for 95\% of the sources. The distribution is double peaked with a dip around zero. This is consistent with the finding of Byun et al. (2012) for a large sample of 284 6.7~GHz maser sources. \rew{In contrast, the distributions} are single peaked around zero for the other masers. The offset dispersion of water maser lines appears to lie between those of class I and class II methanol masers. Half of the sources are within $\pm$2.3~\kms\ and $\pm$3.7~\kms\ for water and 6.7\ghz~masers, respectively. 
For comparison, 
\cite{fontani10} surveyed 6.7\ghz, 44\ghz, and 95\ghz\ methanol masers toward 296 massive YSO candidates, which had been observed in water maser emission.
They found that water and class~II methanol masers have very similar single-peaked distributions of velocity offsets. The bin size in their histograms was 6~\kms. \rew{This large bin size may have impacted their ability to identify a }double-peaked distribution of 6.7\ghz\ masers. \mod{ When we use the same bin size, the distribution of 6.7\ghz\ masers in our sample is also changed to be single peaked. Further detailed studies are required to clarify 
the statistical significance and physical implications of the dip observed in this study.}
\rew{Almost all of the} class I maser sources in their sample showed velocity offsets within $\pm$6\kms.  

Figure \ref{fig:offset6744} compares the velocity offsets of class I and class II masers. Large velocity offsets of both masers appear to be avoided.  
 \cite{slysh94} reported that \rew{class I and class II are anti-correlated in the velocities where they exhibit strong emission.} Some of our data \rew{show} the anti-correlation, but almost do not follow the trend.
We derive the velocity overlap of  {class I and class II methanol} maser spectra to examine the relationship between the two classes. The velocity overlap is defined by the difference between the lowest velocity of the class with higher mean velocity and the highest velocity of the other class with lower mean velocity \citep{ellingsen05}.
Negative values of velocity overlap hence imply that the velocity ranges of the two classes overlap.
Figure \ref{fig:overlap6744} displays a histogram of the velocity overlap.
Nineteen (76\%) of \rew{twenty-five} 44\ghz~maser sources and {fourteen} (74\%) of {nineteen} 95\ghz~maser sources show negative values. This portion is slightly higher than that of \citet{ellingsen05} for 95\ghz\ maser sources, 62\% (23/37).

\subsection{Flux Densities and Luminosities\label{section:flux}}

Tables~\ref{Table:flux} and \ref{Table:MASER_L_BGPS} present the peak flux densities and the isotropic luminosities of the detected masers, respectively. 
Figure~\ref{fig:flux4495} compares the flux densities of 44\ghz~and 95\ghz~methanol masers. A strong correlation exists between the two parameters, as shown below:
\begin{equation}
\rm S_{peak}(95)=(0.71\pm0.08)~S_{peak}(44)-(0.28\pm0.52),~~~r=0.98,
\end{equation}
\begin{equation}
\rm log(S_{peak}(95))=(0.95\pm0.05)~log(S_{peak}(44))-(0.13\pm0.06),~~~r=0.95,
 \end{equation}
 
\noindent
where $r$ is the correlation coefficient.
\cite{valtts00} also reported a quite good correlation between the two with a correlation coefficient of 0.73: 
$\rm S_{peak}(95)=(0.32\pm0.08)\,S_{peak}(95)-(8.1\pm2.7)$. 
\mod{ \cite{jordan15} found a similar relation for a sample of 19 sources:
$\rm S_{peak}(95)=0.31\,S_{peak}(95)+1$.}
However, our estimate of $\rm S_{peak}(95)$/$\rm S{peak}(44)$ is two times higher than theirs.
It should be noted that \mod{the two studies} used data sets obtained with 
different telescopes at \rew{very} different epochs.
Additionally, the sources in the two samples were generally strong ($\gtrsim$5\,Jy).
\mod{In contrast, in this study the data sets were simultaneously taken with the same telescope with a higher sensitivity and 
are compared at the same velocity resolution, 0.2~\kms. }
Our value thus seems to be more reliable. 

\rew{We investigate whether there is any relationship between the peak flux density of the 6.7\ghz~maser and those of the three observed maser transitions,} but there appears to be little correlation between them: 
\begin{equation}
\rm log(S_{peak}(22))=(0.96\pm0.05)~log(S_{peak}(6.7))+(0.42\pm0.12),~~~r=0.18
,
\end{equation}
\begin{equation}
\rm log(S_{peak}(44))=(-0.85\pm0.18)~log(S_{peak}(6.7))+(1.10\pm0.19),~~~r=-0.19
,
\end{equation}
\begin{equation}
\rm log(S_{peak}(95))=(-0.90\pm0.15)~log(S_{peak}(6.7))+(1.12\pm0.22),~~~r=-0.20.
\end{equation}
  
\noindent
\rew{\cite{beuther02} did not find any} correlation between water and 6.7\ghz~methanol masers. \cite{slysh94} reported anti-correlation between 44\ghz~class I and 6.7\ghz~class II methanol masers in velocity and flux density. However, \cite{ellingsen05} found no anti-correlation between 95\ghz\ and 6.7\ghz\ masers. Our results are \rew{consistent} with the suggestions of \cite{ellingsen05} and \cite{beuther02}. 

We also investigate the relationship between water and class I methanol masers, but there \rew{seems} to be no correlation between them.  The correlation coefficients between water maser and the 44\ghz~and 95\ghz~masers are $-0.05$ and $-0.05$, respectively. 

Using the distance estimated by \cite{pand09}, we calculate the isotropic \rew{luminosities} from the integrated flux densities in Table~3 (see, e.g., \citealt{bae11}), and investigate relationships between them. As in the cases of flux densities, there is a strong correlation between the luminosities of the two class I masers (Figure \ref{fig:lmaser}).
\begin{equation}
\rm log(L_{95}/L_{\sun})=(0.89\pm0.08)~log(L_{44}/L_{\sun})-(0.14\pm0.34),~~~r=0.89,
\end{equation}
\begin{equation}
\rm L_{95}/L_{\sun}=(1.84\pm0.19)~L_{44}/L_{\sun},~~~r=0.96.
\end{equation}
We could not find any significant correlation between the luminosities of the water and methanol masers.

\subsection{Associated BGPS Clumps\label{sec:BGPS}}
For targeted searches of 95\ghz~class~I methanol maser, \cite{chen12} proposed criteria of the integrated flux density and beam-averaged H$_2$ column density of the BGPS clumps: log(S$_{\rm int}$) $\lesssim -38.0+1.72$ log($N^{\rm beam}_{\rm H_{2}}$) and log($N^{\rm beam}_{\rm H_{2}}$) $\gtrsim$ 22.1.
The BGPS is a 1.1~mm continuum survey of the inner Galaxy region of $-10\arcdeg < l < 90\arcdeg$ and $|b|\lesssim0.5\arcdeg$ \citep{rosolowsky10}. The survey was undertaken using the Caltech Submillimeter Observatory (CSO) with an HPBW of 33$\arcsec$.
We test the validity of the Chen et al.'s criteria with 44 BGPS clumps associated with the sources in our survey \citep{pand12}. For simplicity, we used the same parameters as in \cite{chen12} in deriving the beam-averaged H$_{2}$ column density: mean mass per particle of $\mu$=2.37, kinetic temperature of 20\,K, and beam solid angle of $2\times10^{-8}$\,Sr. 
Figure \ref{fig:bgps_int_n} shows that all the 14 BGPS clumps with 95\ghz~maser meet the criteria. It is worth noting that they are \rew{only} 39\% of the 36 BGPS clumps fitting the criteria. This detection rate is considerably lower than the prediction of \cite{chen12}, $>$ 60\%.

We estimate the BGPS clump mass in a similar manner to \rew{that of} \cite{chen12} except \rew{for} the kinetic temperature measured {by \cite{pand12} from their NH$_{3}$ line observations.} The relationship between the BGPS mass and the four maser luminosities \rew{is} derived below and displayed in Figure \ref{fig:BGPSml}:
\begin{equation}
\rm log(L_{6.7}/L_{\sun})=(1.46\pm0.22)~log(M_{BGPS}/M_{\sun})-(10.41\pm0.68),~~~r=0.29,
\end{equation}
\begin{equation}
\rm log(L_{22}/L_{\sun})=(1.50\pm0.25)~log(M_{BGPS}/M_{\sun})-(9.50\pm0.82),~~~r=0.50,
\end{equation}
\begin{equation}
\rm log(L_{44}/L_{\sun})=(1.01\pm0.18)~log(M_{BGPS}/M_{\sun})-(7.80\pm0.50),~~~r=0.67,
\end{equation}
\begin{equation}
\rm log(L_{95}/L_{\sun})=({1.21}\pm0.18)~log(M_{BGPS}/M_{\sun})-({8.15}\pm0.51),~~~r={0.59}.
\end{equation}
Our results suggest that the luminosities of class I methanol masers are more closely related to the BGPS clump mass than those of water and 
6.7\ghz~class II methanol masers. 
  
\subsection{Methanol Column Densities and Abundances}

\cite{valtts00} investigated the physical condition of \rew{a} class~I maser emitting region using a large velocity gradient (LVG) code with the observed flux densities of 44\ghz\ and 95\ghz~methanol masers.
Under the assumption of a methanol density divided by velocity gradient of $\rm 0.67\times 10^{-2}\,cm^{-3}$ $(\kms~pc^{-1})^{-1}$, 
they obtained a gas temperature of about 20\,K and an H$_{2}$ number density of less than $\rm 10^{6}\,cm^{-3}$. 
Using the LVG model, we estimate the methanol column density with the measured flux density ratio of 44\ghz~and 95\ghz~masers. 
We utilize the A-CH$_{3}$OH molecular data from the LAMDA database\footnote{See http://home.strw.leidenuniv.nl/$\sim$moldata/} 
\citep{schoier05}, and the model code modified using RADEX\footnote{See http://www.sron.rug.nl/$\sim$vdtak/radex/radex.php} \citep{van07}. 
\cite{van07} suggested that the calculated values may be inaccurate at $-1\lesssim\tau\lesssim-0.1$ and meaningless at $\tau\lesssim-1$. 
The H$_{2}$ number density is calculated from the BGPS clump mass estimated in the last section, assuming that the clump is a sphere with uniform density. {It should be noted that the number density in the maser-emitting region may be significantly higher than this estimate, because the BGPS flux density is the beam-averaged value.} The kinetic temperature is taken {from \cite{pand12} as in the mass estimation of the BGPS clumps.} The line width is adopted from {the mean value} between the FWHMs of the 44\ghz~and 95\ghz~maser spectra in Table 3. We consider only the cosmic microwave background at 2.73\,K, and assume that the 44\ghz~and 95\ghz~maser emissions emanate from the same region. 

As mentioned above, 14 of the 19 sources with both 44\ghz~and 95\ghz~masers have associated BGPS clumps. 
The methanol column densities are derived to be {between} ${1.3}\times10^{15}$ {and} ${3.5}\times10^{16}~\rm cm^{-2}$. \del{with one exception.}
Table \ref{Table:lvv} lists the values together with the calculated optical depths. {Most optical depths are greater than -0.1\rew{, although} a few values are about -0.1.}
We estimate the methanol abundance relative to $\rm H_{2}$ by comparing  {the calculated methanol column density with the $\rm H_{2}$ column density {determined by \citet{pand12} using the BGPS data.}
The fractional methanol abundances range from ${4.2}\times10^{-8}$ to ${2.3}\times10^{-6}$ with a median value of $3.3\pm2.7\times10^{-7}$.
These values are close to the typical abundance in the interaction regions between the outflows and molecular cloud gas \citep{plambeck90, menten88}.

\section{Conclusions}
We simultaneously surveyed 22\ghz~water and 44\ghz~and 95\ghz~class I methanol masers toward 77 6.7\ghz~class II methanol maser sources. Our main results are as follows.

(1) The water maser was most frequently detected with a rate of  {51\,\%}, while the class I methanol masers were detected with lower rates of 32\,\% at 44\ghz\ and 25\,\% at 95\ghz. There are 15, 21, and 13 new detections at the 22\ghz, 44\ghz, and 95\ghz~bands, respectively. 
We identified 4 high-velocity ($>$30~\kms) water maser sources, of which two are dominant blueshifted outflows and one is \rew{a} dominant redshifted outflow.
\rew{The} 95\ghz~class I maser \rew{always} accompanies 44\ghz~maser \rew{emission} and frequently with water maser \rew{emission}.

(2) The peak velocities of class I methanol masers are always within $\pm$4~\kms\ of the systemic velocities, while those of 6.7\ghz~class II methanol masers are significantly offset. The velocity offset dispersion of water masers \rew{appears} to lie between those of class I and II masers. Interestingly, class~II masers show a double-peaked distribution of velocity offsets while water and class~I masers show single-peaked distributions. 

(3) 44\ghz~and 95\ghz~masers exhibit strong correlations in velocity and flux density. The flux density of \rew{the} 95\ghz~maser is usually as strong as that of \rew{the} 44\ghz~maser. We could not find any \rew{anti-correlation} between class~I and class~II methanol masers in velocity and flux density. 

(4) The isotropic luminosities of class I masers are correlated with the \rew{associated} BGPS clump \rew{masses}.
All the 14 BGPS clumps with both 44\ghz~and 95\ghz~masers satisfy the criteria of \citet{chen12}.

(5) Assuming an unsaturated system, we applied the LVG model to derive the 
column densities and the fractional abundances of methanol for the 14 BGPS clumps except one. The calculated \rew{column} densities range from ${1.3}\times10^{15}$ to ${3.5}\times10^{16}~\rm cm^{-2}$. The estimated abundances are between ${4.2}\times10^{-8}$ to ${2.3}\times10^{-6}$, which are comparable to the {typical} value in the interface of outflows with the surrounding region.
 
\rew{\section*{Acknowledgement}}

\rew{We are grateful to all staff members in KVN and TRAO who helped to operate the systems. The KVN and the TRAO are facilities operated by KASI (Korea Astronomy and Space Science Institute). The KVN operations are supported by KREONET (Korea Research Environment Open NETwork), which is managed and operated by KISTI (Korea Institute of Science and Technology Information).}  We thank Chang-Hee Kim, Won-Ju Kim, and Jeong-Seop Kim for helpful discussions.

\clearpage
\begin{deluxetable}{ccccccccc}	
\tabletypesize{\tiny} 
\tablecaption{Source Summary with Detection Results\label{Table:list}}
\tablewidth{0pt}

\tablehead{
Source{\dag} & R.A. (J2000)\tablenotemark{\dag} & Decl. (J2000)\tablenotemark{\dag} & $V_{sys}$\tablenotemark{a} & Date & 22\ghz~& 44\ghz~&95\ghz~& Fig.\\
(l, b) & (h m s) & (\arcdeg \arcmin \arcsec) & (\kms ) & (yyyymmdd) & water& class I& class I&No.\\
}
\startdata
G34.82+0.35 & 18~53~37.4\tablenotemark{\ddag} & 01~50~32\tablenotemark{\ddag}       & 56.6  & 20111216 & &   &    &\\
G35.03+0.35 & 18~54~00.658 & 02~01~19.23        & 53.1  &  20111216 & (4) & {(9)}  & New    &\ref{fig:all}\\
G35.25-0.24 & 18~56~30.388 & 01~57~08.88    & 61.6  &   20111216 &  & {New} &   & \ref{fig:44only} \\
G35.39+0.02\tablenotemark{*} & 18~55~51.2\tablenotemark{\ddag} & 02~11~37\tablenotemark{\ddag}  & 93.9  &   20111216 &  & New     & & \ref{fig:44only}\\
G35.40+0.03\tablenotemark{*} & 18~55~50.799 & 02~12~19.08    & 94.6  &  20111216 &   &    &    &\\
G35.59+0.06 & 18~56~04.219 & 02~23~28.34    & 48.8  &  20111216 & {(10)} & New   & New    & \ref{fig:all}\\
G35.79-0.17 & 18~57~16.892 & 02~27~58.05    & 61.3  & 20111217 &  New   & New   &   &\ref{fig:2244} \\
G36.02-0.20 & 18~57~45.868 & 02~39~05.67    & 86.8  &   20111217 &  &   &    &\\
G36.64-0.21 & 18~58~55.236 & 03~12~04.72    & 74.8  & 20111217 &  New   &     & & \ref{fig:22only} \\
G36.70+0.09 & 18~57~59.123 & 03~24~06.12    & 59.4  &  20120127 & New$^{(8)}$   &   & & \ref{fig:22only}\\
G36.84-0.02 & 18~58~39.214 & 03~28~00.89    & 59.0\tablenotemark{c}  &   20111217 &  &   &    &\\
G36.90-0.41 & 19~00~08.6\tablenotemark{\ddag} & 03~20~35\tablenotemark{\ddag}   & 79.6  &  20111218 & New      &   &  & \ref{fig:22only}\\
G36.92+0.48 & 18~56~59.786 & 03~46~03.60    & -30.7\tablenotemark{b}  &  20111218 &   &   &   & \\
G37.02-0.03\tablenotemark{*} & 18~59~03.642 & 03~37~45.08    & 80.1  &  20120227 & New{$^{(8)}$} &   &  & \ref{fig:22only}  \\
G37.04-0.04\tablenotemark{*} & 18~59~04.406 & 03~38~32.77    & 80.8  &  20120227 & New{$^{(8)}$} &   &   & \ref{fig:22only} \\
G37.38-0.09 & 18~59~51.586 & 03~55~18.02    & 57.1  &   20111217 &  & New   &    &\ref{fig:44only}\\
G37.47-0.11 & 19~00~07.144 & 03~59~53.08    & 58.4  &  20111120 & (8) &   &    & \ref{fig:22only}\\
G37.53-0.11 & 19~00~16.056 & 04~03~16.09    & 51.7  &   20120227 &  &   &    &\\
G37.55+0.19 & 18~59~09.985 & 04~12~15.54    & 84.6  &  20111120 & (8){(10)}   & New   & New$^{{(11)}}$   &\ref{fig:all}\\
G37.60+0.42 & 18~58~26.799 & 04~20~45.47    & 89.2  &  20111217 & (8) & New   & New    &\ref{fig:all}\\
G37.74-0.12 & 19~00~36.841 & 04~13~19.98    & 45.5  &  20111218 & (2)(5) &   &    &\ref{fig:22only}\\
G37.76-0.19 & 19~00~55.421 &  04~12~12.56    & 59.6  &  20111218 & (2)(5) &   &    &\ref{fig:22only}\\
G37.77-0.22 & 19~01~02.268 & 04~12~16.55    & 64.1  &  20111218 & (2)(5) & New   & (13)    &\ref{fig:all}\\
G38.03-0.30 & 19~01~50.470 & 04~24~18.94    & 61.6  & 20111120 &  New$^{(8)}$ &   &    &\ref{fig:22only}\\
G38.08-0.27 & 19~01~47.317 & 04~27~20.90    & 64.7  &    20120128 & &   &    &\\
G38.12-0.24 & 19~01~44.152 & 04~30~37.42    & 82.7  &    20120128 & &   &    &\\
G38.20-0.08 & 19~01~18.730 & 04~39~34.32    & 82.9  &    20120128 & &   &    &\\
G38.26-0.08 & 19~01~26.233 & 04~42~17.26    & 11.6\tablenotemark{b}   &  20120128 & {(10)}    &   &  &\ref{fig:22only}\\
G38.26-0.20 & 19~01~52.956 & 04~38~39.47    & 65.4  &    20120128 & &   &    &\\
G38.56+0.15 & 19~01~08.345 & 05~04~36.71    & 28.2  &  20120128 & {(10)} &   &  &\ref{fig:22only}  \\
G38.60-0.21 & 19~02~33.461 & 04~56~36.37    & 66.1  &  20120128 & New   &      &  &\ref{fig:22only}\\
G38.66+0.08 & 19~01~35.244 & 05~07~47.36    & -39.1 &  20120227 &   &   &    &\\
G38.92-0.36 & 19~03~38.659 & 05~09~42.49    & 37.9  &  20120128 & (3)(5)   & New   & New &\ref{fig:all}   \\
G39.39-0.14 & 19~03~45.312 & 05~40~42.68    & 66.0    &  20120129 &   & New   & {(11)}    &\ref{fig:4495}\\
G39.54-0.38 & 19~04~53.5\tablenotemark{\ddag} & 05~41~59\tablenotemark{\ddag}   & 60.4     &    20120130 & &   &  &\\
G40.28-0.22 & 19~05~41.215 & 06~26~12.69    & 73.1  &  20120129 & New$^{{(10)}}$ & New   & {(11)}    & \ref{fig:all}\\
G40.62-0.14 & 19~06~01.630 &  06~46~36.18    & 32.5  &  20120129 & (4){(10)}   &   &    &\ref{fig:22only}\\
G40.94-0.04 & 19~06~15.378 & 07~05~54.49    & 39.3  &   20111226 &  &   &    &\\
G41.08-0.13 & 19~06~49.047 & 07~11~06.57    & 63.3  &  20111226 & New   &   &   &\ref{fig:22only} \\
G41.12-0.11 & 19~06~50.248 & 07~14~01.49    & 37.8  &   20111226 &  &   &    &\\
G41.12-0.22 & 19~07~14.856 & 07~11~00.69    & 59.7  &   20111226 &  & New      & &\ref{fig:44only}\\
G41.16-0.20 & 19~07~14.369 & 07~13~18.08    & 60.4  &   20111226 &  &   &    &\\
G41.23-0.20 & 19~07~21.378 & 07~17~08.17    & 58.9  &  20111230 &   &   &    &\\
G41.27+0.37 & 19~05~23.606 & 07~35~05.25    & 14.9  &  20111230 &   &   &   & \\
G41.34-0.14 & 19~07~21.842 & 07~25~17.27    & 13.2  &  20111120 & New   &  &    &\ref{fig:22only} \\
G41.58+0.04 & 19~07~09.178 & 07~42~25.24    & 12.4\tablenotemark{b}   &   20111230 &  &   &  &\\
G42.03+0.19 & 19~07~28.185 & 08~10~53.47    & 17.7  &  20111230 & New$^{{(10)}}$ &      &  &\ref{fig:22only}\\
G42.30-0.30 & 19~09~43.592 & 08~11~41.41    & 27.8  & 20111230 &  New   &   &    &\ref{fig:22only}\\
G42.43-0.26 & 19~09~49.858 & 08~19~45.40    & 64.7  &  20111230 & {(10)} & New   & New    &\ref{fig:all}\\
G42.70-0.15 & 19~09~55.069 & 08~36~53.45    & -44.4 &    20111230 & &   &    &\\
G43.04-0.46 & 19~11~38.984 & 08~46~30.71    & 57.4  &  20120130 & (5) & (7)(12)   & (13)  &\ref{fig:all} \\
G43.08-0.08 & 19~10~22.050 & 08~58~51.49    & 12.6  &    20120130 & &   &   &\\
G43.80-0.13 & 19~11~53.990 & 09~35~50.61    & 43.7  &  20120130 & (1)(4)(5){(10)}   & New   & New  &\ref{fig:all}\\
G44.31+0.04 & 19~12~15.816 & 10~07~53.52   & 56.2  &  20111218 & {(10)} & New   & New   &\ref{fig:all}\\
G44.64-0.52 & 19~14~53.766 & 10~10~07.69   & 46.0\tablenotemark{b}   &  20120129 &   &   &   &\\
G45.07+0.13 & 19~13~22.129 & 10~50~53.11   & 59.2\tablenotemark{b}   &  20120129 & (1)(4)(5){(10)}   & (7)(12)   & New &\ref{fig:all} \\
G45.44+0.07 & 19~14~18.291 & 11~08~58.97   & 59.0\tablenotemark{c}    &  20120129 & (2)(4)(6){(10)}   &   &   &\ref{fig:22only}\\
G45.47+0.05 & 19~14~24.147 & 11~09~43.43   & 57.1  & 20120130 &  (2)(4)(6){(10)}   &   &  &\ref{fig:22only}\\
G45.47+0.13 & 19~14~07.362 & 11~12~15.98   & 61.6\tablenotemark{c}  &  20120130 & (4)(6) &   &  &\ref{fig:22only} \\
G45.49+0.13 & 19~14~11.357 & 11~13~06.41   & 59.5\tablenotemark{c}  &  20120130 & (4)(6) & New   & New$^{{(11)}}$  &\ref{fig:all}  \\
G45.57-0.12 & 19~15~13.152 & 11~10~16.54   & 4.5   & 20120130 &  New   &   &    &\ref{fig:22only}\\
G45.81-0.36 & 19~16~31.081 & 11~16~12.01   & 58.7  &  20120130 & New   & New   & {(11)} &\ref{fig:all} \\
G46.07+0.22 & 19~14~56.077 & 11~46~12.98   & 18.8  &   20111216 &  &   &    &\\
G46.12+0.38 & 19~14~25.520 & 11~53~25.99   & 55.0    &  20111225 & New   &      & &\ref{fig:22only} \\
G48.89-0.17 & 19~21~47.5\tablenotemark{\ddag} & 14~04~58\tablenotemark{\ddag}  & 55.7     &  20111225 &   &   & &\\
G48.90-0.27 & 19~22~10.330 & 14~02~43.51   & 68.4  & 20120130 &  (2)(5) & New      & &\ref{fig:2244}\\
G48.99-0.30 & 19~22~26.134 & 14~06~39.78   & 67.4  &  20111225 & (2)(5){(10)}   & New   & New    &\ref{fig:all}\\
G49.27+0.31 & 19~20~44.859 & 14~38~26.91   & 3.3   &   20111225 &  & New   & New    &\ref{fig:4495}\\
G49.35+0.41 & 19~20~32.449 & 14~45~45.44   & 66.2  &  20111225 &   &   &    &\\
G49.41+0.33 & 19~20~59.211 & 14~46~49.66     & -21.3 &   20111225 &  &      & &\\
G49.60-0.25 & 19~23~26.611 & 14~40~16.99   & 56.7  &   20111225 &  &   &   & \\
G49.62-0.36 & 19~23~52.805 & 14~38~03.25   & 54.5  &   20120128 &  &   &    &\\
G50.78+0.15 & 19~24~17.411 & 15~54~01.60   & 42.1  &    20120130 & &   &   & \\
G52.92+0.41 & 19~27~34.960 & 17~54~38.14   & 44.7  &   20111120 &  &   &   & \\
G53.04+0.11 & 19~28~55.494 & 17~52~03.11   & 4.8   &   20111218 &  & {(7)}(12) & New  &\ref{fig:4495}  \\
G53.14+0.07 & 19~29~17.581 & 17~56~23.21   & 21.7  &  20120129 & {(10)} & New   & (13)  &\ref{fig:all}  \\
G53.62+0.04 & 19~30~23.016 & 18~20~26.68   & 22.8  &   20111217 &  &   &    &\\
\enddata
\tablenotetext{\dag}{The position and name from the accurate astrometry \citep{pand11}}
\tablenotetext{\ddag}{The position from the AMGPS catalog \citep{pand07}}
\tablenotetext{a}{The systemic velocity from the NH$_{3}$ observation \citep{pand12}}
\tablenotetext{b}{The systemic velocity from the  $^{13}$CO ($J$ = 2$-$1) or CS ($J$ = 5$-$4) lines \citep{pand09}}
\tablenotetext{c}{The systemic velocity from the NH$_{3}$ observation \citep{pand12} that is near the previous result\citep{pand09}}
\tablenotetext{*}{Overlapped sources at 22 and 44\ghz}
\tablerefs{(1)  \citealt{palagi93}; (2)  \citealt{codella94}; (3)  \citealt{han98}; (4)  \citealt{forster99}; (5)  \citealt{valdettaro01}; (6)  \citealt{codella04}; (7)  \citealt{kurtz04}; (8)  \citealt{szymczak05}; {(9)}  \citealt{cyg09}; {(10)}  \citealt{urquhart11}; {(11)} \citealt{chen11}; {(12) } \citealt{bayandina12}; (13) \citealt{chen12}.}
\end{deluxetable}

\clearpage
\begin{deluxetable}{ccccc}
\tabletypesize{\scriptsize} 
\tablecaption{{KVN Antenna Parameters\label{Table:units}}}
\tablewidth{0pt}
\tablehead{
\multirow{3}{*}{Site}
& \multirow{2}{*}{Frequency} & Aperture & \multirow{2}{*}{HPBW} &Conversion\\
& & efficiency &  &factor\\
& (GHz)  &(\%)&($\arcsec$)&(Jy K$^{-1}$)}
\startdata
\multirow{3}{*}{Yonsei}
&
22 & 65 &119& {12.3}\\
&44 & 63 &62&12.7\\
&95 & 48 &32& {16.6}\\
\hline
\multirow{3}{*}{Ulsan}
&
22 & 62 &120& {12.9}\\
&44 & 62& 62& 12.9\\
&95 & 49 &33& {16.3}\\
\hline
\enddata
\end{deluxetable}

\clearpage
\begin{deluxetable}{cccccccccccc}	
\rotate
\tabletypesize{\tiny} 
\tablecaption{{Line Parameters of Detected Masers\label{Table:flux}}}
\tablewidth{0pt}

\tablehead{
&&\multicolumn{4}{c}{Gaussian Fit} &
&
\multicolumn{5}{c}{Integrated Fit}\\
\cline{3-6} \cline{8-12}
\colhead{Source} &
\colhead{Frequency} &
\colhead{{$\rm\bf S_{int}$}} &
\colhead{$\rm V_{peak}$} &
\colhead{FWHM} &
\colhead{{$\rm\bf S_{peak}$}} &
&
\colhead{{$\rm\bf S_{int}$}} &
\colhead{$\rm v_{min}$} &
\colhead{$\rm v_{max}$} &
\colhead{{$\rm\bf S_{peak}$}} &
\colhead{$\rm V_{peak}$} \\
&
\colhead{(GHz)}&
\colhead{($\rm Jy\,km\,s^{-1}$)} &
\colhead{$\rm (km\,s^{-1})$} &
\colhead{$\rm (km\,s^{-1})$} &
\colhead{$\rm (Jy)$} &
&
\colhead{($\rm Jy\,km\,s^{-1}$)} &
\colhead{$\rm (km\,s^{-1})$} &
\colhead{$\rm (km\,s^{-1})$} &
\colhead{$\rm (Jy)$} &
\colhead{$\rm (km\,s^{-1})$}
 \\
}
\startdata
G35.03+0.35	&	22	&	6.5	$\pm$	0.8	&	40.6	$\pm$	0.0	&	0.7	$\pm$	0.1	&	9.2	$\pm$	0.9	&	&	7.2	&	39.3	&	41.1	&	9.0	&	40.7	\\
	&	22	&	7.5	$\pm$	0.8	&	41.7	$\pm$	0.1	&	0.9	$\pm$	0.1	&	7.9	$\pm$	0.9	&	&	7.4	&	41.1	&	42.7	&	7.5	&	41.7	\\
	&	22	&	47.7	$\pm$	0.5	&	44.7	$\pm$	0.0	&	0.7	$\pm$	0.0	&	62.5	$\pm$	0.9	&	&	49.7	&	43.3	&	46.4	&	64.1	&	44.7	\\
	&	22	&	3.1	$\pm$	4.6	&	55.0	$\pm$	0.6	&	0.9	$\pm$	1.6	&	3.2	$\pm$	4.9	&	&	3.5	&	53.7	&	55.6	&	3.7	&	55.0	\\
	&	22	&	4.6	$\pm$	5.4	&	56.6	$\pm$	0.7	&	1.5	$\pm$	2.3	&	2.9	$\pm$	4.9	&	&	4.8	&	55.6	&	58.2	&	3.0	&	56.7	\\
	&	22	&	3.9	$\pm$	3.0	&	59.2	$\pm$	0.3	&	0.7	$\pm$	0.7	&	5.1	$\pm$	4.9	&	&	4.1	&	58.2	&	59.9	&	5.8	&	59.2	\\
	&	44	&	0.7	$\pm$	0.4	&	50.7	$\pm$	0.1	&	0.4	$\pm$	0.2	&	1.5	$\pm$	0.6	&	&	0.9	&	50.1	&	51.0	&	1.6	&	50.6	\\
	&	44	&	1.8	$\pm$	0.5	&	51.5	$\pm$	0.1	&	0.8	$\pm$	0.3	&	2.1	$\pm$	0.6	&	&	1.7	&	51.0	&	52.1	&	2.1	&	51.5	\\
	&	44	&	1.7	$\pm$	0.2	&	52.8	$\pm$	0.0	&	0.4	$\pm$	0.1	&	4.4	$\pm$	0.6	&	&	1.8	&	52.1	&	53.3	&	4.3	&	52.7	\\
	&	95	&	2.7	$\pm$	0.5	&	51.7	$\pm$	0.1	&	1.0	$\pm$	0.2	&	2.4	$\pm$	0.7	&	&	2.9	&	50.7	&	52.4	&	3.1	&	51.4	\\
	&	95	&	2.3	$\pm$	0.4	&	52.9	$\pm$	0.0	&	0.5	$\pm$	0.1	&	4.2	$\pm$	0.7	&	&	2.2	&	52.4	&	53.6	&	3.9	&	53.0	\\
G35.25-0.24	&	44	&	1.6	$\pm$	0.3	&	62.2	$\pm$	0.0	&	0.5	$\pm$	0.1	&	3.0	$\pm$	0.6	&	&	1.5	&	61.5	&	62.8	&	2.8	&	62.2	\\
	&	44	&	1.5	$\pm$	0.4	&	80.9	$\pm$	0.1	&	0.7	$\pm$	0.3	&	1.9	$\pm$	0.6	&	&	1.4	&	80.2	&	81.8	&	1.9	&	80.9	\\
G35.39+0.02	&	44	&	1.0	$\pm$	0.2	&	93.7	$\pm$	0.0	&	0.4	$\pm$	0.1	&	2.3	$\pm$	0.5	&	&	1.3	&	93.2	&	94.1	&	2.4	&	93.7	\\
	&	44	&	1.1	$\pm$	0.2	&	94.7	$\pm$	0.0	&	0.3	$\pm$	0.1	&	3.4	$\pm$	0.5	&	&	1.1	&	94.1	&	95.6	&	3.5	&	94.7	\\
G35.59+0.06	&	22	&	8.0	$\pm$	0.4	&	-11.5	$\pm$	0.0	&	1.6	$\pm$	0.1	&	4.8	$\pm$	0.5	&	&	8.0	&	-13.5	&	-9.0	&	4.9	&	-11.7	\\
	&	22	&	7.9	$\pm$	0.5	&	-3.4	$\pm$	0.1	&	1.9	$\pm$	0.1	&	3.9	$\pm$	0.5	&	&	7.9	&	-5.9	&	-1.2	&	4.4	&	-3.5	\\
	&	44	&	4.8	$\pm$	0.5	&	47.9	$\pm$	0.1	&	1.2	$\pm$	0.1	&	3.7	$\pm$	0.5	&	&	5.5	&	45.8	&	48.8	&	4.4	&	47.8	\\
	&	44	&	6.1	$\pm$	0.4	&	49.4	$\pm$	0.0	&	0.8	$\pm$	0.1	&	6.8	$\pm$	0.5	&	&	7.0	&	48.8	&	51.4	&	8.4	&	49.3	\\
	&	95	&	7.7	$\pm$	0.9	&	49.2	$\pm$	0.2	&	3.0	$\pm$	0.5	&	2.4	$\pm$	0.7	&	&	7.9	&	45.7	&	51.6	&	3.3	&	49.5	\\
G35.79-0.17	&	22	&	12.5	$\pm$	0.6	&	59.3	$\pm$	0.0	&	2.0	$\pm$	0.1	&	5.9	$\pm$	0.6	&	&	11.1	&	56.9	&	61.2	&	5.7	&	59.9	\\
	&	22	&	7.8	$\pm$	0.5	&	62.9	$\pm$	0.0	&	1.4	$\pm$	0.1	&	5.3	$\pm$	0.6	&	&	7.2	&	61.2	&	64.7	&	4.9	&	63.1	\\
	&	44	&	2.3	$\pm$	0.5	&	59.9	$\pm$	0.1	&	0.7	$\pm$	0.2	&	3.1	$\pm$	0.7	&	&	3.0	&	59.0	&	60.7	&	4.2	&	59.9	\\
	&	44	&	7.7	$\pm$	0.7	&	61.9	$\pm$	0.1	&	1.9	$\pm$	0.2	&	3.7	$\pm$	0.7	&	&	7.2	&	60.7	&	64.4	&	3.9	&	62.1	\\
G36.64-0.21	&	22	&	12.3	$\pm$	0.3	&	77.5	$\pm$	0.0	&	0.8	$\pm$	0.0	&	14.5	$\pm$	0.5	&	&	13.1	&	75.9	&	79.1	&	15.0	&	77.4	\\
G36.70+0.09	&	22	&	4.2	$\pm$	0.5	&	52.9	$\pm$	0.1	&	1.2	$\pm$	0.2	&	3.2	$\pm$	0.7	&	&	4.6	&	51.4	&	54.8	&	3.8	&	52.6	\\
G36.90-0.41	&	22	&	2.9	$\pm$	0.4	&	84.9	$\pm$	0.0	&	0.6	$\pm$	0.1	&	4.7	$\pm$	0.7	&	&	2.4	&	84.1	&	86.1	&	4.5	&	85.1	\\
G37.02-0.03	&	22	&	1.9	$\pm$	0.3	&	80.7	$\pm$	0.1	&	0.9	$\pm$	0.2	&	2.1	$\pm$	0.4	&	&	2.4	&	79.7	&	81.4	&	2.1	&	80.7	\\
	&	22	&	0.7	$\pm$	0.2	&	81.6	$\pm$	0.1	&	0.4	$\pm$	0.2	&	1.6	$\pm$	0.4	&	&	0.8	&	81.4	&	82.3	&	1.7	&	81.5	\\
G37.04-0.04	&	22	&	1.1	$\pm$	0.3	&	80.8	$\pm$	0.1	&	0.9	$\pm$	0.3	&	1.2	$\pm$	0.3	&	&	1.1	&	79.7	&	81.2	&	1.5	&	80.7	\\
	&	22	&	0.7	$\pm$	0.2	&	81.8	$\pm$	0.0	&	0.3	$\pm$	0.1	&	2.1	$\pm$	0.3	&	&	0.9	&	81.2	&	82.4	&	2.1	&	81.8	\\
G37.38-0.09	&	44	&	2.8	$\pm$	0.4	&	56.0	$\pm$	0.1	&	0.8	$\pm$	0.1	&	3.2	$\pm$	0.6	&	&	3.5	&	54.7	&	57.6	&	3.3	&	55.9	\\
G37.47-0.11	&	22	&	1.5	$\pm$	0.2	&	56.8	$\pm$	0.0	&	0.5	$\pm$	0.1	&	3.1	$\pm$	0.4	&	&	1.1	&	55.4	&	58.0	&	3.2	&	56.8	\\
G37.55+0.19	&	22	&	7.5	$\pm$	0.2	&	83.7	$\pm$	0.0	&	0.7	$\pm$	0.0	&	10.6	$\pm$	0.4	&	&	7.8	&	82.5	&	84.4	&	10.3	&	83.7	\\
	&	22	&	16.9	$\pm$	0.3	&	85.7	$\pm$	0.0	&	1.1	$\pm$	0.0	&	14.6	$\pm$	0.4	&	&	17.2	&	84.4	&	87.7	&	14.5	&	85.6	\\
	&	22	&	3.2	$\pm$	0.3	&	94.5	$\pm$	0.1	&	1.4	$\pm$	0.2	&	2.1	$\pm$	0.4	&	&	3.2	&	93.1	&	96.0	&	2.3	&	94.5	\\
	&	44	&	1.1	$\pm$	0.3	&	84.0	$\pm$	0.1	&	0.5	$\pm$	0.2	&	1.9	$\pm$	0.5	&	&	1.3	&	83.3	&	84.6	&	1.8	&	84.0	\\
	&	44	&	1.6	$\pm$	0.4	&	85.1	$\pm$	0.1	&	0.8	$\pm$	0.2	&	1.8	$\pm$	0.5	&	&	2.0	&	84.6	&	85.8	&	1.8	&	85.5	\\
	&	44	&	4.7	$\pm$	0.4	&	86.6	$\pm$	0.0	&	1.2	$\pm$	0.1	&	3.8	$\pm$	0.5	&	&	4.7	&	85.8	&	88.6	&	4.4	&	86.5	\\
	&	95	&	0.9	$\pm$	0.3	&	84.0	$\pm$	0.1	&	0.5	$\pm$	0.2	&	1.5	$\pm$	0.4	&	&	1.2	&	83.0	&	84.5	&	1.6	&	84.1	\\
	&	95	&	1.3	$\pm$	0.6	&	85.2	$\pm$	0.2	&	1.3	$\pm$	0.5	&	1.0	$\pm$	0.4	&	&	1.5	&	84.5	&	86.3	&	1.7	&	85.5	\\
	&	95	&	1.7	$\pm$	0.6	&	87.0	$\pm$	0.3	&	1.9	$\pm$	0.7	&	0.8	$\pm$	0.4	&	&	1.3	&	86.3	&	88.6	&	1.3	&	86.7	\\
G37.60+0.42	&	22	&	2.1	$\pm$	0.7	&	97.4	$\pm$	0.1	&	0.6	$\pm$	0.1	&	3.5	$\pm$	0.5	&	&	3.5	&	96.0	&	98.1	&	3.9	&	97.4	\\
	&	22	&	2.9	$\pm$	1.0	&	98.5	$\pm$	0.2	&	1.3	$\pm$	0.6	&	2.0	$\pm$	0.5	&	&	2.5	&	98.1	&	100.0	&	2.8	&	98.7	\\
	&	22	&	7.4	$\pm$	0.4	&	137.0	$\pm$	0.0	&	1.3	$\pm$	0.1	&	5.3	$\pm$	0.5	&	&	7.4	&	135.3	&	139.1	&	5.2	&	137.2	\\
	&	44	&	3.0	$\pm$	0.3	&	89.8	$\pm$	0.0	&	0.7	$\pm$	0.1	&	4.2	$\pm$	0.5	&	&	2.8	&	87.4	&	90.5	&	4.5	&	89.9	\\
	&	95	&	1.7	$\pm$	0.4	&	89.8	$\pm$	0.1	&	0.7	$\pm$	0.2	&	2.5	$\pm$	0.6	&	&	2.3	&	88.1	&	91.6	&	2.8	&	89.7	\\
G37.74-0.12	&	22	&	7.5	$\pm$	0.5	&	42.0	$\pm$	0.2	&	1.0	$\pm$	0.2	&	7.2	$\pm$	0.7	&	&	11.1	&	40.3	&	43.1	&	7.7	&	42.2	\\
	&	22	&	18.7	$\pm$	0.5	&	43.8	$\pm$	0.2	&	1.0	$\pm$	0.2	&	17.2	$\pm$	0.7	&	&	20.5	&	43.1	&	45.2	&	18.3	&	43.7	\\
	&	22	&	10.2	$\pm$	0.5	&	46.0	$\pm$	0.2	&	1.8	$\pm$	0.2	&	5.4	$\pm$	0.7	&	&	8.3	&	45.2	&	47.2	&	6.7	&	46.0	\\
	&	22	&	7.8	$\pm$	0.5	&	52.7	$\pm$	0.2	&	1.4	$\pm$	0.2	&	5.3	$\pm$	0.7	&	&	7.7	&	51.4	&	54.2	&	6.3	&	52.8	\\
G37.76-0.19	&	22	&	1.9	$\pm$	0.4	&	54.6	$\pm$	0.1	&	0.5	$\pm$	0.1	&	3.8	$\pm$	0.8	&	&	2.1	&	53.3	&	55.2	&	3.5	&	54.7	\\
	&	22	&	23.8	$\pm$	0.6	&	60.1	$\pm$	0.0	&	1.1	$\pm$	0.0	&	19.9	$\pm$	0.8	&	&	24.5	&	58.0	&	62.1	&	20.0	&	60.1	\\
G37.77-0.22	&	22	&	2.4	$\pm$	0.4	&	58.1	$\pm$	0.1	&	1.6	$\pm$	0.3	&	1.5	$\pm$	0.4	&	&	2.1	&	56.6	&	59.0	&	1.4	&	57.8	\\
	&	22	&	3.3	$\pm$	0.4	&	60.2	$\pm$	0.1	&	1.3	$\pm$	0.2	&	2.4	$\pm$	0.4	&	&	3.2	&	59.0	&	61.8	&	2.5	&	60.1	\\
	&	22	&	1.1	$\pm$	0.3	&	64.5	$\pm$	0.1	&	0.6	$\pm$	0.2	&	1.6	$\pm$	0.4	&	&	0.9	&	63.7	&	65.7	&	1.7	&	64.5	\\
	&	22	&	1.6	$\pm$	0.3	&	66.7	$\pm$	0.1	&	0.8	$\pm$	0.2	&	2.0	$\pm$	0.4	&	&	1.7	&	65.7	&	67.1	&	1.9	&	66.9	\\
	&	22	&	2.4	$\pm$	0.4	&	67.8	$\pm$	0.1	&	1.0	$\pm$	0.2	&	2.4	$\pm$	0.4	&	&	2.1	&	67.1	&	68.8	&	2.5	&	67.7	\\
	&	44	&	4.2	$\pm$	0.6	&	63.5	$\pm$	0.1	&	1.8	$\pm$	0.3	&	2.2	$\pm$	0.4	&	&	3.6	&	62.6	&	64.5	&	2.5	&	64.0	\\
	&	44	&	3.1	$\pm$	0.4	&	65.3	$\pm$	0.0	&	0.9	$\pm$	0.1	&	3.4	$\pm$	0.4	&	&	3.5	&	64.5	&	66.3	&	3.9	&	65.2	\\
	&	95	&	3.1	$\pm$	0.7	&	63.5	$\pm$	0.1	&	1.3	$\pm$	0.2	&	2.3	$\pm$	0.6	&	&	4.1	&	61.8	&	64.9	&	2.8	&	63.7	\\
	&	95	&	4.0	$\pm$	0.9	&	65.6	$\pm$	0.2	&	2.2	$\pm$	0.6	&	1.7	$\pm$	0.6	&	&	5.2	&	64.9	&	68.8	&	3.1	&	65.3	\\
G38.03-0.30	&	22	&	9.3	$\pm$	1.2	&	63.6	$\pm$	0.1	&	2.0	$\pm$	0.3	&	4.3	$\pm$	0.5	&	&	8.3	&	60.6	&	64.4	&	4.6	&	63.4	\\
	&	22	&	1.7	$\pm$	0.9	&	65.1	$\pm$	0.1	&	0.7	$\pm$	0.2	&	2.2	$\pm$	0.5	&	&	3.9	&	64.4	&	66.3	&	3.4	&	64.6	\\
G38.26-0.08	&	22	&	49.5	$\pm$	0.7	&	9.8	$\pm$	0.0	&	1.2	$\pm$	0.0	&	39.7	$\pm$	0.9	&	&	47.4	&	7.8	&	12.1	&	38.7	&	9.8	\\
G38.56+0.15	&	22	&	4.3	$\pm$	0.6	&	29.8	$\pm$	0.1	&	1.5	$\pm$	0.2	&	2.7	$\pm$	0.6	&	&	4.4	&	28.2	&	31.5	&	4.2	&	29.8	\\
G38.60-0.21	&	22	&	2.3	$\pm$	0.4	&	62.9	$\pm$	0.1	&	1.0	$\pm$	0.2	&	2.3	$\pm$	0.6	&	&	3.1	&	62.1	&	64.2	&	2.7	&	63.1	\\
	&	22	&	6.9	$\pm$	0.6	&	65.3	$\pm$	0.1	&	1.5	$\pm$	0.1	&	4.4	$\pm$	0.6	&	&	6.4	&	64.2	&	66.2	&	5.3	&	65.7	\\
	&	22	&	8.5	$\pm$	0.5	&	67.3	$\pm$	0.0	&	1.0	$\pm$	0.1	&	8.0	$\pm$	0.6	&	&	8.9	&	66.2	&	69.1	&	7.6	&	67.1	\\
G38.92-0.36	&	22	&	47.4	$\pm$	0.4	&	33.2	$\pm$	0.0	&	0.8	$\pm$	0.0	&	54.9	$\pm$	0.6	&	&	48.9	&	32.1	&	35.5	&	58.6	&	33.1	\\
	&	22	&	2.1	$\pm$	0.3	&	37.6	$\pm$	0.0	&	0.6	$\pm$	0.1	&	3.4	$\pm$	0.6	&	&	2.1	&	36.8	&	38.1	&	3.5	&	37.5	\\
	&	22	&	2.1	$\pm$	0.3	&	38.7	$\pm$	0.0	&	0.5	$\pm$	0.1	&	3.8	$\pm$	0.6	&	&	2.1	&	38.1	&	39.2	&	3.5	&	38.6	\\
	&	44	&	12.6	$\pm$	1.7	&	37.7	$\pm$	0.2	&	3.9	$\pm$	0.7	&	3.1	$\pm$	1.1	&	&	10.1	&	35.7	&	41.9	&	5.5	&	37.4	\\
	&	95	&	14.0	$\pm$	1.0	&	37.9	$\pm$	0.1	&	3.8	$\pm$	0.4	&	3.5	$\pm$	0.7	&	&	13.5	&	34.1	&	41.3	&	4.9	&	38.1	\\
G39.39-0.14	&	44	&	2.6	$\pm$	0.6	&	66.3	$\pm$	0.1	&	1.0	$\pm$	0.3	&	2.4	$\pm$	0.6	&	&	2.5	&	64.6	&	67.6	&	2.5	&	66.5	\\
	&	95	&	3.5	$\pm$	0.3	&	66.3	$\pm$	0.1	&	1.2	$\pm$	0.1	&	2.8	$\pm$	0.5	&	&	3.3	&	65.0	&	67.6	&	3.0	&	66.5	\\
G40.28-0.22	&	22	&	41.5	$\pm$	0.6	&	71.6	$\pm$	0.0	&	1.0	$\pm$	0.0	&	40.3	$\pm$	0.9	&	&	46.5	&	70.4	&	72.5	&	41.1	&	71.4	\\
	&	22	&	90.7	$\pm$	0.7	&	73.7	$\pm$	0.0	&	1.2	$\pm$	0.0	&	71.3	$\pm$	0.9	&	&	89.5	&	72.5	&	75.2	&	76.2	&	73.7	\\
	&	22	&	7.4	$\pm$	0.8	&	84.2	$\pm$	0.1	&	1.5	$\pm$	0.2	&	4.6	$\pm$	0.9	&	&	7.5	&	83.2	&	85.7	&	6.6	&	84.1	\\
	&	44	&	58.1	$\pm$	2.2	&	72.7	$\pm$	0.2	&	0.6	$\pm$	0.2	&	89.2	$\pm$	1.4	&	&	88.7	&	71.6	&	73.3	&	101.8	&	72.7	\\
	&	44	&	67.8	$\pm$	2.2	&	73.6	$\pm$	0.2	&	1.2	$\pm$	0.2	&	53.6	$\pm$	1.4	&	&	49.4	&	73.3	&	74.5	&	53.6	&	73.4	\\
	&	44	&	14.4	$\pm$	2.2	&	75.2	$\pm$	0.2	&	0.6	$\pm$	0.2	&	21.8	$\pm$	1.4	&	&	15.6	&	74.5	&	75.7	&	21.2	&	75.1	\\
	&	95	&	59.1	$\pm$	1.7	&	72.8	$\pm$	0.2	&	0.8	$\pm$	0.2	&	72.7	$\pm$	1.0	&	&	78.4	&	71.6	&	73.6	&	78.4	&	72.8	\\
	&	95	&	34.7	$\pm$	1.7	&	73.8	$\pm$	0.2	&	1.1	$\pm$	0.2	&	30.7	$\pm$	1.0	&	&	23.8	&	73.6	&	74.9	&	28.9	&	73.8	\\
	&	95	&	11.9	$\pm$	1.7	&	75.3	$\pm$	0.2	&	0.8	$\pm$	0.2	&	14.6	$\pm$	1.0	&	&	10.9	&	74.9	&	75.8	&	15.1	&	75.1	\\
G40.62-0.14	&	22	&	13.6	$\pm$	0.7	&	33.9	$\pm$	0.0	&	1.2	$\pm$	0.1	&	10.3	$\pm$	0.8	&	&	14.8	&	32.1	&	36.8	&	10.6	&	34.1	\\
G41.08-0.13	&	22	&	0.8	$\pm$	0.1	&	-16.9	$\pm$	0.2	&	0.9	$\pm$	0.2	&	0.8	$\pm$	0.3	&	&	1.2	&	-18.2	&	-16.6	&	1.2	&	-16.8	\\
	&	22	&	1.3	$\pm$	0.1	&	-15.8	$\pm$	0.2	&	1.0	$\pm$	0.2	&	1.3	$\pm$	0.3	&	&	1.9	&	-16.6	&	-15.5	&	1.8	&	-15.3	\\
	&	22	&	1.4	$\pm$	0.1	&	-15.0	$\pm$	0.2	&	0.7	$\pm$	0.2	&	1.9	$\pm$	0.3	&	&	2.8	&	-15.5	&	-14.4	&	3.0	&	-14.3	\\
	&	22	&	2.7	$\pm$	0.1	&	-14.1	$\pm$	0.2	&	0.9	$\pm$	0.2	&	2.8	$\pm$	0.3	&	&	2.7	&	-14.4	&	-13.2	&	3.0	&	-14.3	\\
	&	22	&	1.0	$\pm$	0.1	&	-13.0	$\pm$	0.2	&	0.8	$\pm$	0.2	&	1.3	$\pm$	0.3	&	&	0.9	&	-13.2	&	-11.9	&	1.5	&	-13.0	\\
G41.12-0.22	&	44	&	2.2	$\pm$	0.5	&	59.0	$\pm$	0.1	&	0.9	$\pm$	0.3	&	2.4	$\pm$	0.7	&	&	2.4	&	57.9	&	59.8	&	3.2	&	59.0	\\
	&	44	&	4.1	$\pm$	0.7	&	60.5	$\pm$	0.1	&	0.8	$\pm$	0.1	&	5.0	$\pm$	0.7	&	&	4.4	&	59.8	&	60.9	&	4.8	&	60.5	\\
	&	44	&	4.4	$\pm$	0.7	&	61.7	$\pm$	0.1	&	0.9	$\pm$	0.2	&	4.4	$\pm$	0.7	&	&	4.8	&	60.9	&	63.3	&	5.3	&	62.0	\\
G41.34-0.14	&	22	&	2.5	$\pm$	0.4	&	60.8	$\pm$	0.1	&	1.9	$\pm$	0.4	&	1.3	$\pm$	0.4	&	&	3.3	&	58.6	&	63.8	&	1.7	&	60.7	\\
G42.03+0.19	&	22	&	2.7	$\pm$	0.3	&	13.7	$\pm$	0.0	&	0.7	$\pm$	0.1	&	3.8	$\pm$	0.5	&	&	2.9	&	12.3	&	14.5	&	3.6	&	13.7	\\
G42.30-0.30	&	22	&	3.4	$\pm$	0.6	&	21.8	$\pm$	0.1	&	0.8	$\pm$	0.1	&	3.8	$\pm$	0.4	&	&	3.7	&	20.7	&	22.5	&	3.9	&	21.9	\\
	&	22	&	2.2	$\pm$	0.8	&	23.1	$\pm$	0.1	&	1.2	$\pm$	0.6	&	1.8	$\pm$	0.4	&	&	2.1	&	22.5	&	24.2	&	1.9	&	23.0	\\
G42.43-0.26	&	22	&	4.7	$\pm$	0.6	&	63.4	$\pm$	0.1	&	2.3	$\pm$	0.3	&	1.9	$\pm$	0.5	&	&	4.7	&	61.0	&	64.9	&	3.0	&	64.0	\\
	&	22	&	1.7	$\pm$	0.4	&	65.7	$\pm$	0.1	&	0.9	$\pm$	0.2	&	1.9	$\pm$	0.5	&	&	2.0	&	64.9	&	66.8	&	2.1	&	65.4	\\
	&	22	&	3.4	$\pm$	0.3	&	68.7	$\pm$	0.0	&	0.8	$\pm$	0.1	&	3.8	$\pm$	0.5	&	&	4.7	&	66.8	&	70.0	&	3.9	&	68.6	\\
	&	44	&	5.1	$\pm$	0.3	&	63.0	$\pm$	0.0	&	0.4	$\pm$	0.0	&	11.2	$\pm$	0.7	&	&	5.0	&	62.1	&	63.7	&	9.7	&	63.0	\\
	&	95	&	2.2	$\pm$	0.3	&	63.1	$\pm$	0.0	&	0.6	$\pm$	0.1	&	3.5	$\pm$	0.6	&	&	2.1	&	62.3	&	64.3	&	4.1	&	63.1	\\
G43.04-0.46	&	22	&	4.6	$\pm$	0.4	&	45.7	$\pm$	0.0	&	0.8	$\pm$	0.1	&	5.6	$\pm$	0.7	&	&	5.4	&	43.8	&	46.6	&	5.4	&	45.8	\\
	&	22	&	3.0	$\pm$	0.6	&	51.4	$\pm$	0.1	&	1.3	$\pm$	0.3	&	2.1	$\pm$	0.7	&	&	3.7	&	49.2	&	52.6	&	2.6	&	51.3	\\
	&	22	&	1.5	$\pm$	0.3	&	58.0	$\pm$	0.1	&	0.7	$\pm$	0.1	&	2.1	$\pm$	0.7	&	&	1.2	&	57.1	&	58.7	&	2.0	&	57.8	\\
	&	44	&	4.5	$\pm$	0.6	&	58.0	$\pm$	0.0	&	0.5	$\pm$	0.1	&	8.7	$\pm$	1.1	&	&	7.1	&	56.5	&	59.2	&	8.3	&	58.1	\\
	&	95	&	4.2	$\pm$	1.0	&	58.1	$\pm$	0.0	&	0.7	$\pm$	0.3	&	5.5	$\pm$	0.8	&	&	9.6	&	55.7	&	61.5	&	7.1	&	58.1	\\
G43.80-0.13	&	22	&	30.0	$\pm$	13.2	&	35.4	$\pm$	0.2	&	1.3	$\pm$	0.2	&	22.0	$\pm$	4.1	&	&	32.2	&	33.7	&	36.1	&	23.5	&	35.4	\\
	&	22	&	88.4	$\pm$	13.2	&	37.4	$\pm$	0.2	&	1.2	$\pm$	0.2	&	70.2	$\pm$	4.1	&	&	88.7	&	36.1	&	38.3	&	71.7	&	37.3	\\
	&	22	&	306.0	$\pm$	13.2	&	39.9	$\pm$	0.2	&	1.5	$\pm$	0.2	&	194.5	$\pm$	4.1	&	&	265.7	&	38.3	&	40.4	&	199.2	&	40.0	\\
	&	22	&	493.9	$\pm$	13.2	&	41.2	$\pm$	0.2	&	0.6	$\pm$	0.2	&	716.9	$\pm$	4.1	&	&	607.8	&	40.4	&	42.0	&	714.2	&	41.1	\\
	&	22	&	69.6	$\pm$	13.2	&	42.2	$\pm$	0.2	&	0.9	$\pm$	0.2	&	73.9	$\pm$	4.1	&	&	51.9	&	42.0	&	42.9	&	80.3	&	42.1	\\
	&	22	&	34.9	$\pm$	1.0	&	43.6	$\pm$	0.2	&	1.4	$\pm$	0.2	&	23.9	$\pm$	0.8	&	&	47.7	&	42.9	&	44.6	&	35.0	&	44.0	\\
	&	22	&	32.4	$\pm$	1.0	&	44.6	$\pm$	0.2	&	2.3	$\pm$	0.2	&	13.4	$\pm$	0.8	&	&	17.8	&	44.6	&	46.9	&	15.3	&	45.1	\\
	&	22	&	23.5	$\pm$	1.0	&	49.2	$\pm$	0.2	&	2.5	$\pm$	0.2	&	8.7	$\pm$	0.8	&	&	22.0	&	46.9	&	51.1	&	9.7	&	49.3	\\
	&	44	&	4.3	$\pm$	0.7	&	44.4	$\pm$	0.4	&	5.4	$\pm$	1.2	&	0.8	$\pm$	0.4	&	&	3.0	&	42.1	&	47.6	&	1.3	&	44.7	\\
	&	95	&	6.5	$\pm$	0.8	&	43.5	$\pm$	0.4	&	6.8	$\pm$	0.9	&	0.9	$\pm$	0.4	&	&	6.1	&	38.3	&	49.2	&	2.1	&	43.5	\\
G44.31+0.04	&	22	&	45.5	$\pm$	2.7	&	52.5	$\pm$	0.2	&	0.5	$\pm$	0.2	&	83.4	$\pm$	0.8	&	&	80.8	&	50.9	&	52.9	&	100.9	&	53.1	\\
	&	22	&	92.0	$\pm$	2.7	&	53.3	$\pm$	0.2	&	0.6	$\pm$	0.2	&	138.2	$\pm$	0.8	&	&	83.1	&	52.9	&	54.1	&	139.3	&	53.3	\\
	&	22	&	59.7	$\pm$	2.7	&	55.2	$\pm$	0.2	&	1.3	$\pm$	0.2	&	43.4	$\pm$	0.8	&	&	107.1	&	53.4	&	65.0	&	96.8	&	53.5	\\
	&	22	&	6.9	$\pm$	2.7	&	64.2	$\pm$	0.2	&	0.6	$\pm$	0.2	&	11.1	$\pm$	0.8	&	&	59.8	&	54.1	&	56.8	&	44.6	&	55.2	\\
	&	44	&	4.6	$\pm$	0.6	&	57.4	$\pm$	0.1	&	1.0	$\pm$	0.2	&	4.4	$\pm$	0.7	&	&	6.1	&	55.0	&	58.9	&	6.0	&	57.4	\\
	&	95	&	10.0	$\pm$	1.2	&	57.7	$\pm$	0.2	&	3.4	$\pm$	0.6	&	2.7	$\pm$	0.8	&	&	8.3	&	54.9	&	59.5	&	4.9	&	58.0	\\
G45.07+0.13	&	22	&	2.9	$\pm$	0.6	&	54.7	$\pm$	0.2	&	0.7	$\pm$	0.2	&	3.9	$\pm$	0.6	&	&	3.6	&	53.6	&	55.3	&	4.3	&	54.7	\\
	&	22	&	3.8	$\pm$	0.6	&	56.7	$\pm$	0.2	&	2.0	$\pm$	0.2	&	1.8	$\pm$	0.6	&	&	3.5	&	55.3	&	57.6	&	2.6	&	56.4	\\
	&	22	&	8.2	$\pm$	0.6	&	58.6	$\pm$	0.2	&	0.8	$\pm$	0.2	&	9.7	$\pm$	0.6	&	&	10.6	&	57.6	&	59.2	&	10.1	&	58.7	\\
	&	22	&	24.2	$\pm$	0.6	&	60.1	$\pm$	0.2	&	1.0	$\pm$	0.2	&	23.2	$\pm$	0.6	&	&	24.9	&	59.2	&	61.2	&	23.7	&	59.9	\\
	&	22	&	5.2	$\pm$	0.6	&	62.0	$\pm$	0.2	&	1.5	$\pm$	0.2	&	3.3	$\pm$	0.6	&	&	5.6	&	61.2	&	62.9	&	4.1	&	62.0	\\
	&	22	&	2.7	$\pm$	0.5	&	63.4	$\pm$	0.1	&	1.2	$\pm$	0.3	&	2.2	$\pm$	0.6	&	&	2.5	&	62.9	&	64.6	&	2.4	&	63.1	\\
	&	22	&	5.1	$\pm$	0.4	&	68.0	$\pm$	0.0	&	0.9	$\pm$	0.1	&	5.4	$\pm$	0.6	&	&	5.9	&	66.6	&	69.0	&	6.1	&	67.9	\\
	&	44	&	0.9	$\pm$	0.3	&	59.2	$\pm$	0.1	&	0.5	$\pm$	0.3	&	1.7	$\pm$	0.5	&	&	0.9	&	58.3	&	60.0	&	1.6	&	59.1	\\
	&	95	&	5.5	$\pm$	0.6	&	58.6	$\pm$	0.2	&	3.5	$\pm$	0.5	&	1.5	$\pm$	0.5	&	&	4.7	&	55.7	&	60.6	&	2.3	&	59.4	\\
G45.44+0.07	&	22	&	1.2	$\pm$	0.3	&	38.8	$\pm$	0.1	&	0.5	$\pm$	0.1	&	2.3	$\pm$	0.6	&	&	1.3	&	38.1	&	39.4	&	2.2	&	38.8	\\
	&	22	&	5.3	$\pm$	0.4	&	41.0	$\pm$	0.0	&	1.0	$\pm$	0.1	&	4.9	$\pm$	0.6	&	&	5.1	&	40.0	&	42.0	&	4.8	&	41.1	\\
	&	22	&	0.8	$\pm$	0.2	&	49.5	$\pm$	0.0	&	0.2	$\pm$	1.1	&	3.8	$\pm$	0.6	&	&	1.0	&	48.9	&	49.9	&	2.5	&	49.5	\\
	&	22	&	14.0	$\pm$	0.4	&	51.4	$\pm$	0.0	&	0.9	$\pm$	0.0	&	14.0	$\pm$	0.6	&	&	13.9	&	49.9	&	52.6	&	14.2	&	51.2	\\
	&	22	&	2.8	$\pm$	0.4	&	56.0	$\pm$	0.1	&	0.8	$\pm$	0.1	&	3.2	$\pm$	0.6	&	&	2.7	&	55.0	&	56.9	&	3.3	&	56.1	\\
G45.47+0.05	&	22	&	3.1	$\pm$	0.5	&	49.4	$\pm$	0.1	&	1.5	$\pm$	0.3	&	1.9	$\pm$	0.5	&	&	2.7	&	48.2	&	50.0	&	2.2	&	49.5	\\
	&	22	&	12.9	$\pm$	0.4	&	51.4	$\pm$	0.0	&	0.9	$\pm$	0.0	&	13.3	$\pm$	0.5	&	&	13.6	&	50.0	&	52.5	&	13.0	&	51.4	\\
G45.47+0.13	&	22	&	4.9	$\pm$	0.5	&	67.1	$\pm$	0.0	&	0.9	$\pm$	0.1	&	5.2	$\pm$	0.7	&	&	4.6	&	66.3	&	68.1	&	5.1	&	67.3	\\
G45.49+0.13	&	22	&	2.7	$\pm$	0.4	&	67.0	$\pm$	0.1	&	1.2	$\pm$	0.2	&	2.1	$\pm$	0.5	&	&	2.9	&	65.1	&	68.2	&	2.5	&	66.9	\\
	&	22	&	5.0	$\pm$	0.6	&	74.4	$\pm$	0.2	&	2.7	$\pm$	0.5	&	1.7	$\pm$	0.5	&	&	5.1	&	71.9	&	76.0	&	2.6	&	74.7	\\
	&	22	&	1.2	$\pm$	0.3	&	77.0	$\pm$	0.1	&	0.6	$\pm$	0.1	&	1.7	$\pm$	0.5	&	&	1.4	&	76.0	&	77.9	&	1.9	&	77.0	\\
	&	44	&	4.1	$\pm$	0.6	&	62.4	$\pm$	0.3	&	3.5	$\pm$	0.6	&	1.1	$\pm$	0.5	&	&	3.5	&	60.5	&	64.5	&	2.2	&	61.5	\\
	&	95	&	3.9	$\pm$	0.5	&	63.0	$\pm$	0.2	&	3.2	$\pm$	0.5	&	1.1	$\pm$	0.5	&	&	4.0	&	59.8	&	66.2	&	1.7	&	62.6	\\
G45.57-0.12	&	22	&	2.7	$\pm$	0.4	&	7.4	$\pm$	0.1	&	0.7	$\pm$	0.1	&	3.8	$\pm$	0.8	&	&	2.4	&	6.5	&	7.9	&	3.5	&	7.6	\\
G45.81-0.36	&	22	&	3.2	$\pm$	0.5	&	59.0	$\pm$	0.1	&	1.1	$\pm$	0.2	&	2.8	$\pm$	0.7	&	&	3.7	&	57.6	&	60.2	&	3.0	&	58.9	\\
	&	22	&	25.1	$\pm$	0.4	&	63.0	$\pm$	0.0	&	0.7	$\pm$	0.0	&	32.7	$\pm$	0.7	&	&	25.5	&	61.8	&	64.2	&	34.3	&	62.9	\\
	&	44	&	6.8	$\pm$	0.8	&	57.7	$\pm$	0.1	&	1.8	$\pm$	0.3	&	3.5	$\pm$	0.7	&	&	7.0	&	55.8	&	59.5	&	4.6	&	57.7	\\
	&	44	&	2.1	$\pm$	0.5	&	60.7	$\pm$	0.1	&	1.0	$\pm$	0.4	&	2.0	$\pm$	0.7	&	&	1.7	&	59.5	&	62.2	&	2.5	&	60.9	\\
	&	95	&	8.3	$\pm$	0.9	&	57.8	$\pm$	0.1	&	2.4	$\pm$	0.3	&	3.2	$\pm$	0.7	&	&	8.4	&	55.6	&	59.7	&	4.8	&	57.7	\\
	&	95	&	2.5	$\pm$	0.7	&	60.6	$\pm$	0.2	&	1.5	$\pm$	0.4	&	1.6	$\pm$	0.7	&	&	3.1	&	59.7	&	62.3	&	2.3	&	60.1	\\
G46.12+0.38	&	22	&	1.9	$\pm$	0.3	&	56.1	$\pm$	0.1	&	0.9	$\pm$	0.1	&	2.1	$\pm$	0.5	&	&	1.4	&	55.2	&	57.0	&	1.7	&	55.9	\\
G48.90-0.27	&	22	&	5.9	$\pm$	0.5	&	66.4	$\pm$	0.0	&	1.1	$\pm$	0.1	&	5.1	$\pm$	0.6	&	&	5.8	&	65.0	&	67.8	&	5.5	&	66.6	\\
	&	44	&	2.6	$\pm$	0.3	&	68.5	$\pm$	0.0	&	0.5	$\pm$	0.1	&	4.7	$\pm$	0.6	&	&	2.0	&	67.7	&	69.7	&	4.4	&	68.6	\\
G48.99-0.30	&	22	&	86.5	$\pm$	0.9	&	66.0	$\pm$	0.0	&	1.3	$\pm$	0.0	&	62.4	$\pm$	0.7	&	&	87.9	&	64.2	&	66.9	&	62.4	&	65.9	\\
	&	22	&	11.8	$\pm$	0.9	&	67.3	$\pm$	0.0	&	0.8	$\pm$	0.1	&	13.4	$\pm$	0.7	&	&	14.9	&	66.9	&	68.3	&	21.9	&	67.0	\\
	&	44	&	10.0	$\pm$	1.7	&	65.5	$\pm$	0.2	&	0.8	$\pm$	0.2	&	12.4	$\pm$	1.2	&	&	8.8	&	64.7	&	65.7	&	11.9	&	65.3	\\
	&	44	&	50.8	$\pm$	1.7	&	66.4	$\pm$	0.2	&	0.5	$\pm$	0.2	&	93.2	$\pm$	1.2	&	&	56.0	&	65.7	&	66.8	&	96.3	&	66.4	\\
	&	44	&	34.9	$\pm$	1.7	&	67.5	$\pm$	0.2	&	0.8	$\pm$	0.2	&	39.1	$\pm$	1.2	&	&	47.7	&	66.8	&	68.1	&	43.2	&	67.7	\\
	&	44	&	24.5	$\pm$	1.7	&	68.6	$\pm$	0.2	&	2.0	$\pm$	0.2	&	11.4	$\pm$	1.2	&	&	16.7	&	68.1	&	70.3	&	14.8	&	68.3	\\
	&	95	&	17.2	$\pm$	1.4	&	65.9	$\pm$	0.2	&	1.5	$\pm$	0.2	&	10.6	$\pm$	0.9	&	&	18.5	&	64.1	&	66.0	&	34.3	&	66.2	\\
	&	95	&	29.5	$\pm$	1.4	&	66.5	$\pm$	0.2	&	0.5	$\pm$	0.2	&	54.2	$\pm$	0.9	&	&	39.7	&	66.0	&	66.9	&	60.8	&	66.4	\\
	&	95	&	43.8	$\pm$	1.4	&	67.6	$\pm$	0.2	&	1.1	$\pm$	0.2	&	38.8	$\pm$	0.9	&	&	46.6	&	66.9	&	68.3	&	40.8	&	67.6	\\
	&	95	&	21.7	$\pm$	1.4	&	69.0	$\pm$	0.2	&	1.7	$\pm$	0.2	&	12.2	$\pm$	0.9	&	&	22.8	&	68.3	&	70.9	&	17.9	&	68.4	\\
G49.27+0.31	&	44	&	5.1	$\pm$	0.5	&	3.2	$\pm$	0.1	&	2.1	$\pm$	0.3	&	2.3	$\pm$	0.5	&	&	3.6	&	1.5	&	5.3	&	2.4	&	3.6	\\
	&	95	&	2.8	$\pm$	0.4	&	3.2	$\pm$	0.2	&	2.5	$\pm$	0.5	&	1.1	$\pm$	0.4	&	&	2.2	&	0.9	&	5.6	&	1.3	&	3.0	\\
G53.04+0.11	&	44	&	11.3	$\pm$	0.7	&	5.1	$\pm$	0.1	&	3.1	$\pm$	0.2	&	3.4	$\pm$	0.6	&	&	11.0	&	2.2	&	7.6	&	4.4	&	4.4	\\
	&	95	&	6.8	$\pm$	0.9	&	5.6	$\pm$	0.2	&	3.2	$\pm$	0.5	&	2.0	$\pm$	0.7	&	&	7.1	&	1.5	&	7.9	&	3.0	&	6.7	\\
G53.14+0.07	&	22	&	2.3	$\pm$	0.5	&	24.1	$\pm$	0.1	&	0.7	$\pm$	0.2	&	3.3	$\pm$	0.7	&	&	2.7	&	23.1	&	25.5	&	3.7	&	24.0	\\
	&	44	&	11.6	$\pm$	0.8	&	21.9	$\pm$	0.0	&	0.7	$\pm$	0.1	&	16.3	$\pm$	1.3	&	&	14.3	&	20.6	&	23.8	&	17.1	&	22.0	\\
	&	95	&	10.9	$\pm$	0.8	&	22.0	$\pm$	0.0	&	0.8	$\pm$	0.1	&	13.5	$\pm$	1.2	&	&	27.3	&	15.6	&	26.1	&	14.7	&	22.0	\\
\enddata
\end{deluxetable}

\clearpage
\begin{deluxetable}{cccccccccc}	
\tabletypesize{\tiny} 
\tablecaption{{Maser Luminosities and associated BGPS Sources\label{Table:MASER_L_BGPS}}}
\tablewidth{0pt}

\tablehead{
&&&&&&
\multicolumn{4}{c}{BGPS Source}
\\
\cline{7-10}
\colhead{Source} &
\colhead{$\rm L_{6.7}$\tablenotemark{*}} &
\colhead{{$\rm\bf L_{22}$}} &
\colhead{{$\rm\bf L_{44}$}} &
\colhead{{$\rm\bf L_{95}$}} &
\colhead{$\rm T_{kin}$\tablenotemark{**}} &
\multirow{2}{*}{Name} &
\colhead{$\rm S_{int}$} &
\colhead{N$^{beam}_{\htwo}$(T=20K)} &
\colhead{$\rm M_{gas}(T_{kin})$} \\ 
&
\colhead{($\rm L_{\Sun}$)} &
\colhead{($\rm L_{\Sun}$)} &
\colhead{($\rm L_{\Sun}$)} &
\colhead{($\rm L_{\Sun}$)} &
\colhead{(K)} &
&
\colhead{($\rm Jy\,km\,s^{-1}$)} &
\colhead{($\rm 10^{22}cm^{-2}$)}  &
\colhead{($\rm M_{\odot}$)}\\
}
\startdata
G34.82+0.35	&	9.0E-09	&		&		&		&	23.4	&	G034.820+00.350	&	3.4	&	6.2	&	6.9E+02	\\
G35.03+0.35	&	2.1E-05	&	1.9E-04	&	2.2E-05	&	5.4E-05	&	40.0	&	G035.026+00.350	&	6.5	&	8.4	&	5.7E+03	\\
G35.25-0.24	&	7.1E-08	&		&	1.9E-06	&		&	17.8	&	G035.247-00.238	&	0.1	&	0.5	&	4.1E+01	\\
G35.39+0.02	&	2.8E-08	&		&	4.1E-06	&		&	16.7	&		&		&		&		\\
G35.40+0.03	&	7.7E-08	&		&		&		&	24.5	&	G035.398+00.026	&	0.9	&	1.2	&	4.8E+02	\\
G35.59+0.06	&	7.3E-07	&	4.1E-05	&	6.3E-05	&	8.6E-05	&	23.5	&		&		&		&		\\
G35.79-0.17	&	4.0E-06	&	6.4E-06	&	7.1E-06	&		&	21.0	&	G035.794-00.176	&	2.6	&	4.5	&	7.2E+02	\\
G36.02-0.20	&	1.6E-08	&		&		&		&	17.6	&	G036.012-00.198	&	0.7	&	1.5	&	4.6E+02	\\
G36.64-0.21	&	2.9E-07	&	2.4E-05	&		&		&		&		&		&		&		\\
G36.70+0.09	&	5.5E-06	&	9.9E-06	&		&		&	16.2	&	G036.704+00.094	&	1.1	&	0.9	&	2.7E+03	\\
G36.84-0.02	&	4.4E-07	&		&		&		&	14.3	&	G036.840-00.022	&	2.1	&	2.2	&	9.2E+02	\\
G36.90-0.41	&	4.5E-08	&	1.4E-06	&		&		&	20.4	&	G036.899-00.410	&	1.3	&	1.6	&	6.1E+02	\\
G36.92+0.48	&	8.8E-07	&		&		&		&		&		&		&		&		\\
G37.02-0.03	&	8.6E-07	&	1.9E-06	&		&		&	17.9	&		&		&		&		\\
G37.04-0.04	&	1.2E-06	&	1.2E-06	&		&		&	18.9	&	G037.042-00.034	&	2.8	&	2.8	&	1.5E+03	\\
G37.38-0.09	&	8.5E-08	&		&	1.5E-05	&		&	48.4	&	G037.381-00.078	&	0.4	&	0.9	&	2.1E+02	\\
G37.47-0.11	&	1.4E-05	&	2.3E-06	&		&		&	32.2	&	G037.475-00.102	&	0.8	&	1.8	&	8.1E+02	\\
G37.53-0.11	&	3.9E-06	&		&		&		&	67.3	&	G037.547-00.112	&	2.6	&	4.5	&	1.1E+03	\\
G37.55+0.19	&	1.4E-06	&	2.0E-05	&	1.2E-05	&	1.3E-05	&	25.4	&	G037.555+00.200	&	3.6	&	6.3	&	1.6E+03	\\
G37.60+0.42	&	6.7E-06	&	1.2E-05	&	4.9E-06	&	8.6E-06	&	18.7	&	G037.599+00.426	&	0.6	&	1.8	&	4.7E+02	\\
G37.74-0.12	&	1.8E-07	&	1.2E-04	&		&		&	28.2	&	G037.737-00.112	&	2.7	&	4.5	&	3.6E+03	\\
G37.76-0.19	&	9.1E-07	&	5.5E-05	&		&		&	22.6	&	G037.753-00.192	&	0.9	&	2.1	&	1.3E+03	\\
G37.77-0.22	&	2.3E-07	&	2.0E-05	&	2.9E-05	&	8.1E-05	&	25.1	&	G037.765-00.216	&	4.6	&	7.7	&	5.8E+03	\\
G38.03-0.30	&	1.9E-06	&	4.3E-06	&		&		&	23.7	&	G038.041-00.300	&	0.4	&	1.1	&	1.0E+02	\\
G38.08-0.27	&	2.1E-08	&		&		&		&		&		&		&		&		\\
G38.12-0.24	&	1.2E-06	&		&		&		&	24.5	&	G038.119-00.232	&	0.7	&	1.6	&	3.1E+02	\\
G38.20-0.08	&	6.7E-06	&		&		&		&	18.6	&	G038.203-00.068	&	1.3	&	2.7	&	1.7E+03	\\
G38.26-0.08	&	5.6E-06	&	1.6E-04	&		&		&		&		&		&		&		\\
G38.26-0.20	&	7.2E-07	&		&		&		&	22.4	&		&		&		&		\\
G38.56+0.15	&	4.0E-09	&	4.5E-07	&		&		&		&		&		&		&		\\
G38.60-0.21	&	7.2E-08	&	7.5E-06	&		&		&	19.1	&	G038.599-00.214	&	0.8	&	1.2	&	3.0E+02	\\
G38.66+0.08	&	1.3E-06	&		&		&		&		&		&		&		&		\\
G38.92-0.36	&	9.9E-07	&	1.4E-04	&	5.1E-05	&	1.5E-04	&	24.8	&		&		&		&		\\
G39.39-0.14	&	9.9E-08	&		&	2.1E-06	&	6.0E-06	&	27.7	&	G039.389-00.143	&	1.4	&	3.2	&	3.2E+02	\\
G39.54-0.38	&	1.3E-07	&		&		&		&	25.4	&		&		&		&		\\
G40.28-0.22	&	1.1E-05	&	8.0E-05	&	1.7E-04	&	2.7E-04	&	37.7	&	G040.283-00.221	&	3.5	&	9.5	&	7.3E+02	\\
G40.62-0.14	&	5.3E-06	&	3.8E-05	&		&		&	26.4	&	G040.622-00.139	&	3.0	&	5.1	&	4.4E+03	\\
G40.94-0.04	&	1.0E-06	&		&		&		&		&		&		&		&		\\
G41.08-0.13	&	1.6E-07	&	1.5E-05	&		&		&	22.8	&	G041.076-00.125	&	0.9	&	1.5	&	1.0E+03	\\
G41.12-0.11	&	4.4E-07	&		&		&		&		&		&		&		&		\\
G41.12-0.22	&	6.5E-07	&		&	4.0E-05	&		&	19.7	&	G041.122-00.223	&	1.8	&	1.8	&	2.7E+03	\\
G41.16-0.20	&	1.0E-07	&		&		&		&	22.0	&		&		&		&		\\
G41.23-0.20	&	4.0E-06	&		&		&		&	26.2	&		&		&		&		\\
G41.27+0.37	&	1.5E-07	&		&		&		&	18.4	&	G041.268+00.373	&	0.5	&	0.8	&	1.4E+03	\\
G41.34-0.14	&	2.4E-05	&	1.0E-05	&		&		&	16.6	&	G041.347-00.139	&	0.2	&	0.8	&	7.6E+02	\\
G41.58+0.04	&	2.0E-07	&		&		&		&		&		&		&		&		\\
G42.03+0.19	&	2.6E-05	&	8.3E-06	&		&		&		&		&		&		&		\\
G42.30-0.30	&	3.9E-06	&	1.5E-05	&		&		&	20.3	&	G042.305-00.301	&	1.4	&	2.0	&	2.8E+03	\\
G42.43-0.26	&	7.0E-07	&	1.6E-05	&	1.4E-05	&	1.3E-05	&	42.5	&	G042.435-00.263	&	2.9	&	2.6	&	1.4E+03	\\
G42.70-0.15	&	7.1E-06	&		&		&		&		&		&		&		&		\\
G43.04-0.46	&	5.0E-06	&	1.6E-05	&	2.2E-05	&	6.5E-05	&	23.5	&	G043.039-00.455	&	1.7	&	4.8	&	1.8E+03	\\
G43.08-0.08	&	3.6E-06	&		&		&		&	20.6	&	G043.073-00.079	&	0.6	&	1.4	&	1.4E+03	\\
G43.80-0.13	&	2.5E-05	&	2.2E-03	&	1.2E-05	&	5.0E-05	&	28.3	&		&		&		&		\\
G44.31+0.04	&	1.9E-07	&	4.8E-04	&	1.7E-05	&	5.1E-05	&	61.3	&	G044.307+00.041	&	3.9	&	4.6	&	1.2E+03	\\
G44.64-0.52	&	1.5E-07	&		&		&		&		&		&		&		&		\\
G45.07+0.13	&	8.4E-06	&	7.2E-05	&	2.4E-06	&	2.5E-05	&		&		&		&		&		\\
G45.44+0.07	&	2.7E-07	&	3.0E-05	&		&		&	25.7	&		&		&		&		\\
G45.47+0.05	&	3.4E-06	&	1.9E-05	&		&		&	37.5	&		&		&		&		\\
G45.47+0.13	&	1.6E-06	&	5.0E-06	&		&		&	34.6	&	G045.477+00.135	&	6.8	&	7.9	&	3.1E+03	\\
G45.49+0.13	&	1.7E-06	&	1.1E-05	&	8.1E-06	&	2.0E-05	&	16.2	&		&		&		&		\\
G45.57-0.12	&	2.6E-07	&	6.9E-06	&		&		&		&		&		&		&		\\
G45.81-0.36	&	2.9E-06	&	3.3E-05	&	1.9E-05	&	5.6E-05	&	24.3	&	G045.805-00.355	&	1.0	&	2.6	&	7.1E+02	\\
G46.07+0.22	&	8.5E-07	&		&		&		&		&		&		&		&		\\
G46.12+0.38	&	4.1E-07	&	1.9E-06	&		&		&	30.5	&		&		&		&		\\
G48.89-0.17	&	3.6E-09	&		&		&		&		&		&		&		&		\\
G48.90-0.27	&	1.0E-07	&	3.5E-06	&	2.4E-06	&		&	17.0	&		&		&		&		\\
G48.99-0.30	&	1.0E-07	&	6.2E-05	&	1.5E-04	&	3.3E-04	&	22.9	&	G048.989-00.300	&	9.6	&	12.1	&	4.1E+03	\\
G49.27+0.31	&	1.0E-05	&		&	1.8E-05	&	2.3E-05	&	35.8	&	G049.264+00.312	&	0.6	&	1.5	&	5.6E+02	\\
G49.35+0.41	&	1.2E-06	&		&		&		&		&		&		&		&		\\
G49.41+0.33	&	2.4E-05	&		&		&		&		&		&		&		&		\\
G49.60-0.25	&	1.4E-05	&		&		&		&	23.3	&	G049.599-00.250	&	0.9	&	2.2	&	3.6E+02	\\
G49.62-0.36	&	2.0E-07	&		&		&		&	23.3	&		&		&		&		\\
G50.78+0.15	&	9.5E-07	&		&		&		&	21.4	&	G050.774+00.151	&	0.7	&	1.6	&	5.9E+02	\\
G52.92+0.41	&	8.4E-07	&		&		&		&	20.3	&	G052.922+00.412	&	1.0	&	2.2	&	4.8E+02	\\
G53.04+0.11	&	4.2E-07	&		&	4.4E-05	&	6.2E-05	&	25.8	&	G053.036+00.112	&	1.2	&	3.4	&	1.5E+03	\\
G53.14+0.07	&	1.4E-08	&	2.2E-07	&	2.4E-06	&	9.7E-06	&	27.9	&	G053.142+00.068	&	5.6	&	9.0	&	2.5E+02	\\
G53.62+0.04	&	2.2E-07	&		&		&		&	20.8	&	G053.616+00.036	&	1.9	&	3.7	&	1.5E+02	\\
\enddata
\tablenotetext{*}{The luminosity is from \cite{pand09}.}
\tablenotetext{**}{The kinetic temperature is from \cite{pand12}.}

\end{deluxetable}

\clearpage
\begin{deluxetable}{cccccccc}
\tabletypesize{\tiny} 
\tablewidth{0pt}\tablecaption{Column Densities and Fractional Abundances of Methanol\label{Table:lvv}}
\tablehead{
Source & {$\rm\bf S_{p,95}/S_{p,44}$} & $\rm n(H_{2})$  & $\rm N(H_{2})\tablenotemark{\dag}$ & {$\rm\bf \tau_{44}$} & {$\rm\bf \tau_{95}$}  & {$\rm\bf N(CH_{3}OH)$} & {$\rm\bf X(CH_{3}OH)$}\\
& &  $\rm (10^{3}cm^{-3})$  &$\rm (10^{21}cm^{-2})$ &$\rm (10^{-2})$ & $\rm (10^{-3})$ & $\rm (10^{15}cm^{-2})$ & $(10^{-7})$\\
}
\startdata
G35.03+0.35 &   1.0$^{+0.3}_{-0.3}$    &    0.3    &   11.6 &          -1.8$^{+0.8}_{-1.1}$        & -3.2$^{+0.9}_{-0.7}$  &    3.8$^{+0.6}_{-0.9}$ &      3.3\\
G37.55+0.19 &   0.2$^{+0.1}_{-0.1}$    &    2.4    &   24.5 &          -26.3$^{+16.3}_{-46.0}$       & -11.9$^{+4.9}_{-16.8}$  &    11.0$^{\tablenotemark{*}}_{-2.6}$ &      4.4{\tablenotemark{\ddag}}\\
G37.60+0.42 &   0.6$^{+0.3}_{-0.2}$    &   19.9    &   17.0 &          -6.7$^{+4.8}_{-12.5}$        & -7.4$^{+4.2}_{-7.6}$ &    1.3$^{+0.8}_{-0.7}$ &      0.8\\
G37.77-0.22 &   0.5$^{+0.3}_{-0.2}$    &    2.0    &   35.7 &          -5.0$^{+2.7}_{-8.1}$        & -4.6$^{+1.4}_{-2.6}$  &    5.4$^{+1.7}_{-1.3}$ &      1.5\\
G39.39-0.14 &   1.2$^{+0.7}_{-0.4}$    &    3.2    &   13.2 &          -1.2$^{\tablenotemark{**}}_{-2.5}$       & -2.6$^{\tablenotemark{**}}_{-2.7}$   &    1.7$^{+1.3}_{\tablenotemark{**}}$ &       1.3\\
G40.28-0.22 &   0.8$^{+0.0}_{-0.0}$    &    5.2    &   25.1 &          -9.5$^{+0.8}_{-0.3}$        & -1.5$^{+0.2}_{-0.7}$ &    1.3$^{+0.0}_{-0.0}$ &      0.5\\
G42.43-0.26 &   0.3$^{+0.1}_{-0.1}$    &    0.3    &   6.5  &          -10.7$^{+3.1}_{-3.9}$        & -6.6$^{+0.8}_{-0.9}$  &    6.7$^{+0.5}_{-0.5}$ &      10.0{\tablenotemark{\ddag}}\\
G43.04-0.46 &   0.6$^{+0.2}_{-0.2}$    &    5.4    &   26.8 &          -5.0$^{+2.3}_{-3.7}$        & -5.8$^{+1.6}_{-2.0}$  &    1.6$^{+0.4}_{-0.4}$ &      0.6\\
G44.31+0.04 &   0.6$^{+0.3}_{-0.3}$    &    0.2    &   15.1 &          -1.5$^{+0.8}_{-1.5}$        & -1.7$^{+0.5}_{-0.4}$  &   35.0$^{+6.6}_{-8.8}$ &      23.0\\
G45.81-0.36 &   0.9$^{+0.4}_{-0.3}$   &    4.3    &   11.9 &          -5.0$^{+4.1}_{-4.5}$       & -6.8$^{+4.6}_{-4.1}$  &    4.2$^{+1.7}_{-2.5}$ &      3.6\\
G48.99-0.30 &   0.6$^{+0.0}_{-0.0}$    &    3.0    &   27.6 &          -11.4$^{+1.1}_{-1.2}$       & -12.8$^{+0.6}_{-0.7}$ &    1.6$^{+0.1}_{-0.1}$ &      0.6{\tablenotemark{\ddag}}\\
G49.27+0.31 &   0.5$^{+0.3}_{-0.2}$    &    1.3    &   10.2 &          -3.5$^{+2.3}_{-6.1}$       & -3.0$^{+1.2}_{-1.5}$  &   16.0$^{+4.8}_{-4.8}$ &      15.0\\
G53.04+0.11 &   0.6$^{+0.4}_{-0.3}$    &    2.3    &   14.6 &          -4.5$^{+2.9}_{-8.4}$        & -4.8$^{+2.0}_{-3.1}$  &   13.0$^{+5.1}_{-4.5}$ &      9.2\\
G53.14+0.07 &   0.8$^{+0.2}_{-0.1}$    &    6.7    &   30.4 &          -5.3$^{+2.2}_{-2.4}$        & -8.1$^{+2.3}_{-1.9}$  &    1.3$^{+0.2}_{-0.3}$ &      0.4\\
\enddata
\tablenotetext{\dag}{The H$_{2}$ column density from \cite{pand12}}
\tablenotetext{\ddag}{Obtained values were calculated with $\tau < -0.1$ at 44\ghz.}
\tablenotetext{*}{ Result of LVG code is not searched for high opacity values.}
\tablenotetext{**}{ Result of LVG code is not converged at maximum value.}
\end{deluxetable}

\clearpage
\begin{figure}
\begin{center}
\begin{tabular}{c}
\includegraphics[width=60mm,angle=-90]{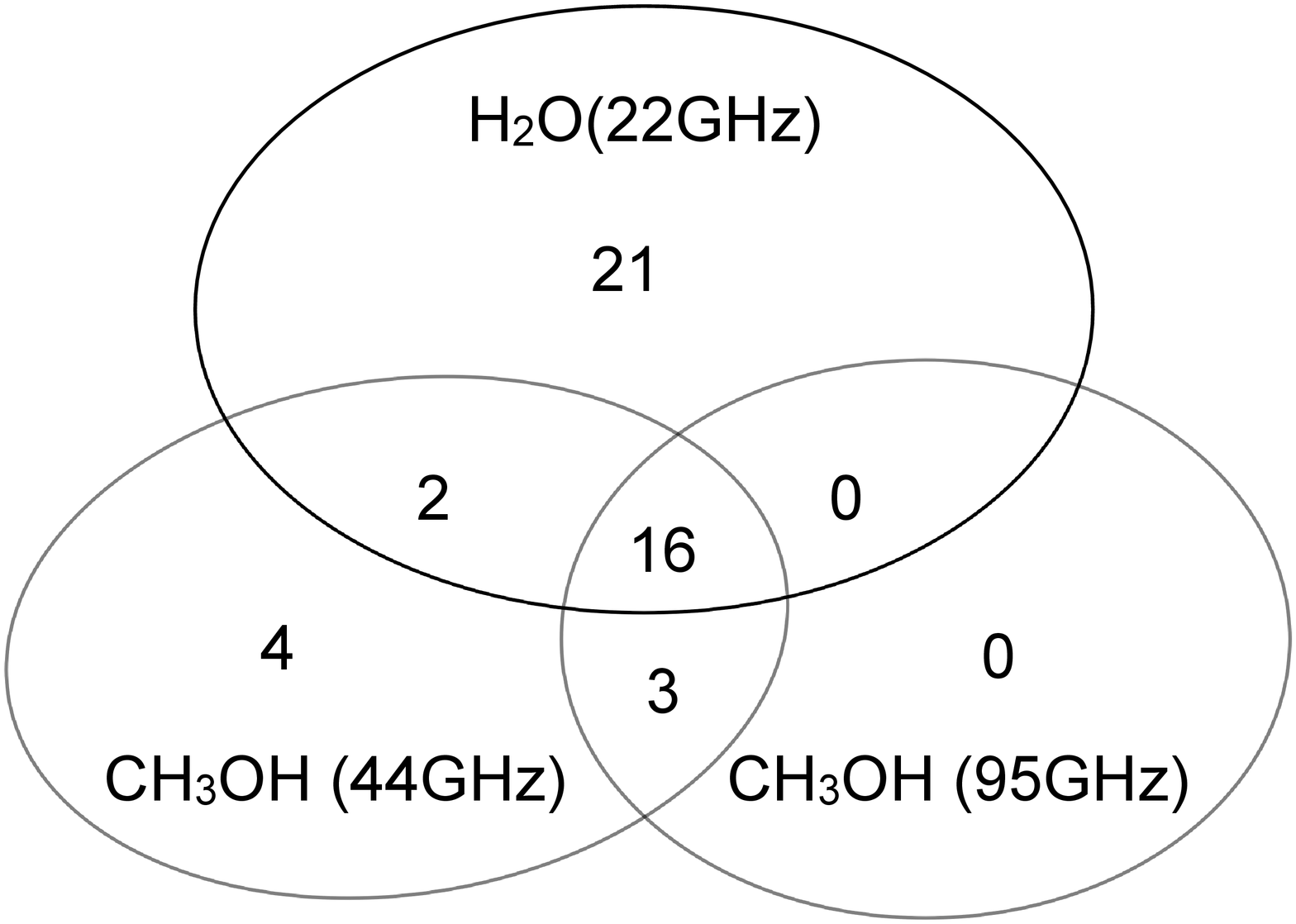} 
\\
\end{tabular}
\caption{{Venn} diagram displaying the number of sources detected in the three observed maser transitions.\label{fig:results}}
\end{center}
\end{figure}

\clearpage
\begin{figure}
\begin{tabular}{cccc}
&
\includegraphics[width=60mm]{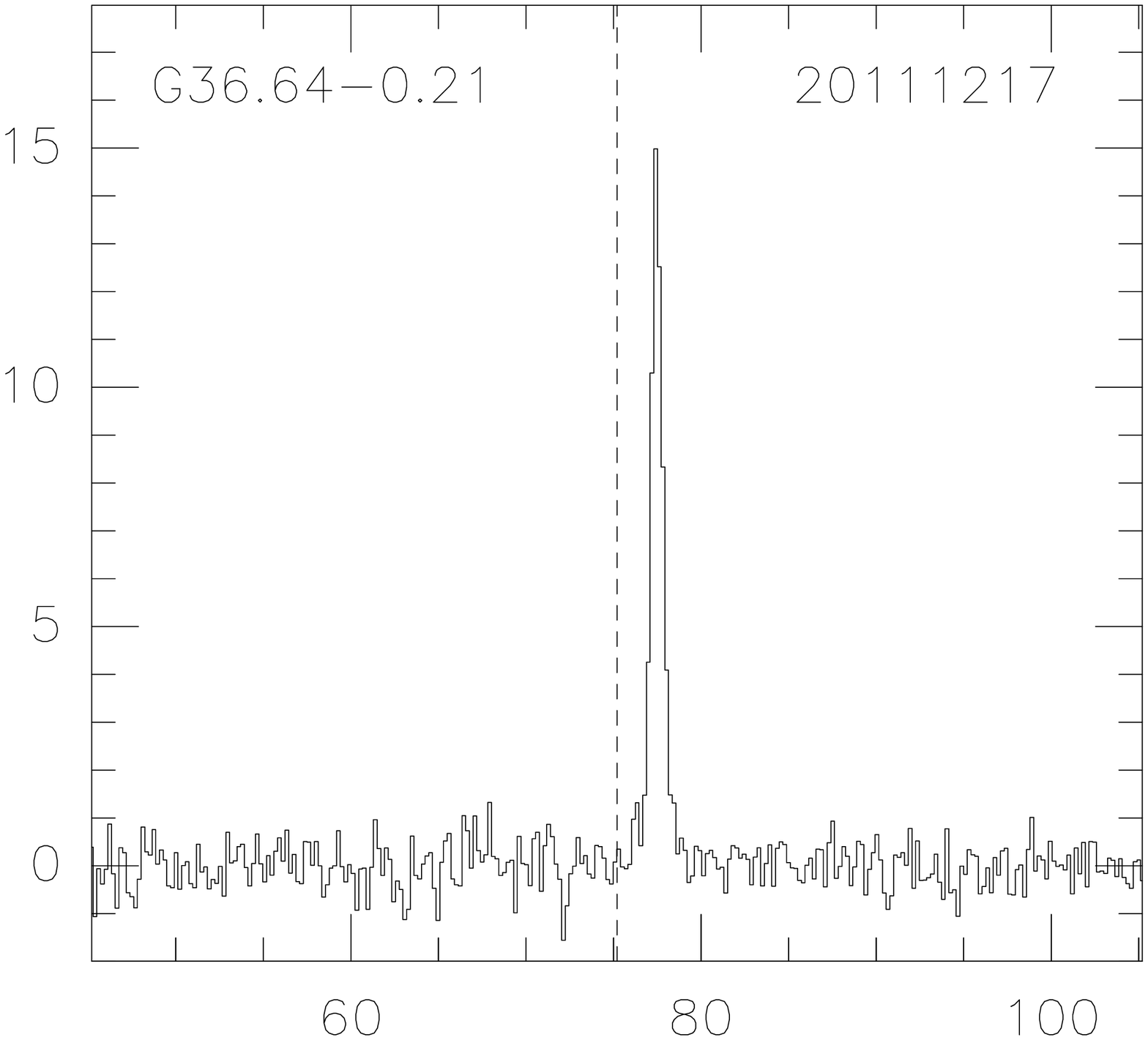} 
& 
&
\includegraphics[width=61mm]{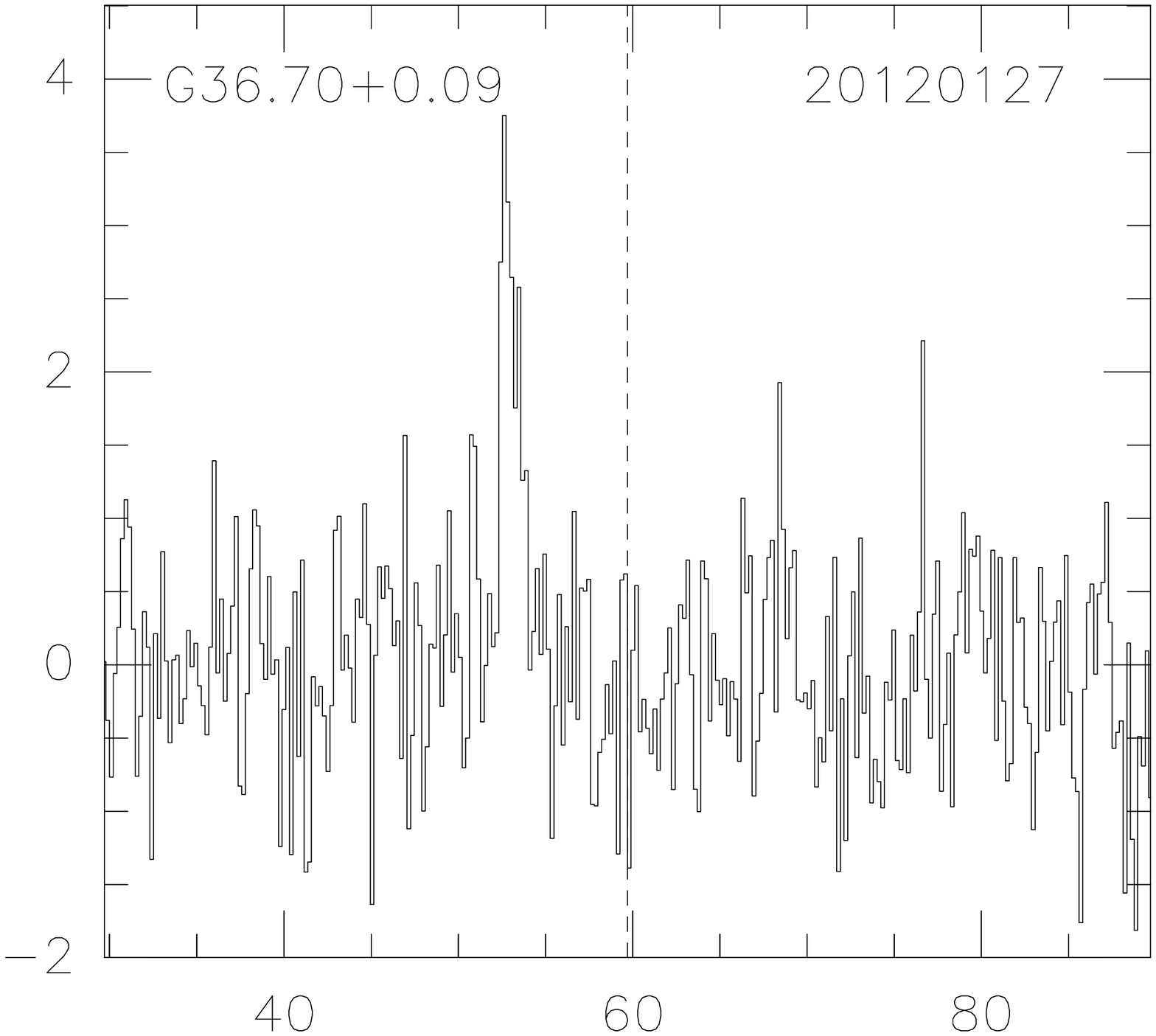} 
\\
{\rotatebox{90}{\qquad\qquad\qquad S (Jy)}}
&
\includegraphics[width=58mm]{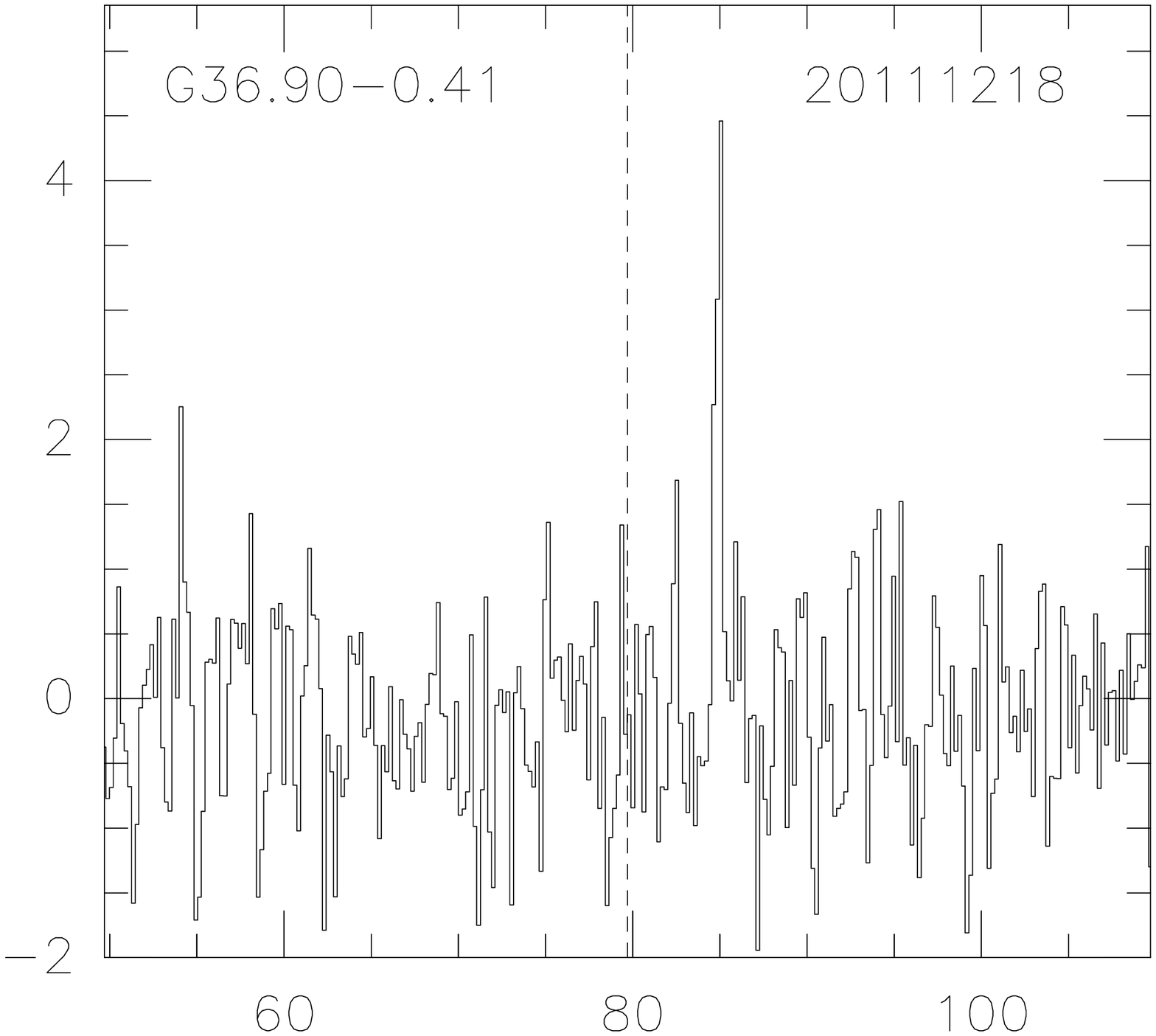} 
&
& 
\includegraphics[width=61mm]{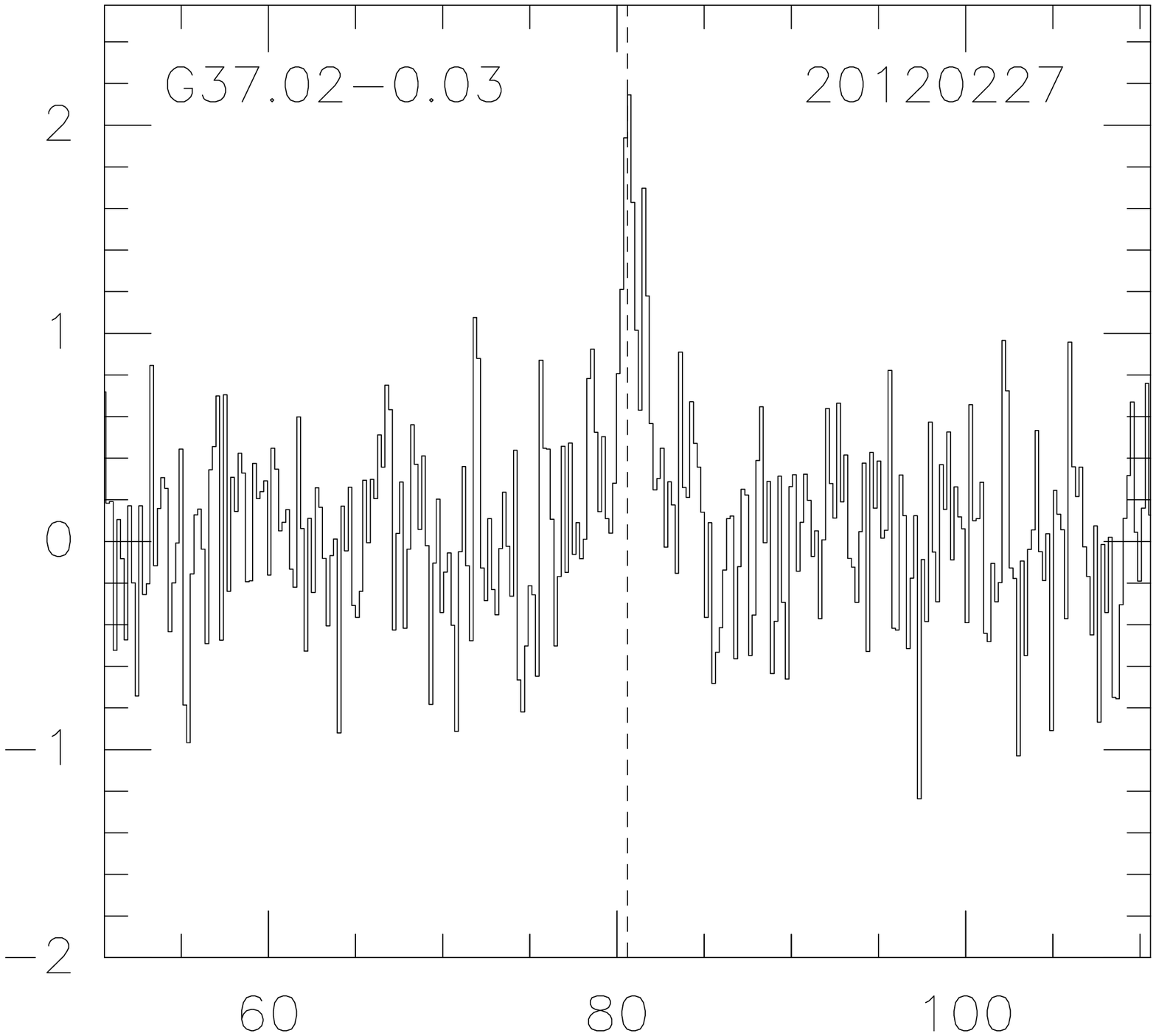} 
\\
&
\includegraphics[width=60mm]{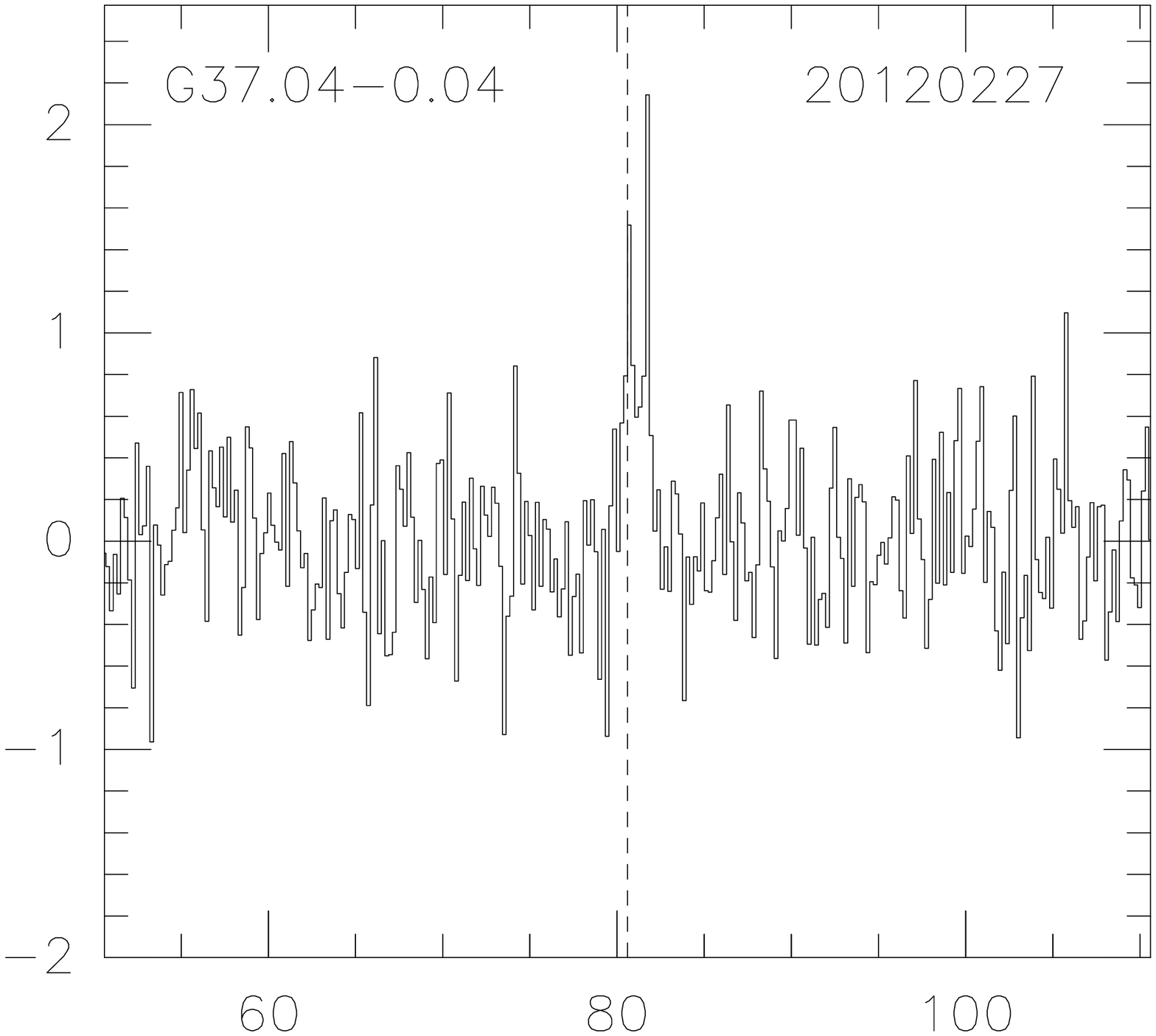} 
&
& 
\includegraphics[width=60mm]{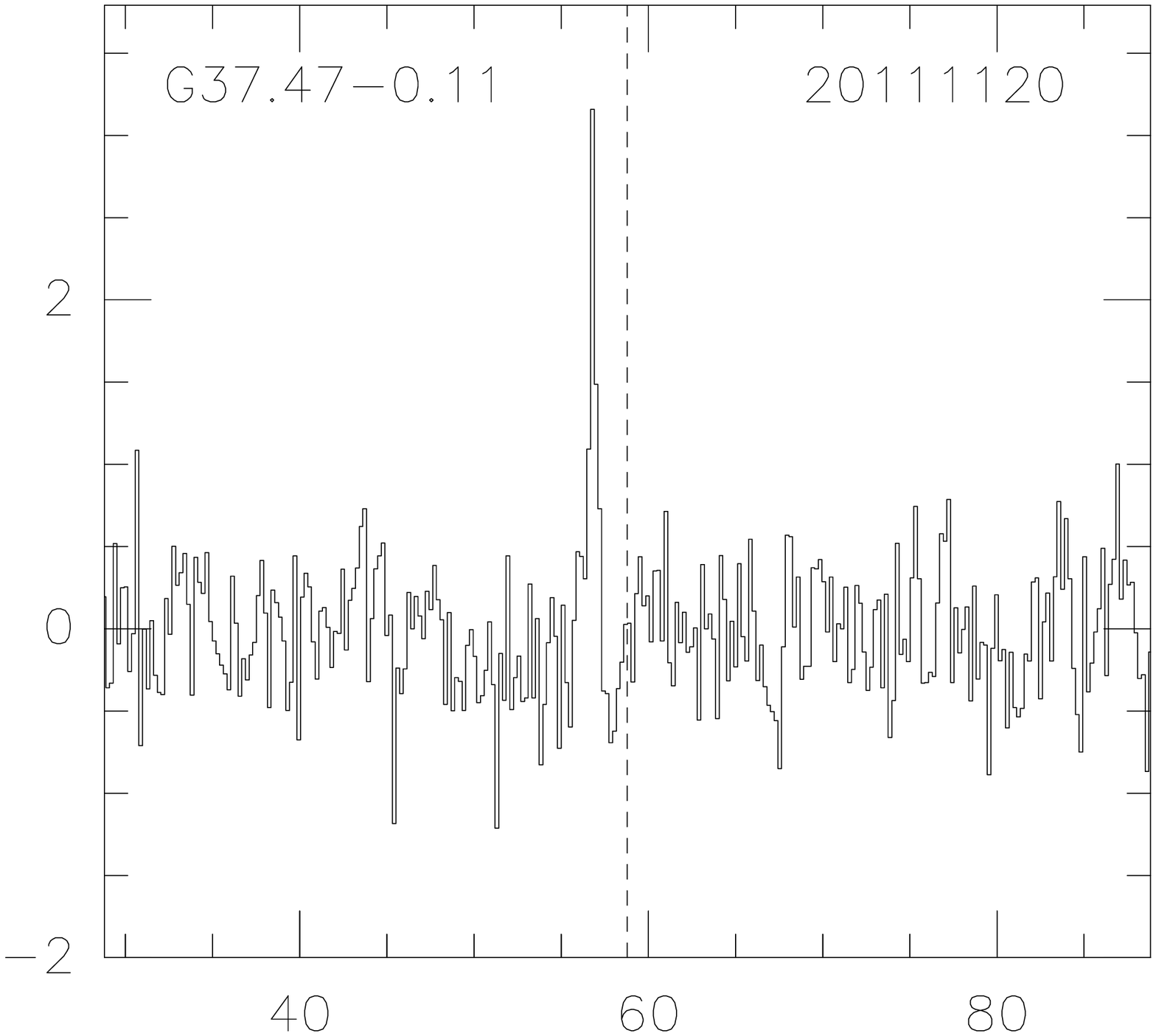}\\ 
&\multicolumn{3}{c}{V$_{lsr}$ (\kms)} \\
\end{tabular}
\caption{Sources detected only in the 22\ghz\ water maser. In each panel, the source name is presented at the top right corner and the dashed line shows the systemic velocity (Table 1).\label{fig:22only}}
\end{figure}
\clearpage
\begin{figure}
\ContinuedFloat
\begin{tabular}{cccc}
&
\includegraphics[width=60mm]{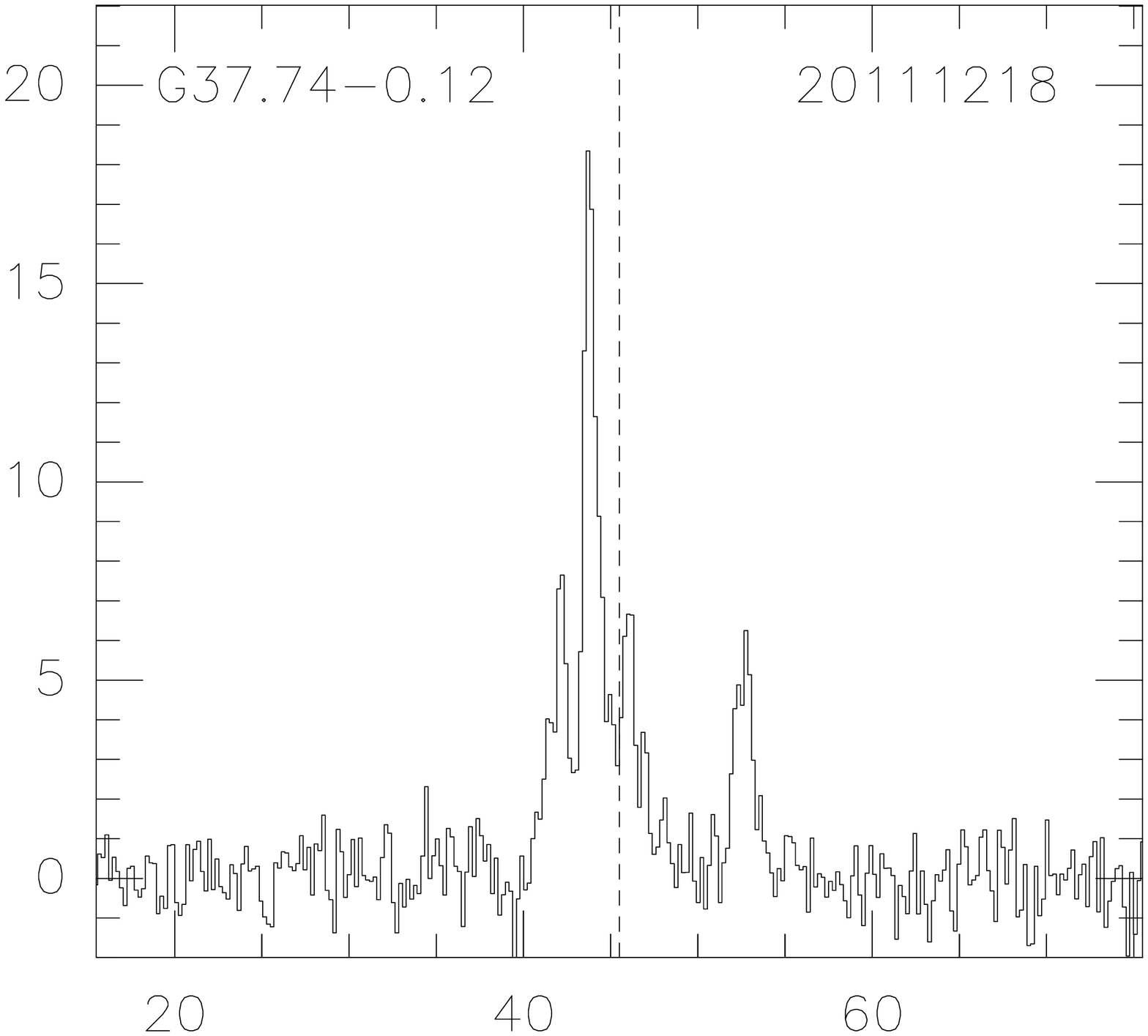} 
& 
&
\includegraphics[width=60mm]{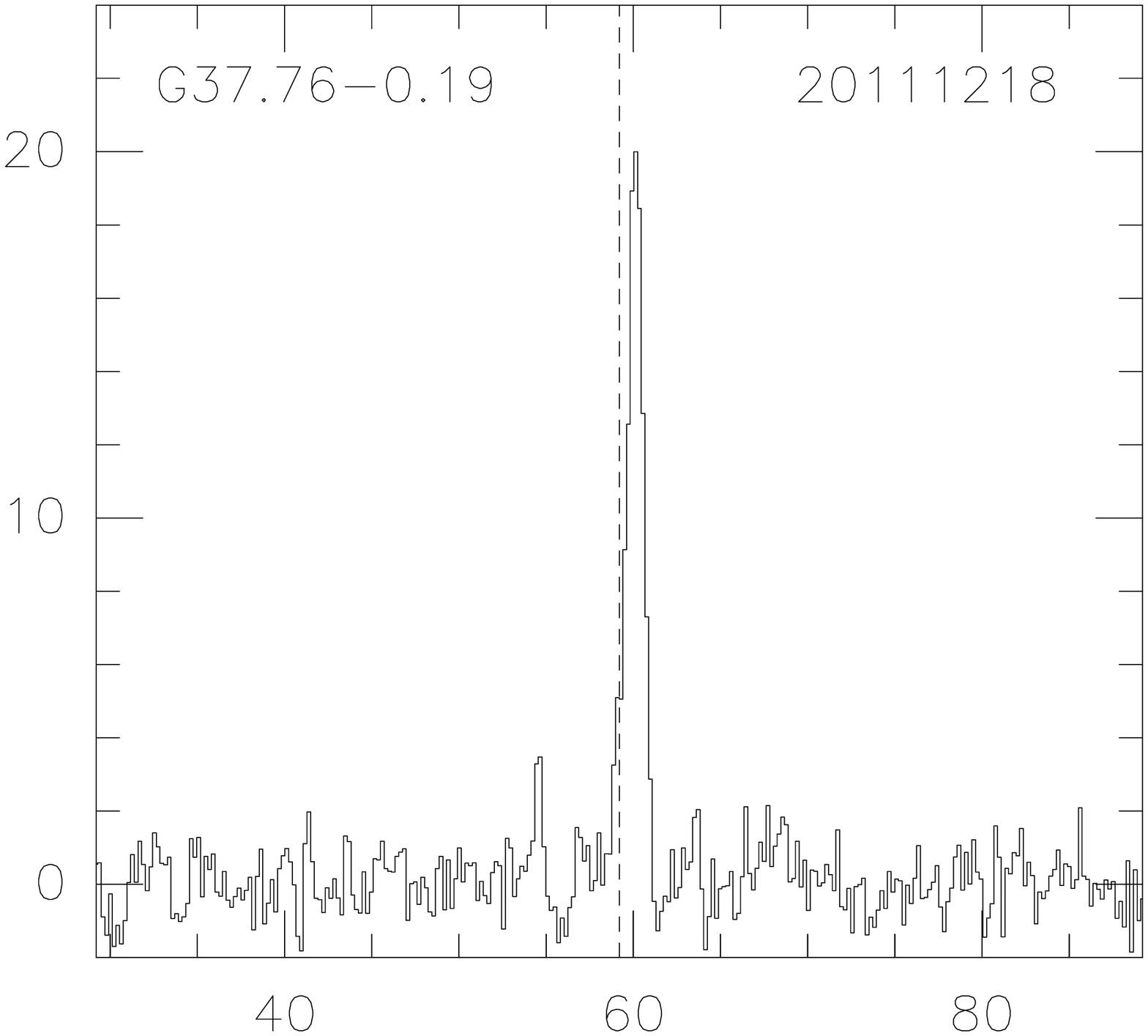} 
\\
{\rotatebox{90}{\qquad\qquad\qquad S (Jy)}}
&
\includegraphics[width=59mm]{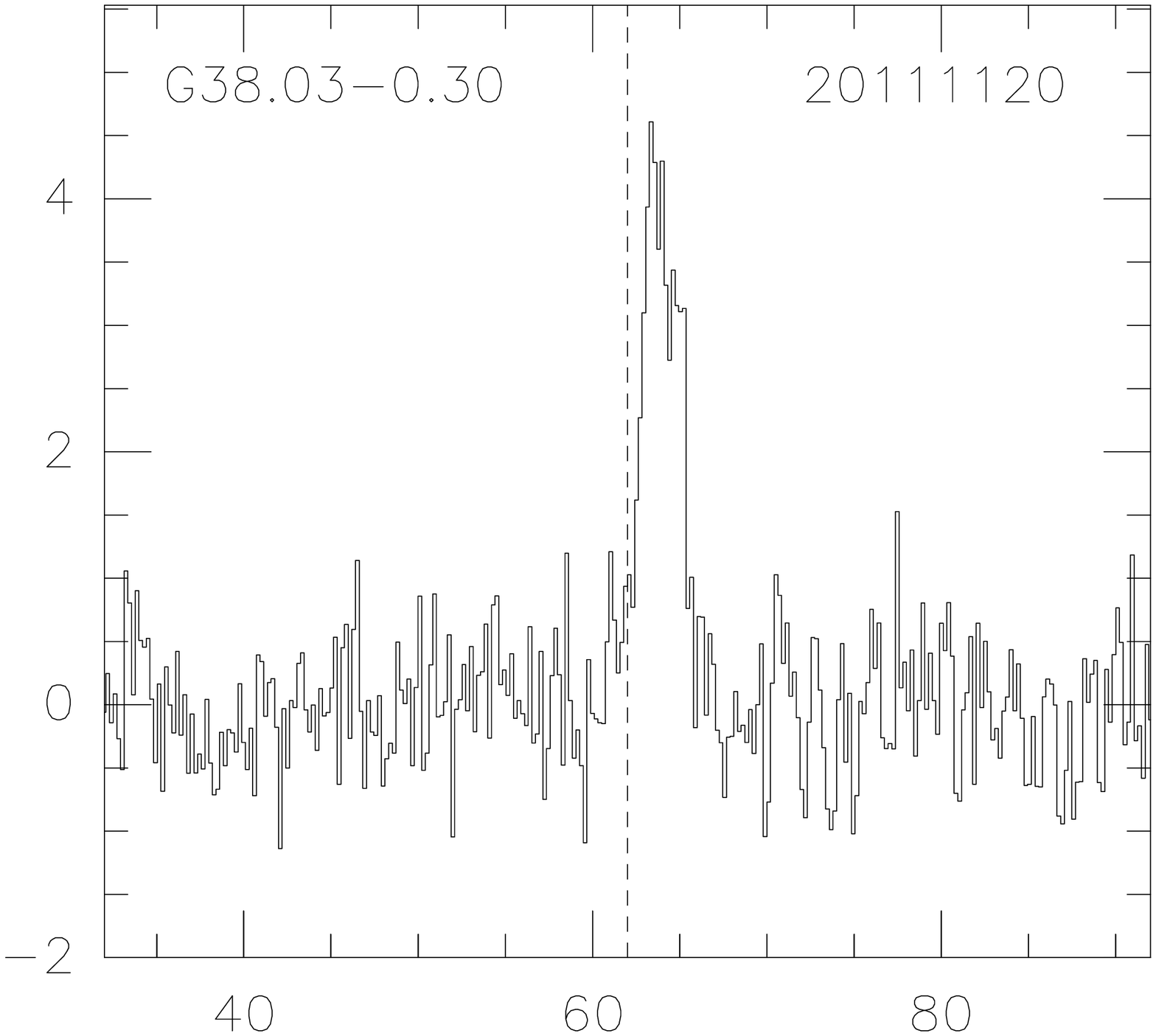} 
&
& 
\includegraphics[width=60mm]{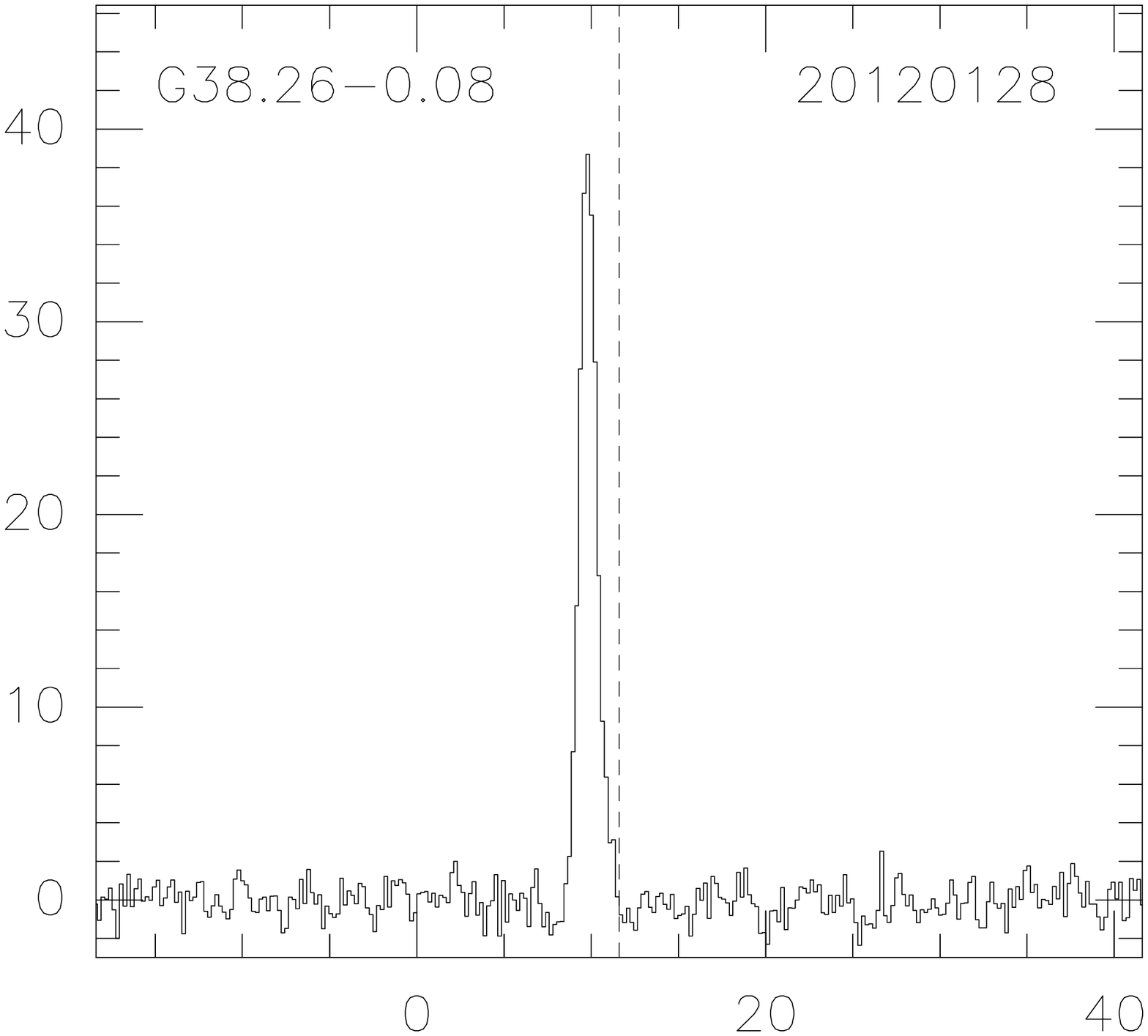} 
\\
&
\includegraphics[width=60mm]{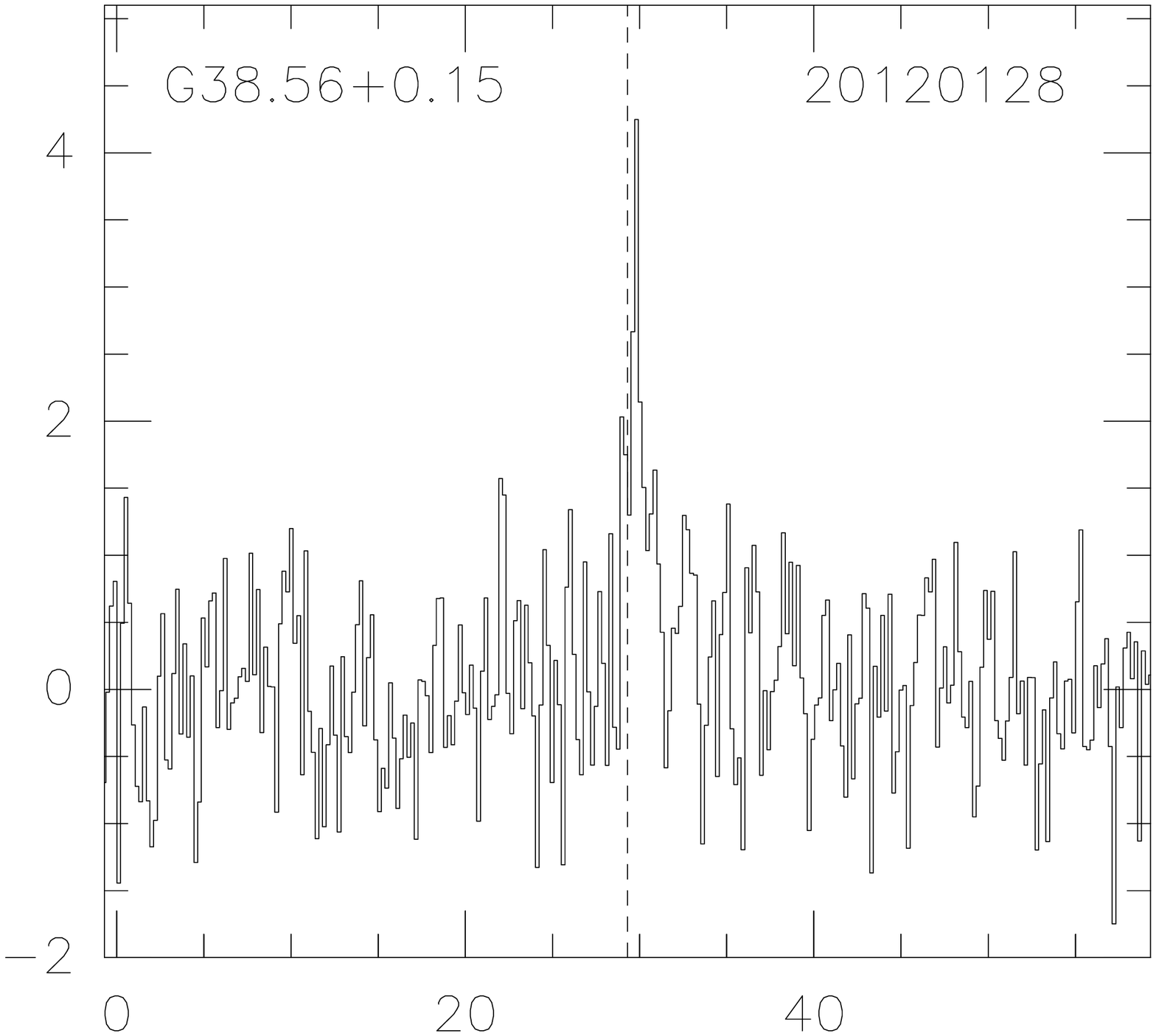} 
&
& 
\includegraphics[width=60mm]{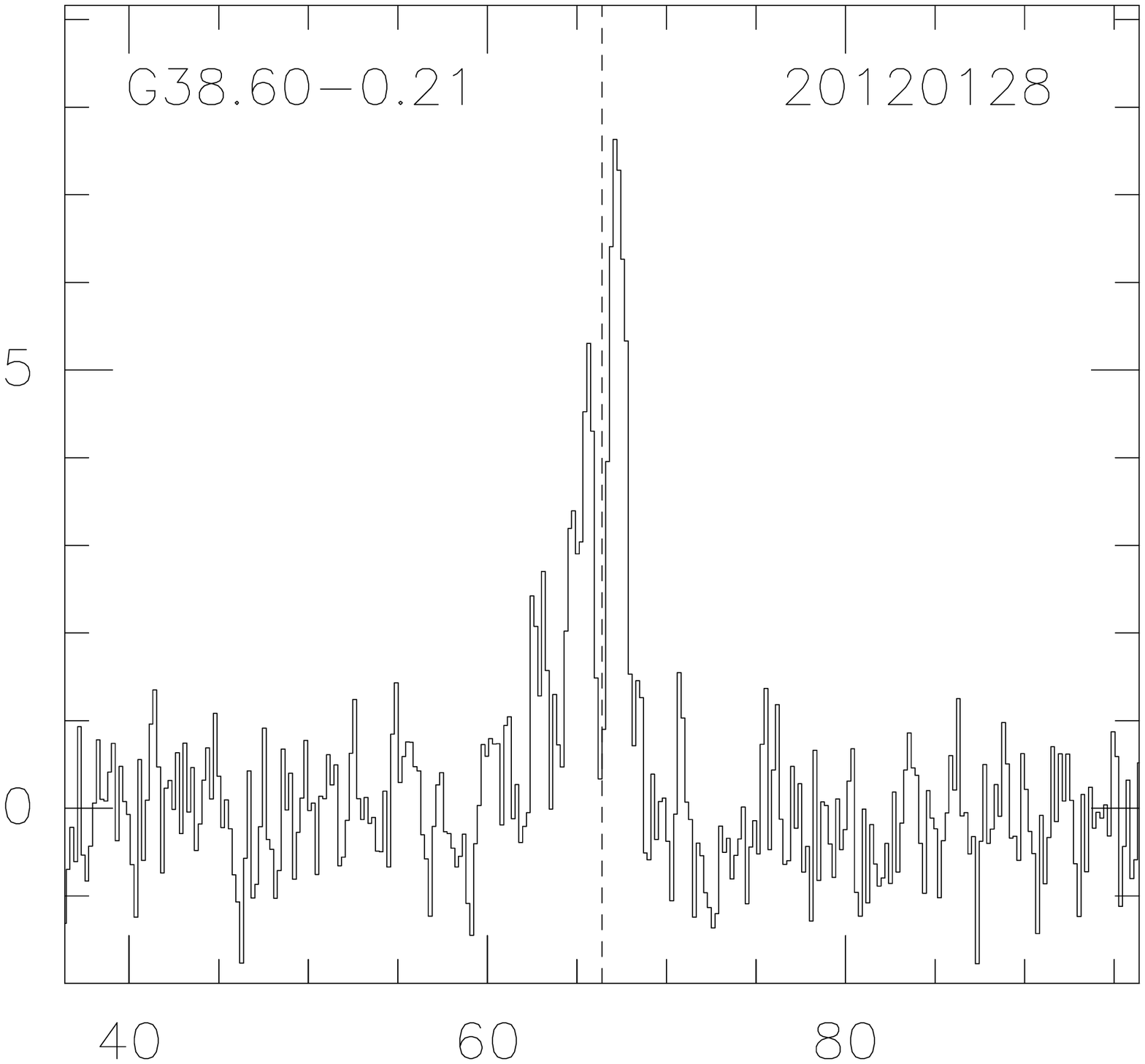} 
\\
&\multicolumn{3}{c}{V$_{lsr}$ (\kms)} \\
\end{tabular}
\caption{continued} 
\end{figure}
\clearpage
\begin{figure}
\ContinuedFloat
\begin{tabular}{cccc}
&
\includegraphics[width=58mm]{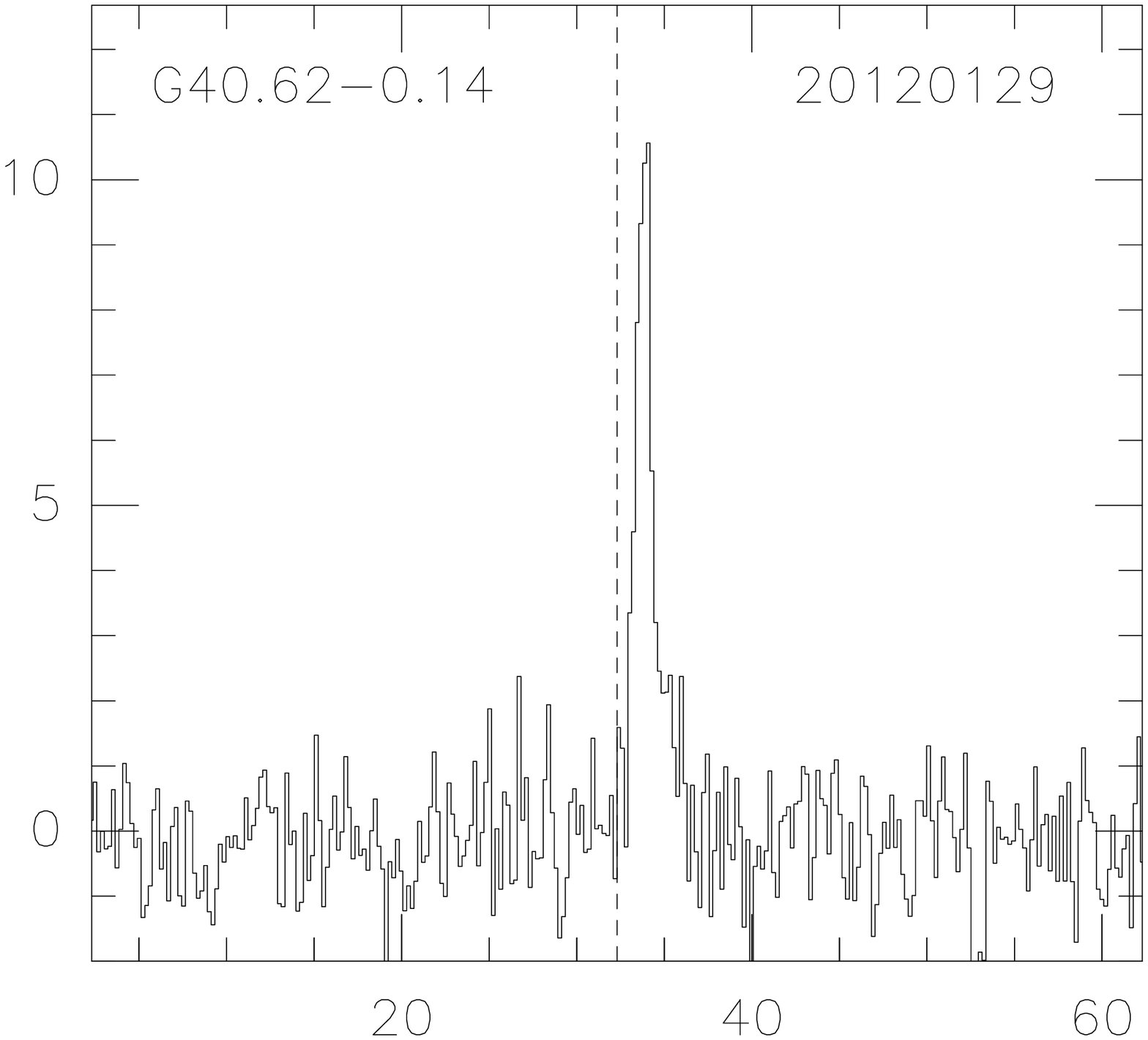} 
& 
&
\includegraphics[width=60mm]{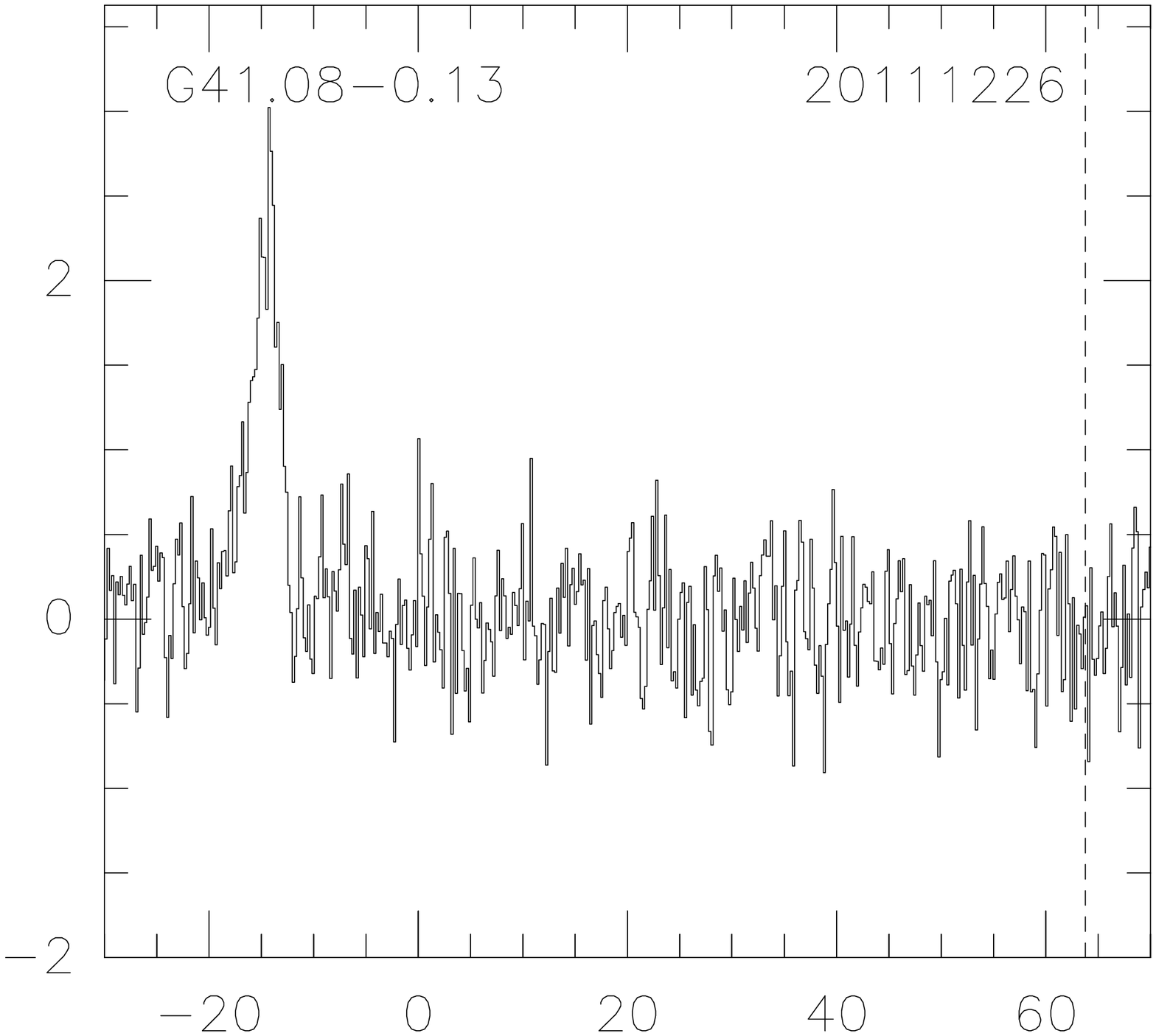} 
\\
{\rotatebox{90}{\qquad\qquad\qquad S (Jy)}}
&
\includegraphics[width=61mm]{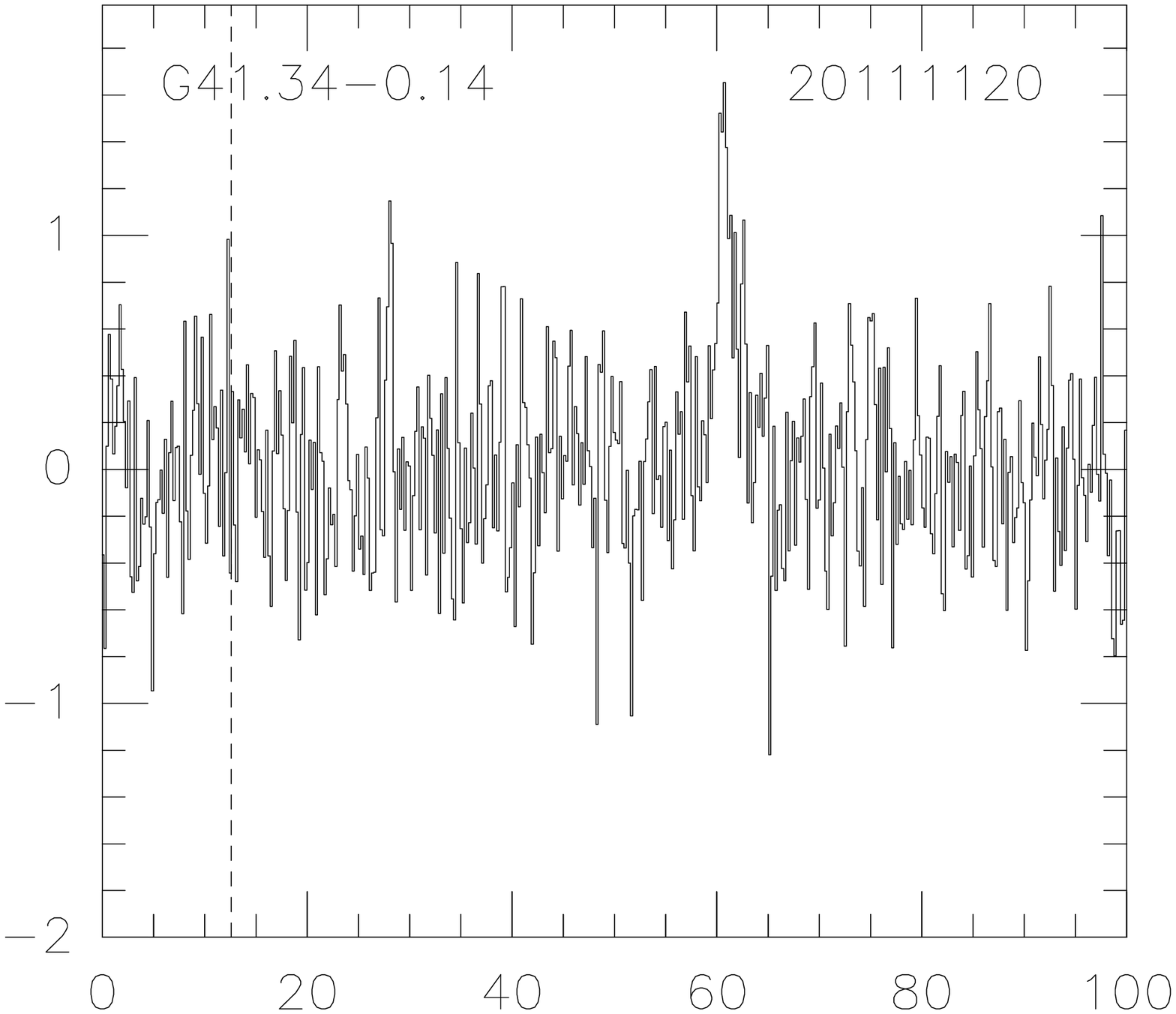} 
&
& 
\includegraphics[width=59mm]{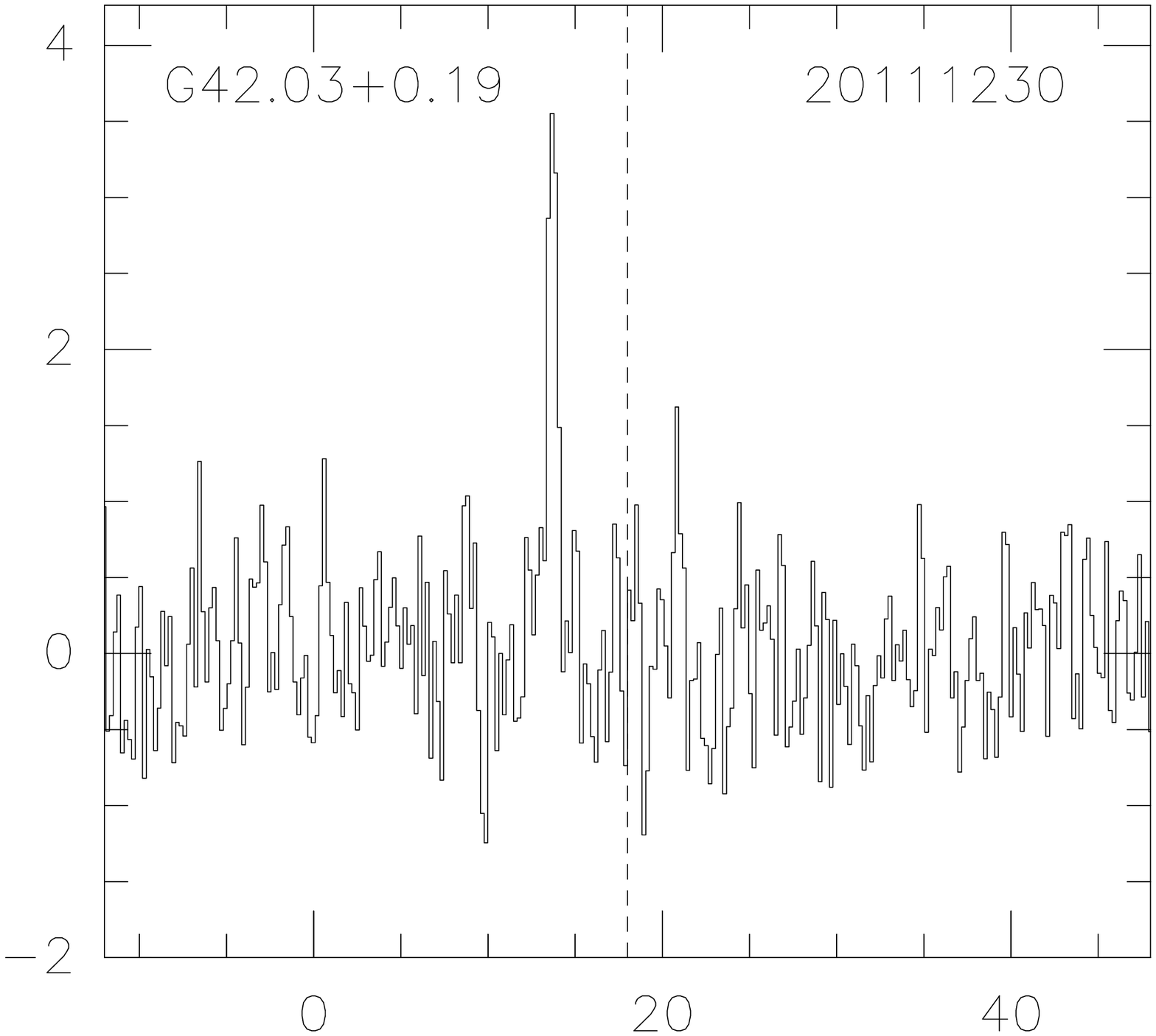} 
\\
&
\includegraphics[width=59mm]{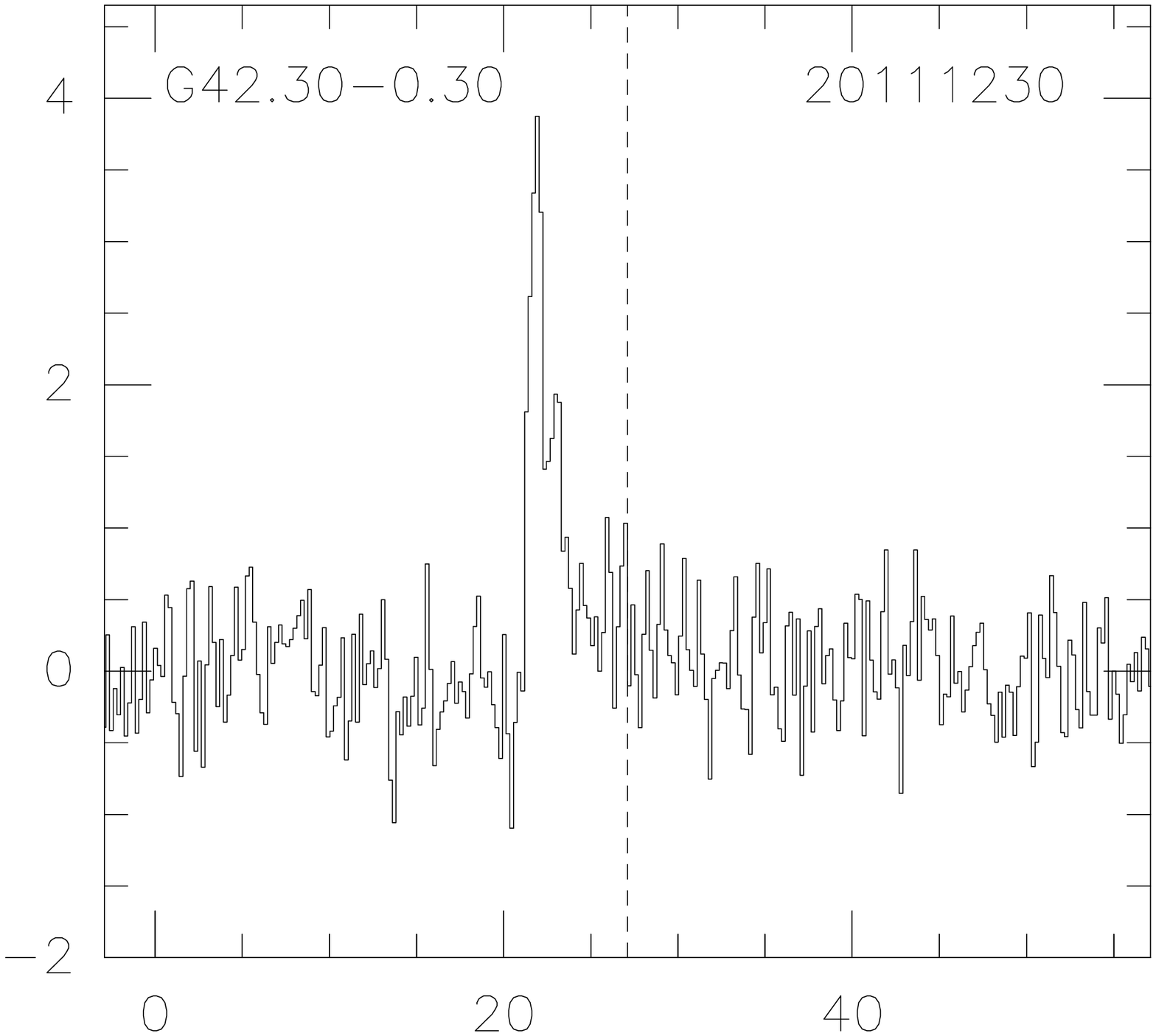} 
& 
&
\includegraphics[width=59mm]{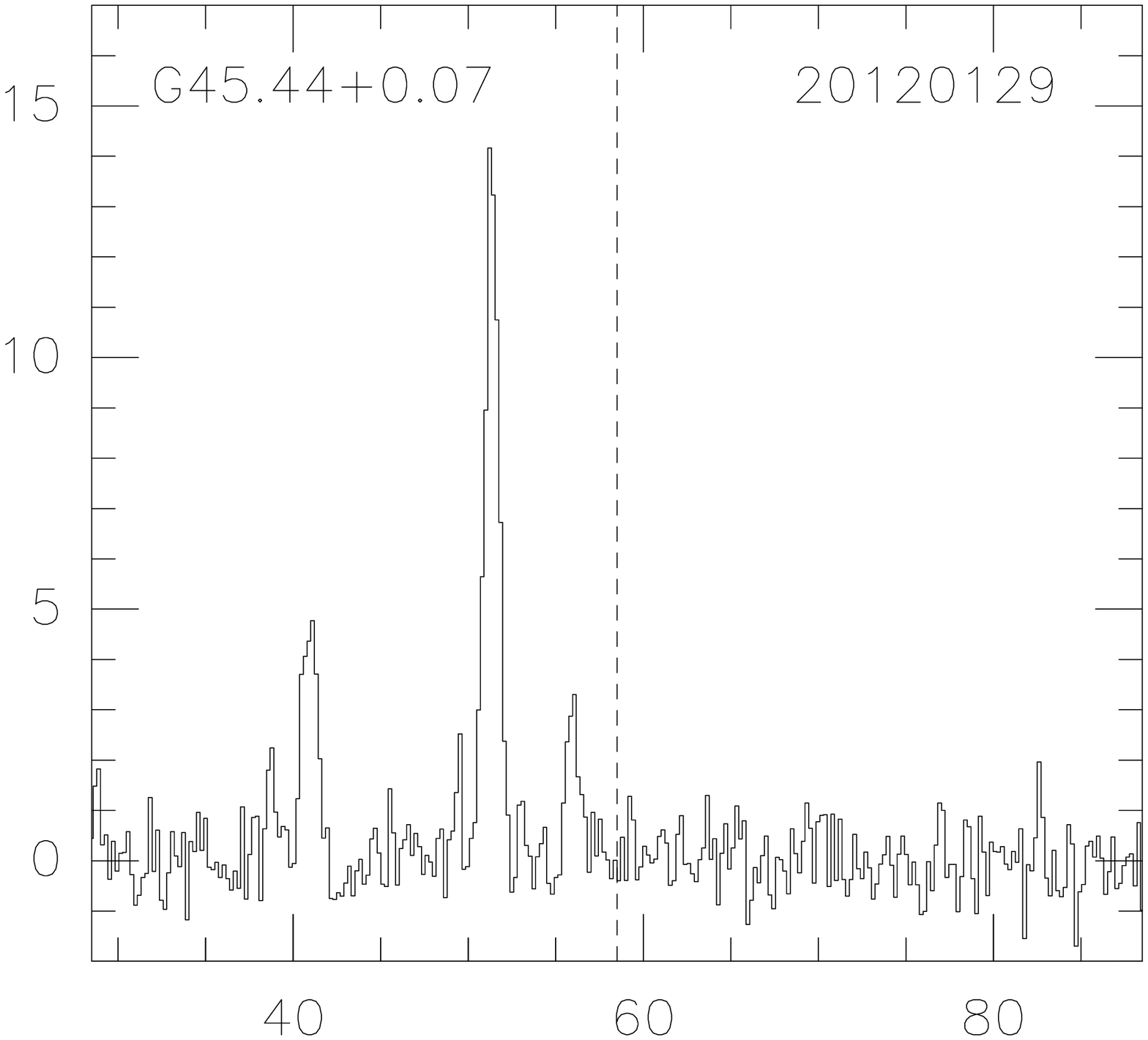} 
\\
&\multicolumn{3}{c}{V$_{lsr}$ (\kms)} \\
\end{tabular}
\caption{continued} 
\end{figure}
\clearpage
\begin{figure}
\ContinuedFloat
\begin{tabular}{cccc}
\multirow{3}*[4ex]{\rotatebox{90}{S (Jy)}}
&
\includegraphics[width=60mm]{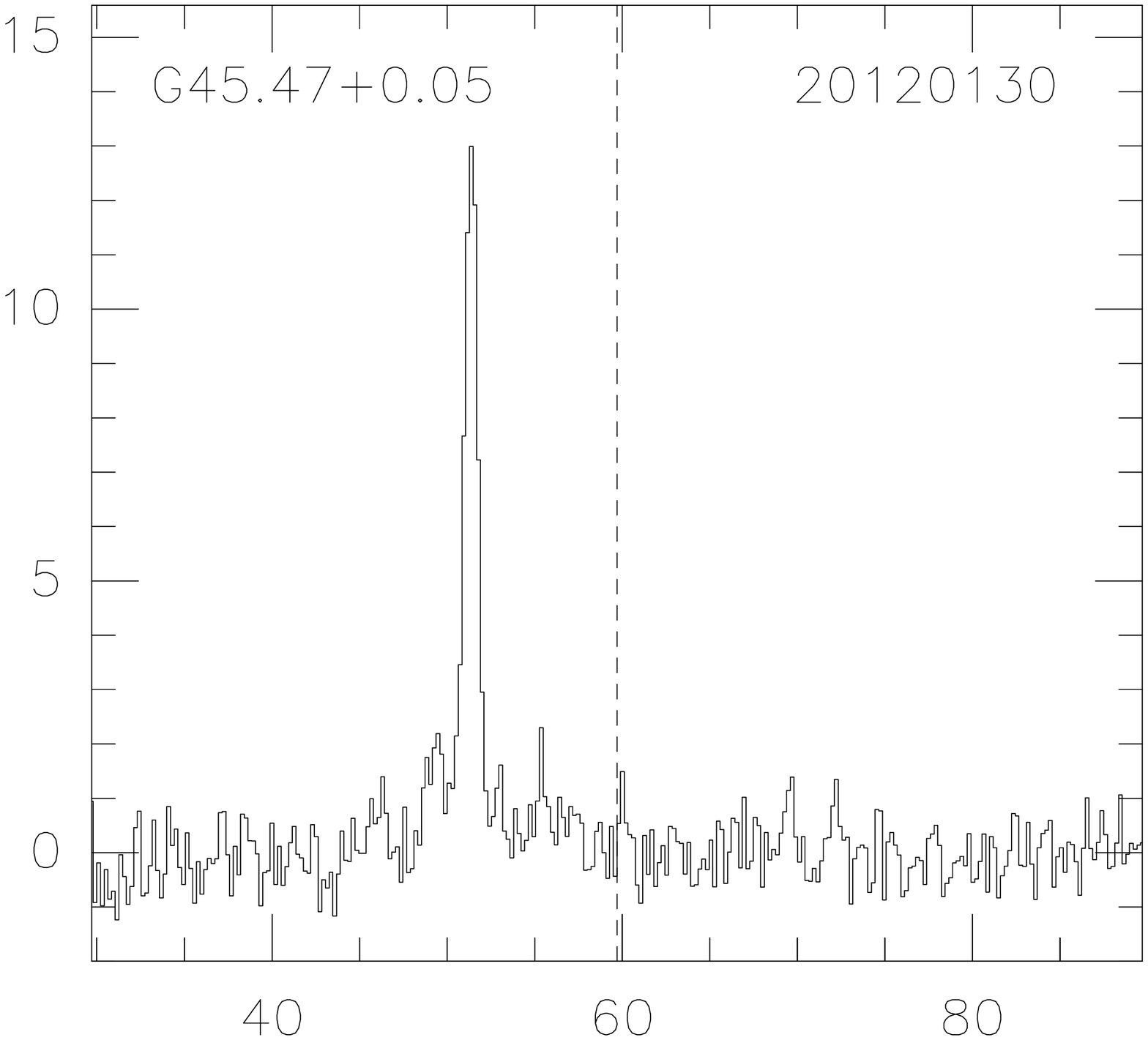} 
&
&
\includegraphics[width=59mm,height=54mm]{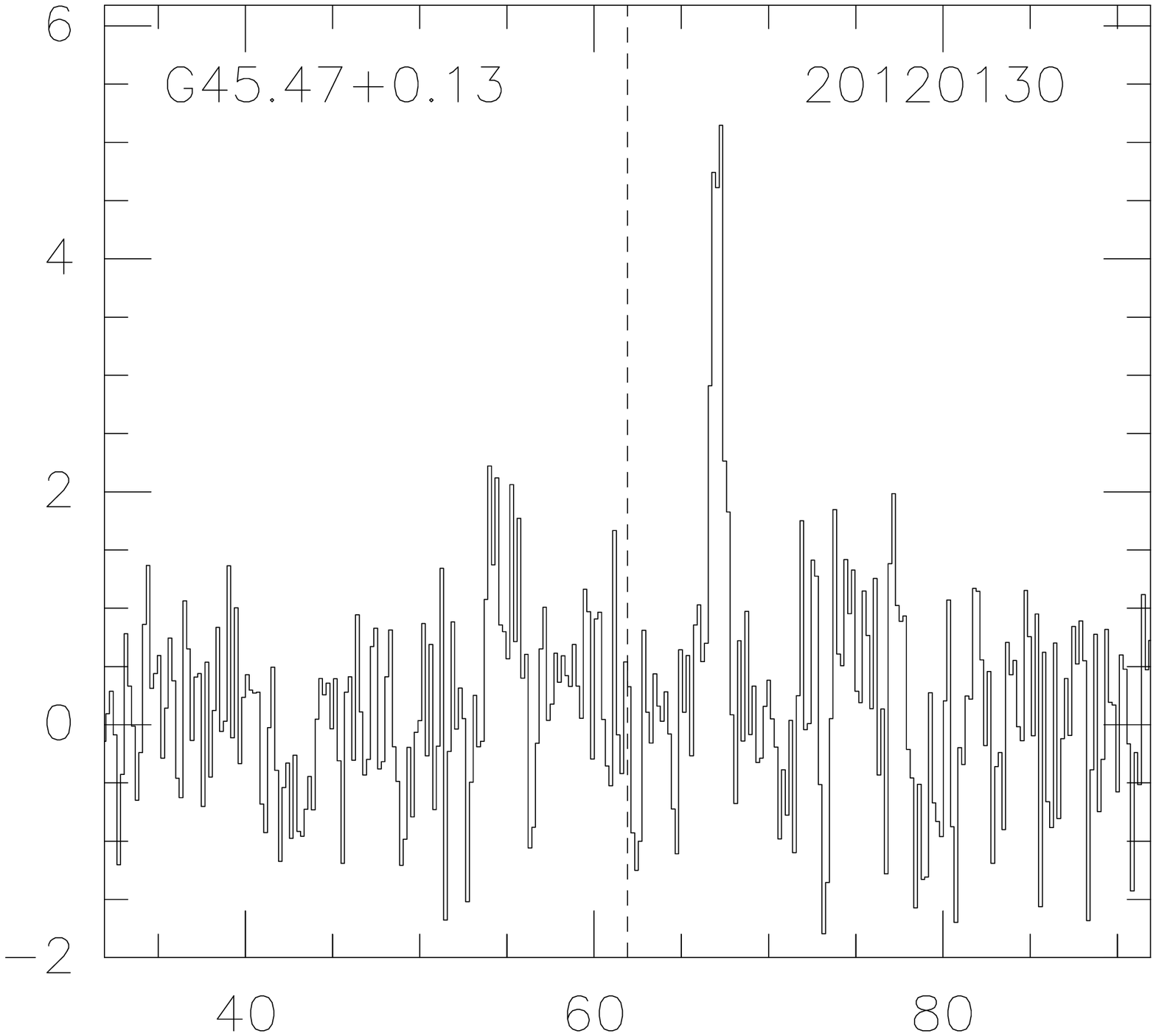} 
\\
& 
\includegraphics[width=60mm]{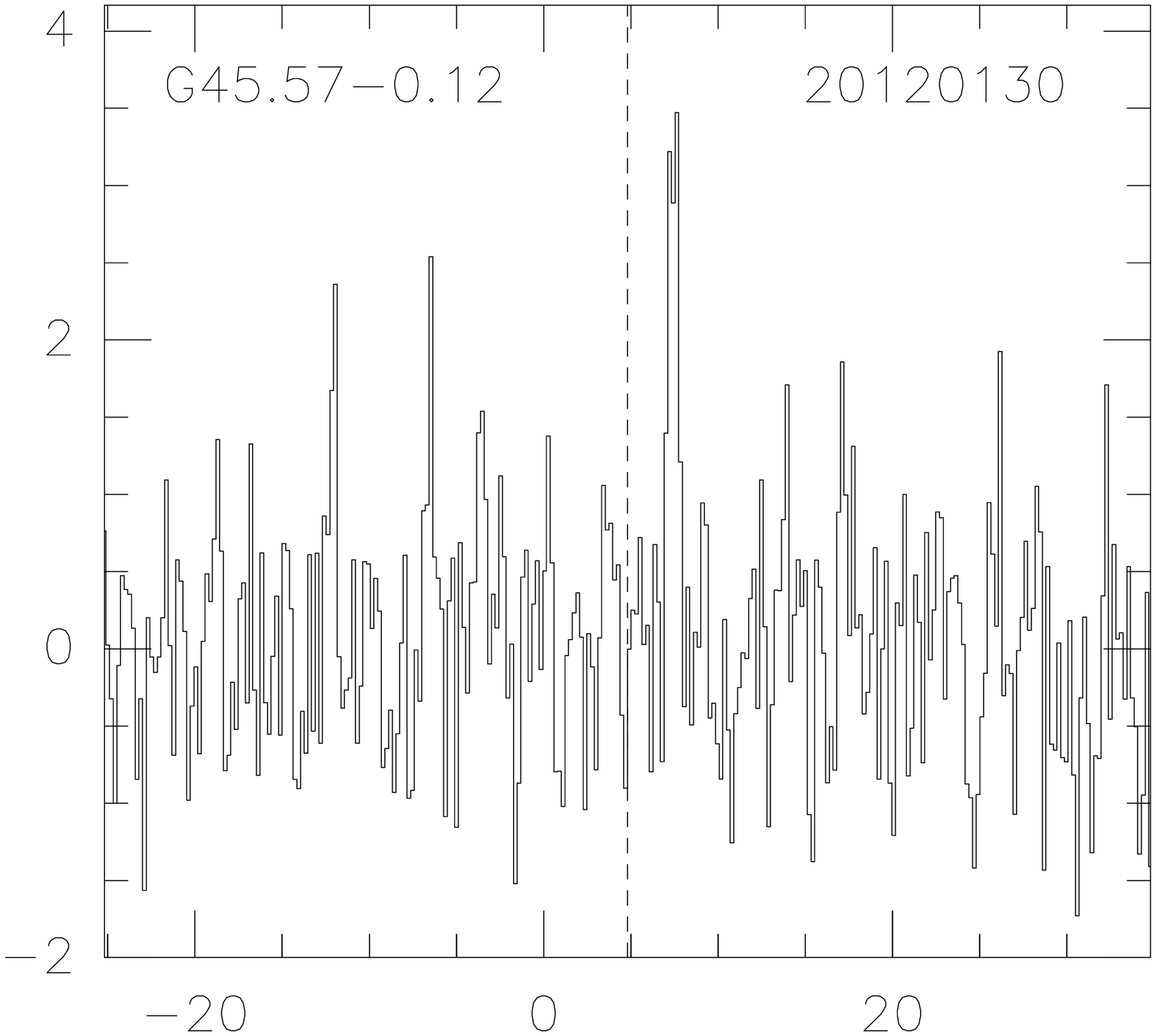} 
&
&
\includegraphics[width=61mm,height=54mm]{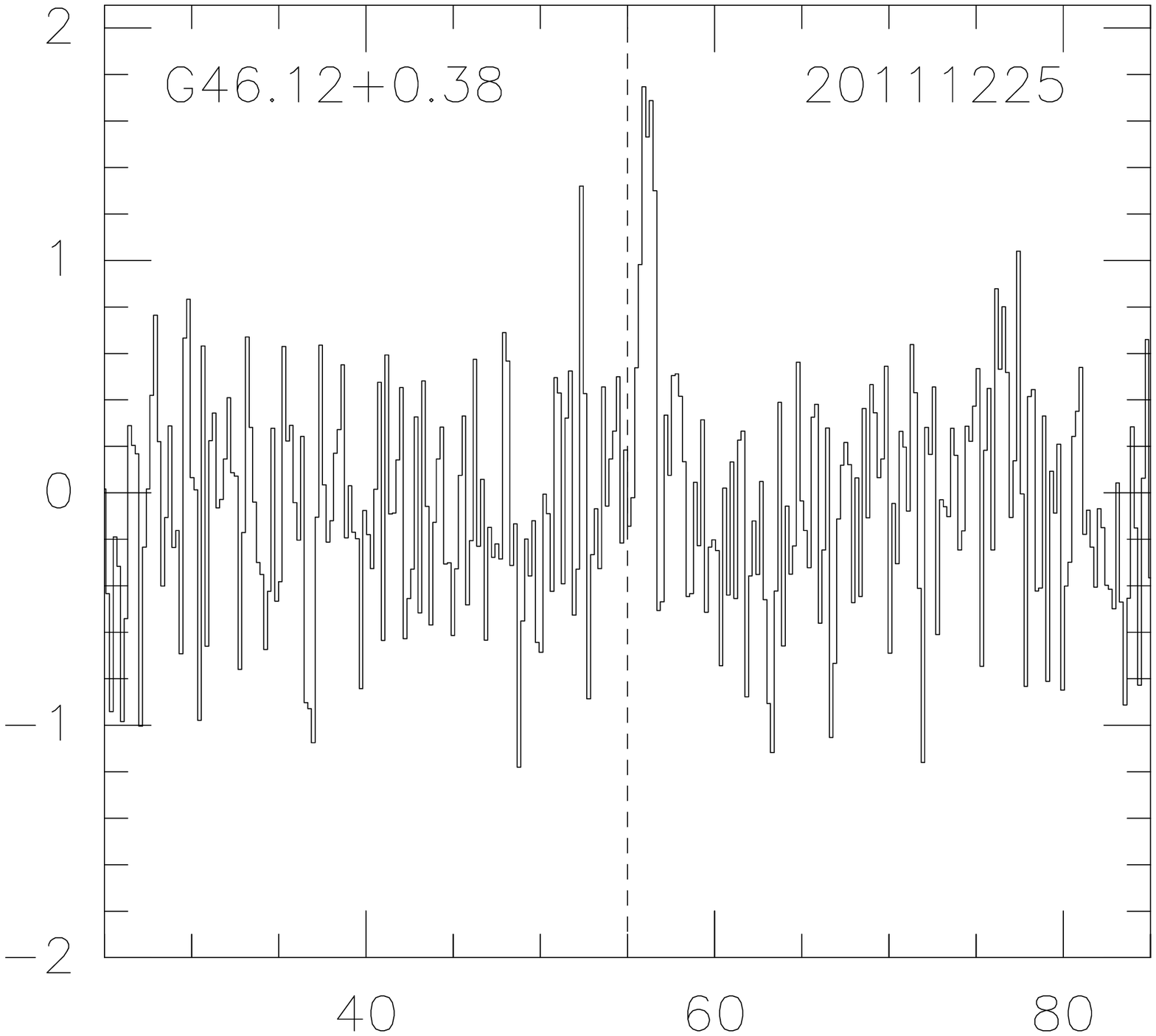} 
\\
&\multicolumn{3}{c}{V$_{lsr}$ (\kms)} \\
\end{tabular}
\caption{continued} 
\end{figure}

\begin{figure}
\begin{tabular}{cccc}
\multirow{3}*[4ex]{\rotatebox{90}{S (Jy)}}
&
\includegraphics[width=60mm]{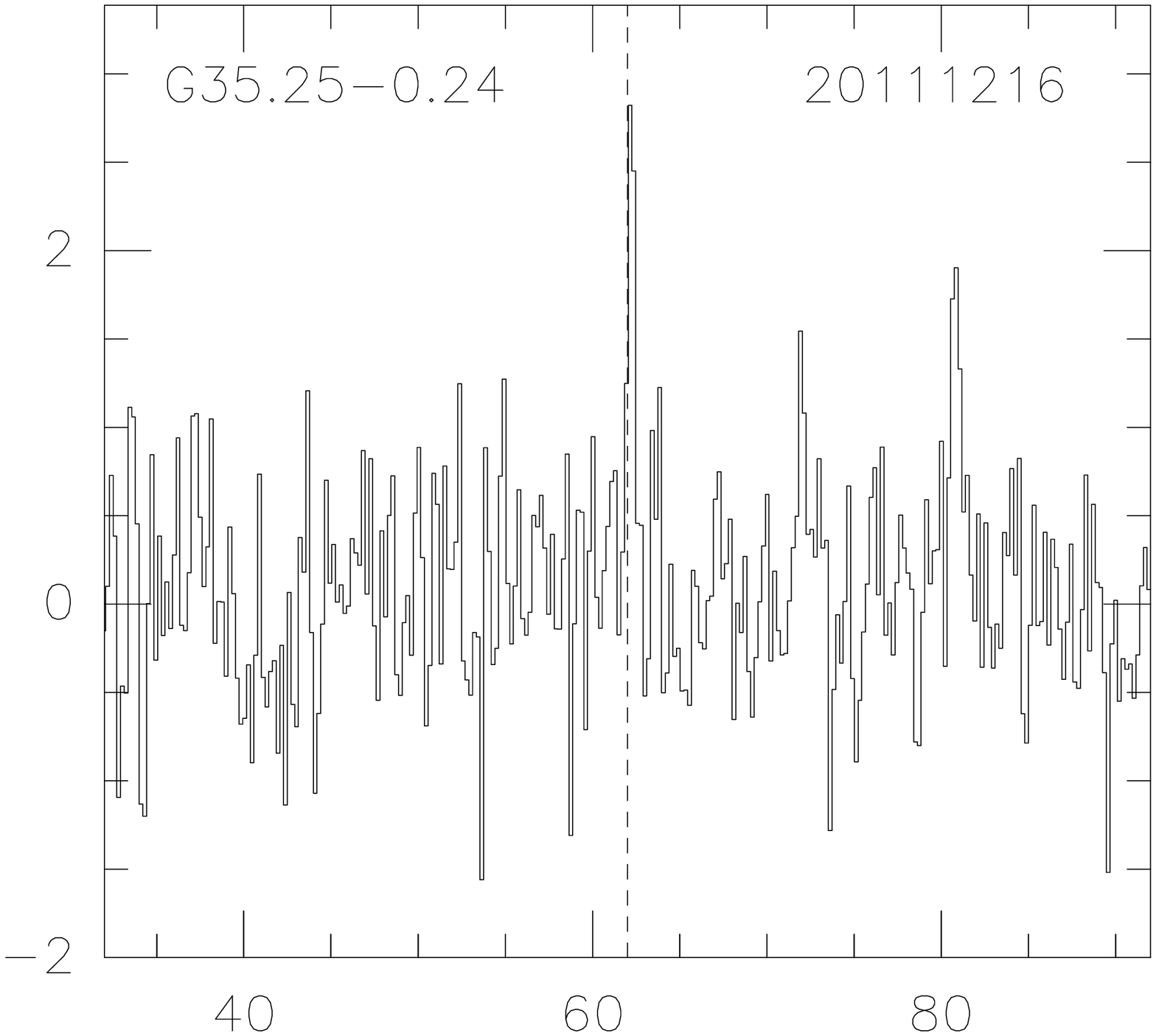} 
& 
&
\includegraphics[width=60mm]{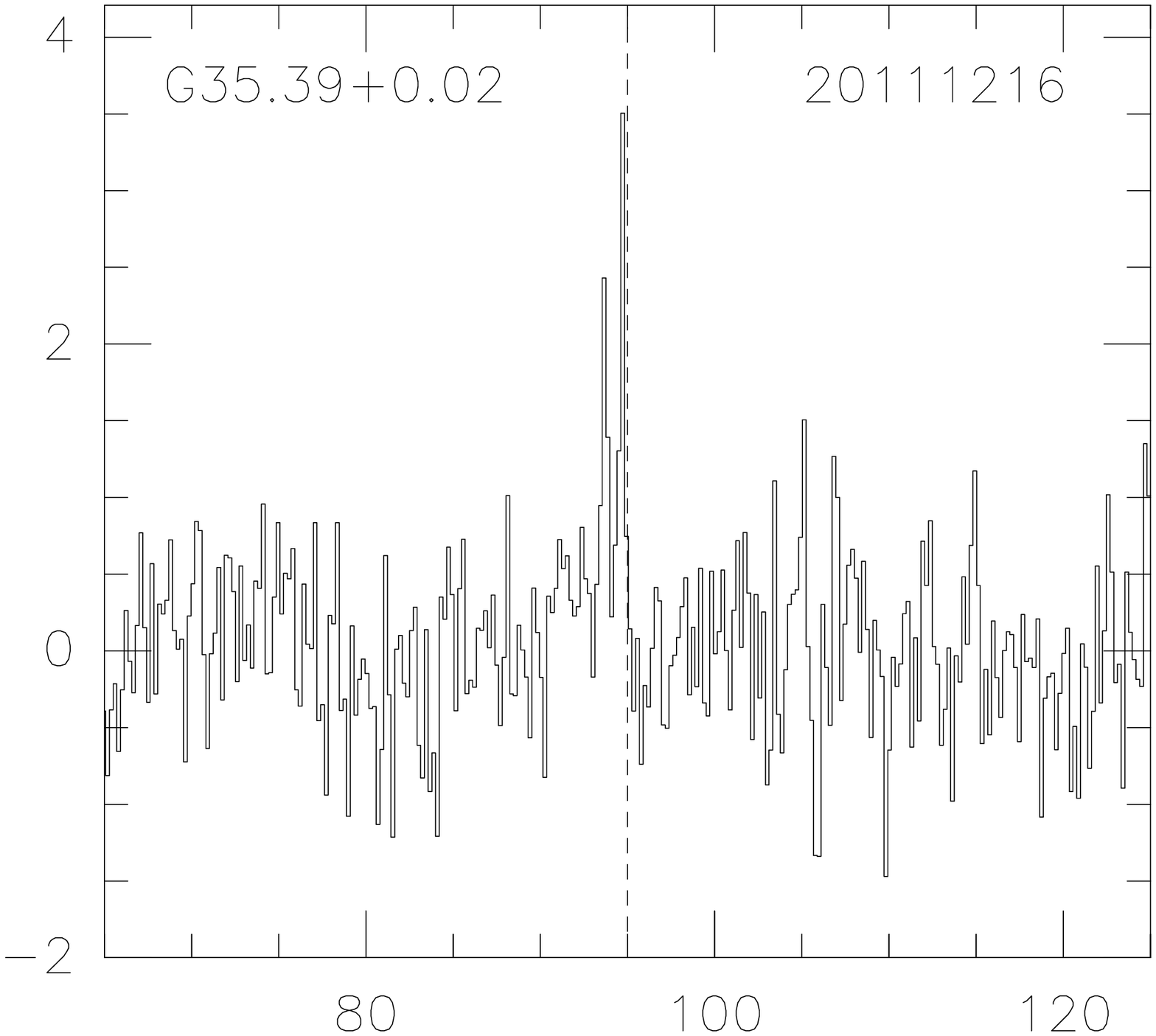} 
\\
&
\includegraphics[width=60mm]{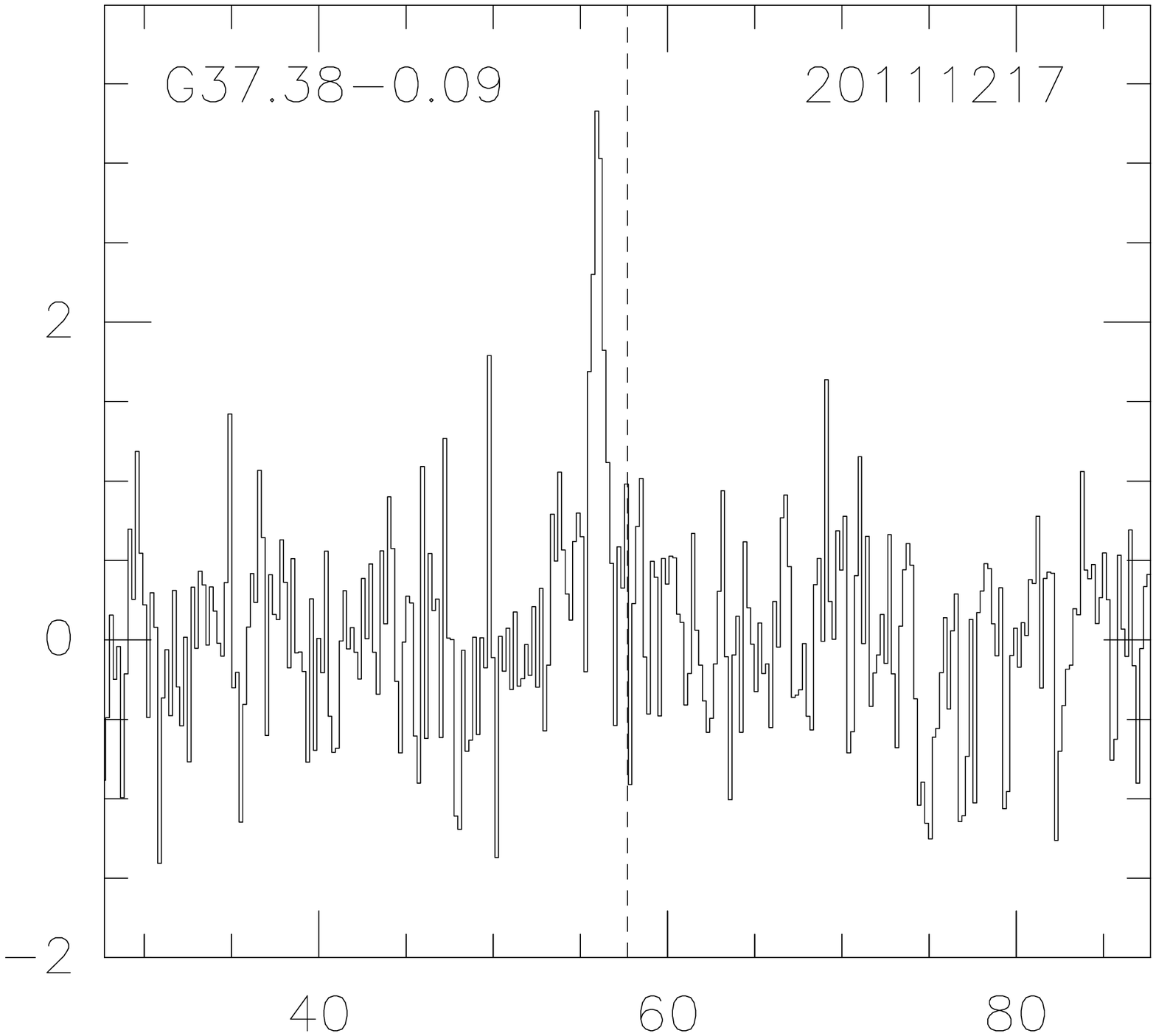} 
&
& 
\includegraphics[width=60mm]{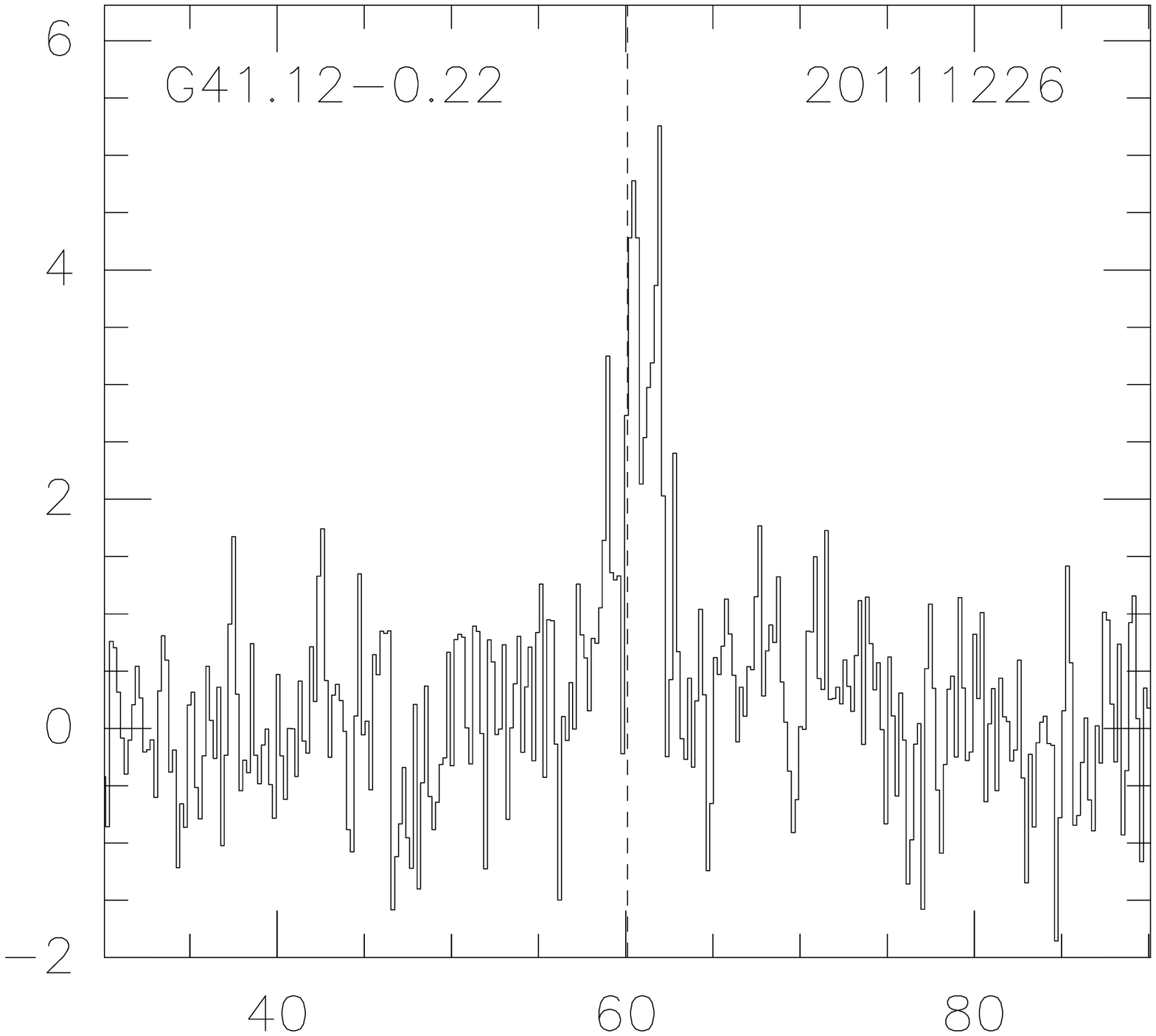} 
\\
&\multicolumn{3}{c}{V$_{lsr}$ (\kms)} \\
\end{tabular}
\caption{Same as in Figure 2 except for sources detected only in the 44\ghz\ class I methanol maser.\label{fig:44only}}
\end{figure}
\clearpage
\begin{figure}
\begin{tabular}{cccc}
&
\includegraphics[width=60mm]{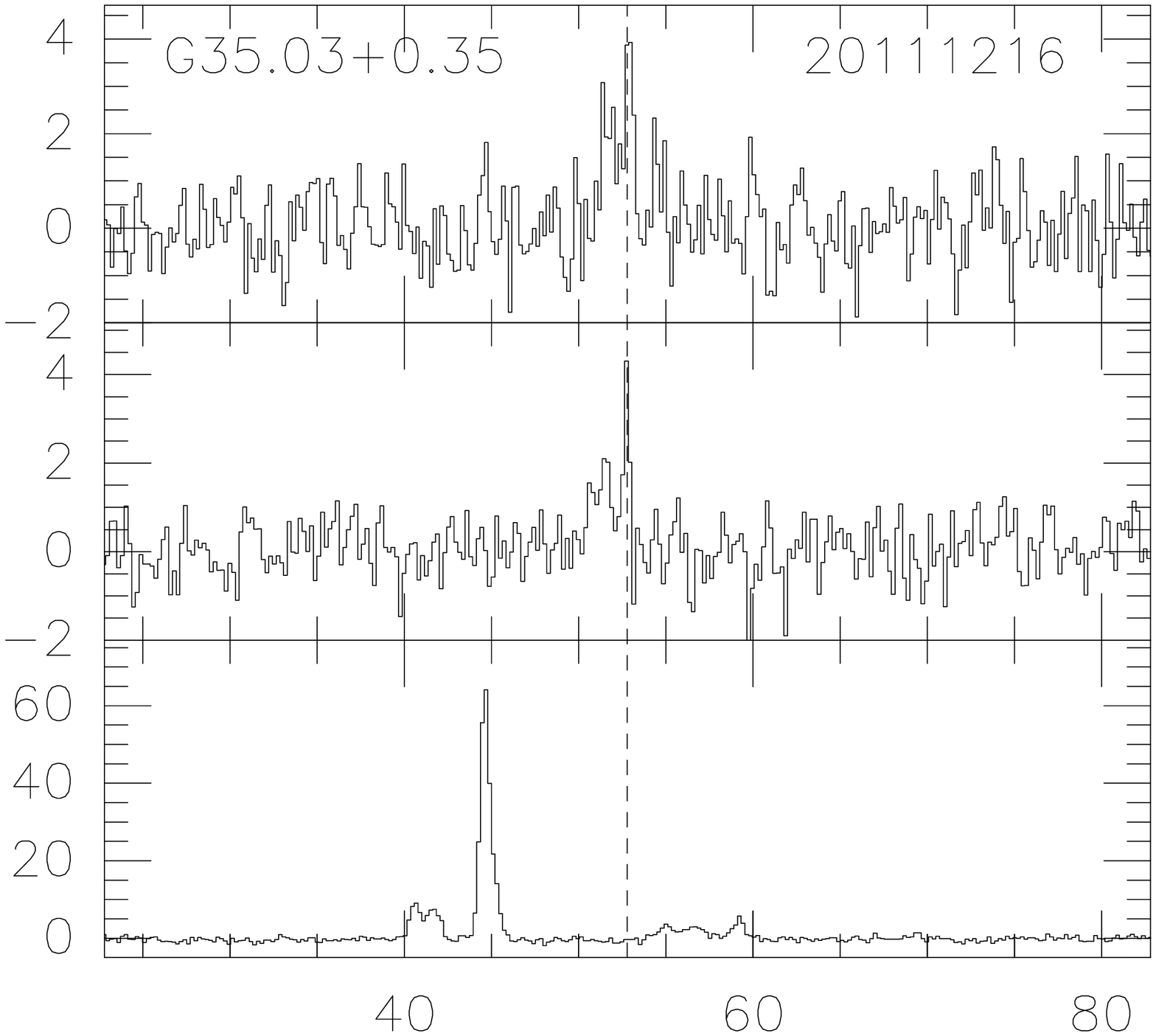} 
& 
&
\includegraphics[width=60mm]{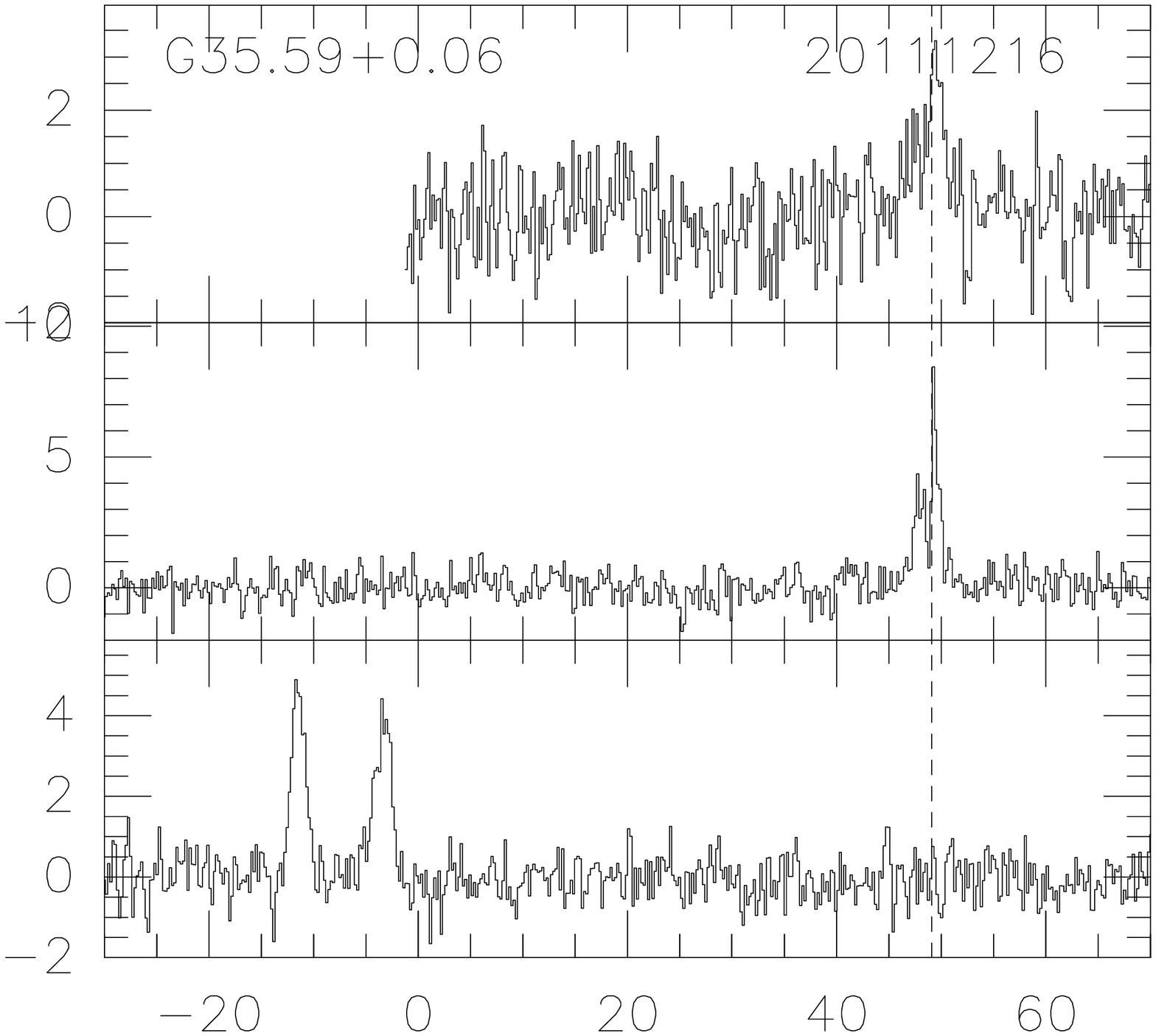} 
\\
{\rotatebox{90}{\qquad\qquad\qquad S (Jy)}}
&
\includegraphics[width=60mm]{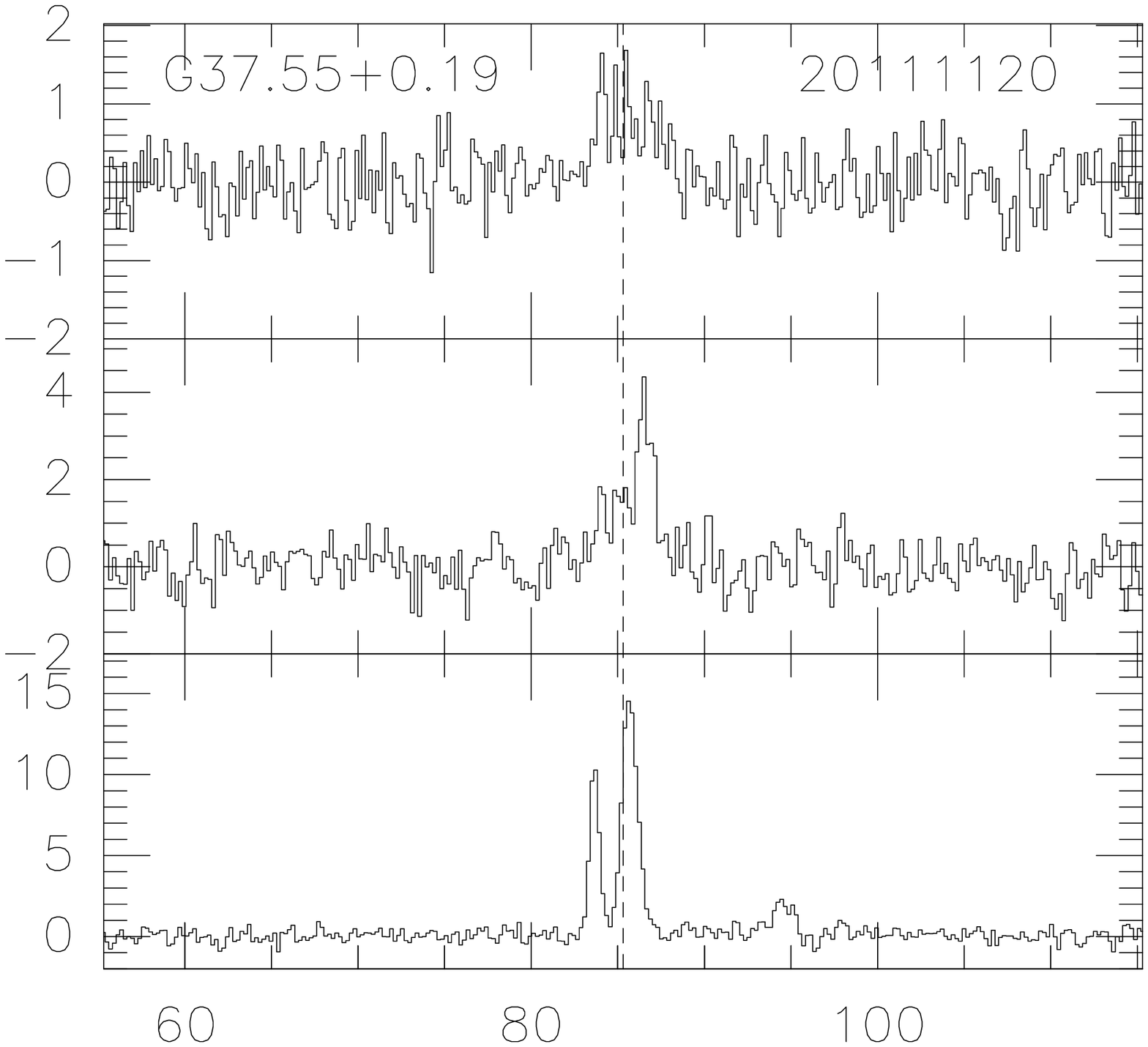} 
&
& 
\includegraphics[width=62mm]{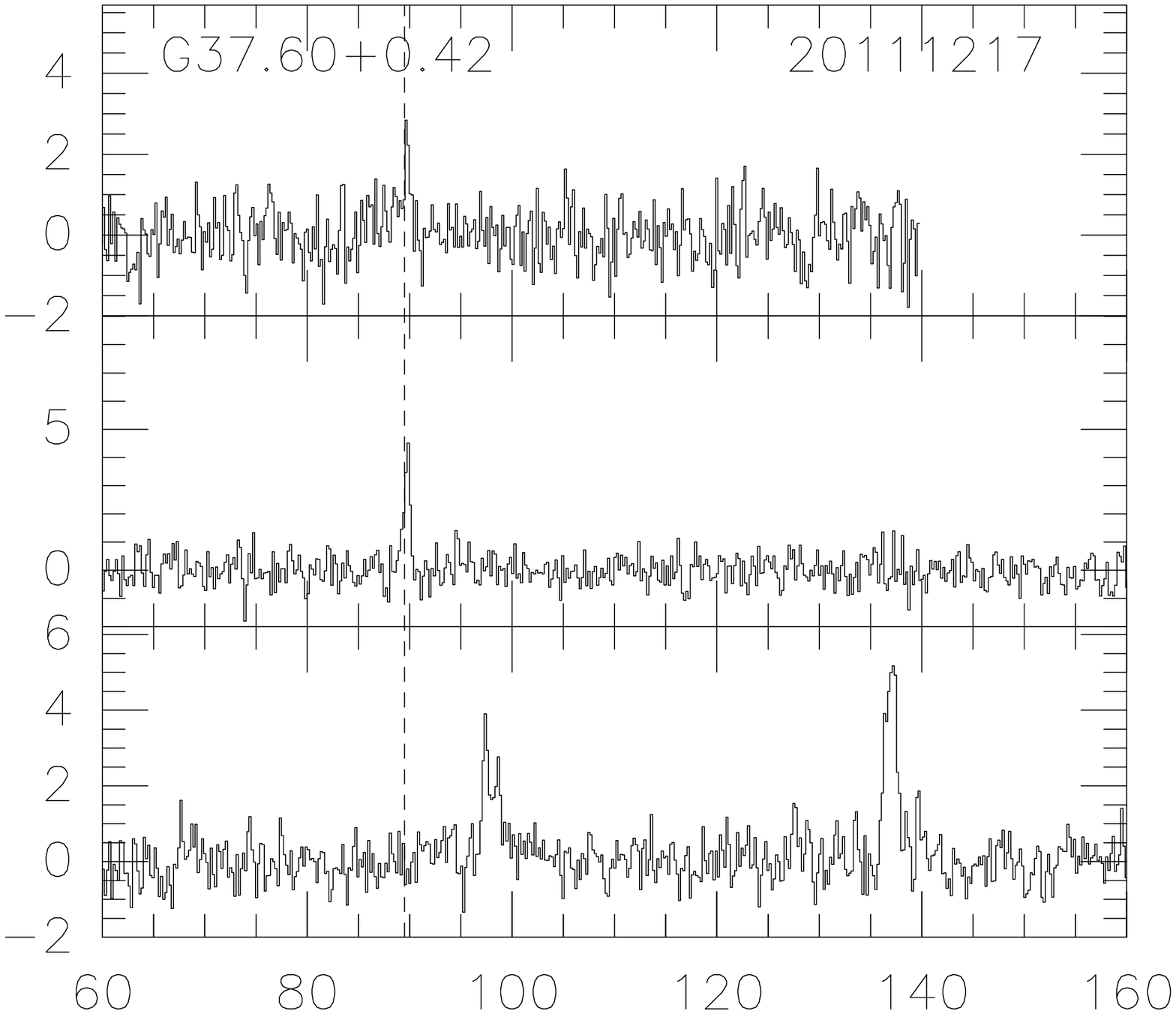} 
\\
&
\includegraphics[width=60mm]{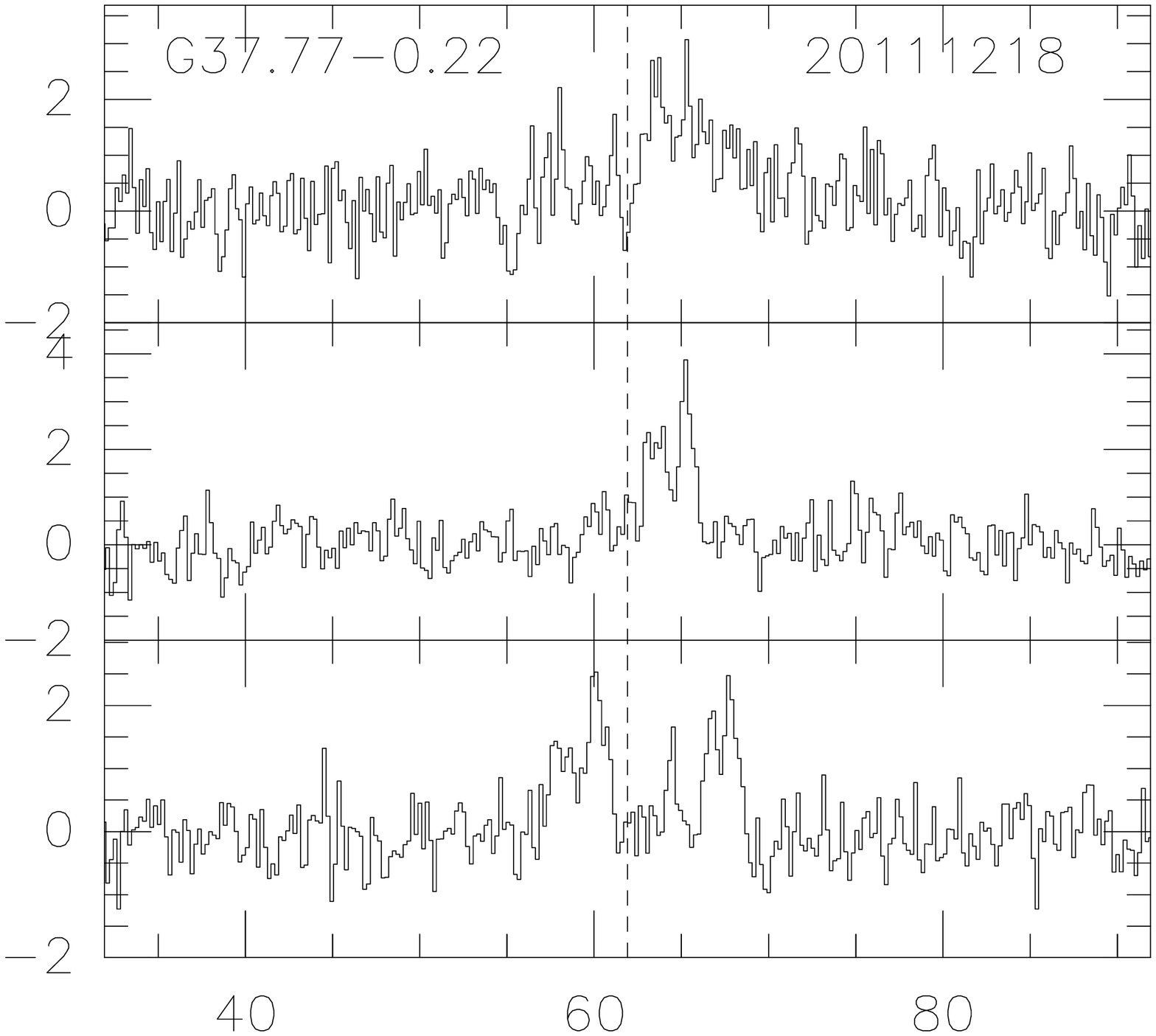} 
&
& 
\includegraphics[width=60mm]{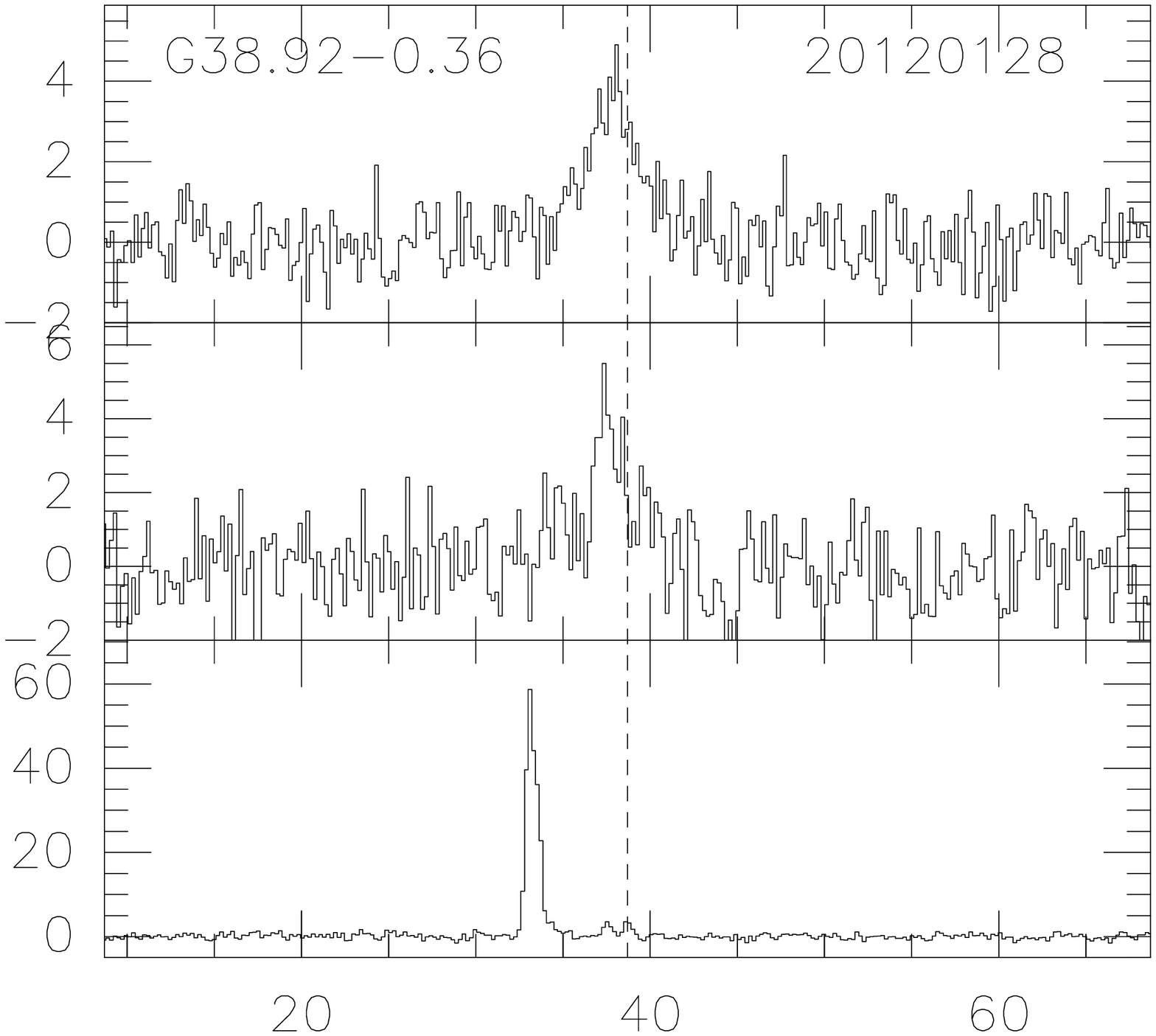} 
\\
&\multicolumn{3}{c}{V$_{lsr}$ (\kms)} \\
\end{tabular}
\caption{Same as in Figure 2 except for sources detected in all the three maser transitions: 22\ghz\ (bottom), 44\ghz\ (middle), and 95\ghz\ (top).\label{fig:all}}
\end{figure}
\clearpage
\begin{figure}
\ContinuedFloat
\begin{tabular}{cccc}
&
\includegraphics[width=60mm]{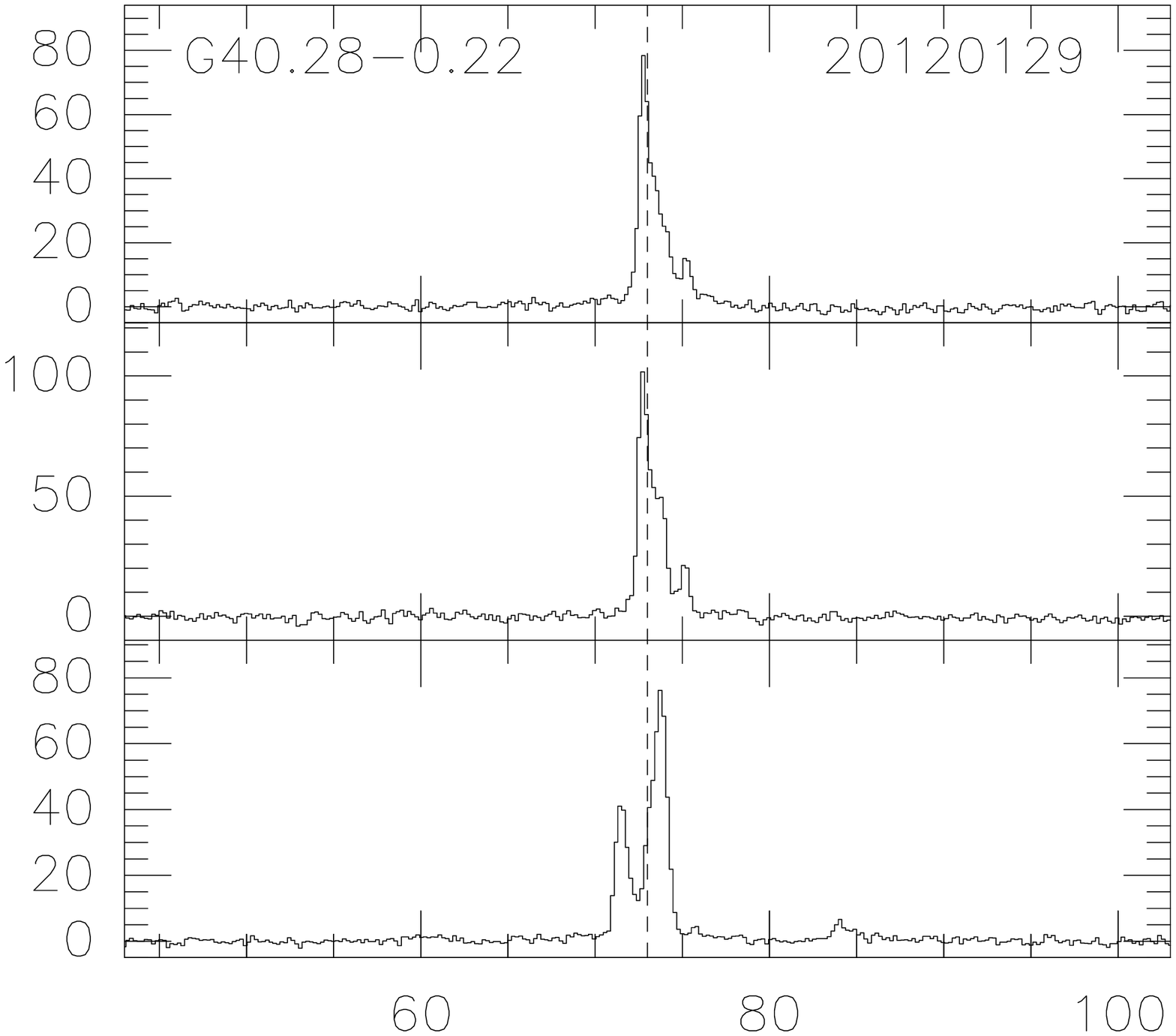} 
&
& 
\includegraphics[width=60mm]{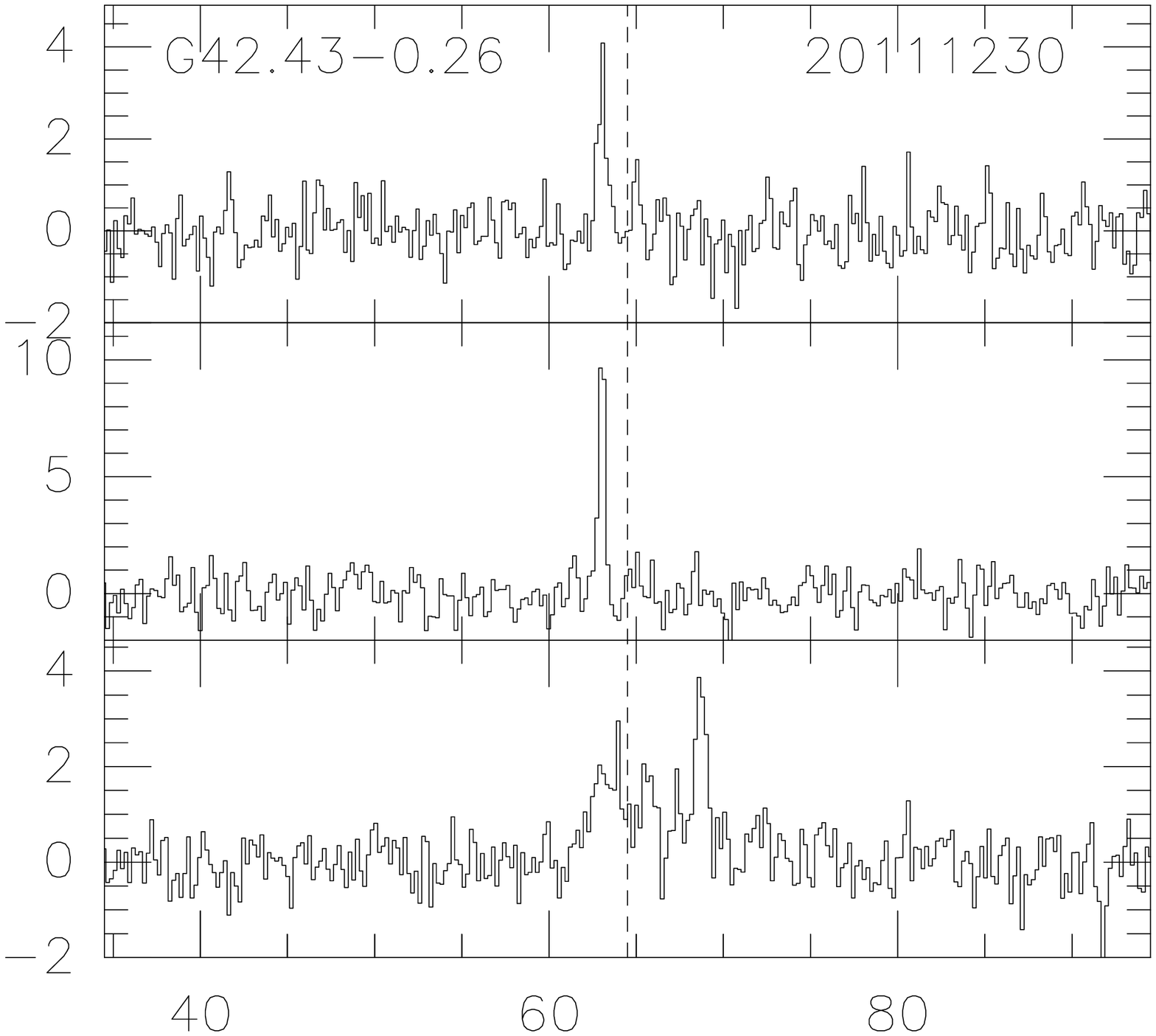} 
\\
{\rotatebox{90}{\qquad\qquad\qquad S (Jy)}}
&
\includegraphics[width=57mm]{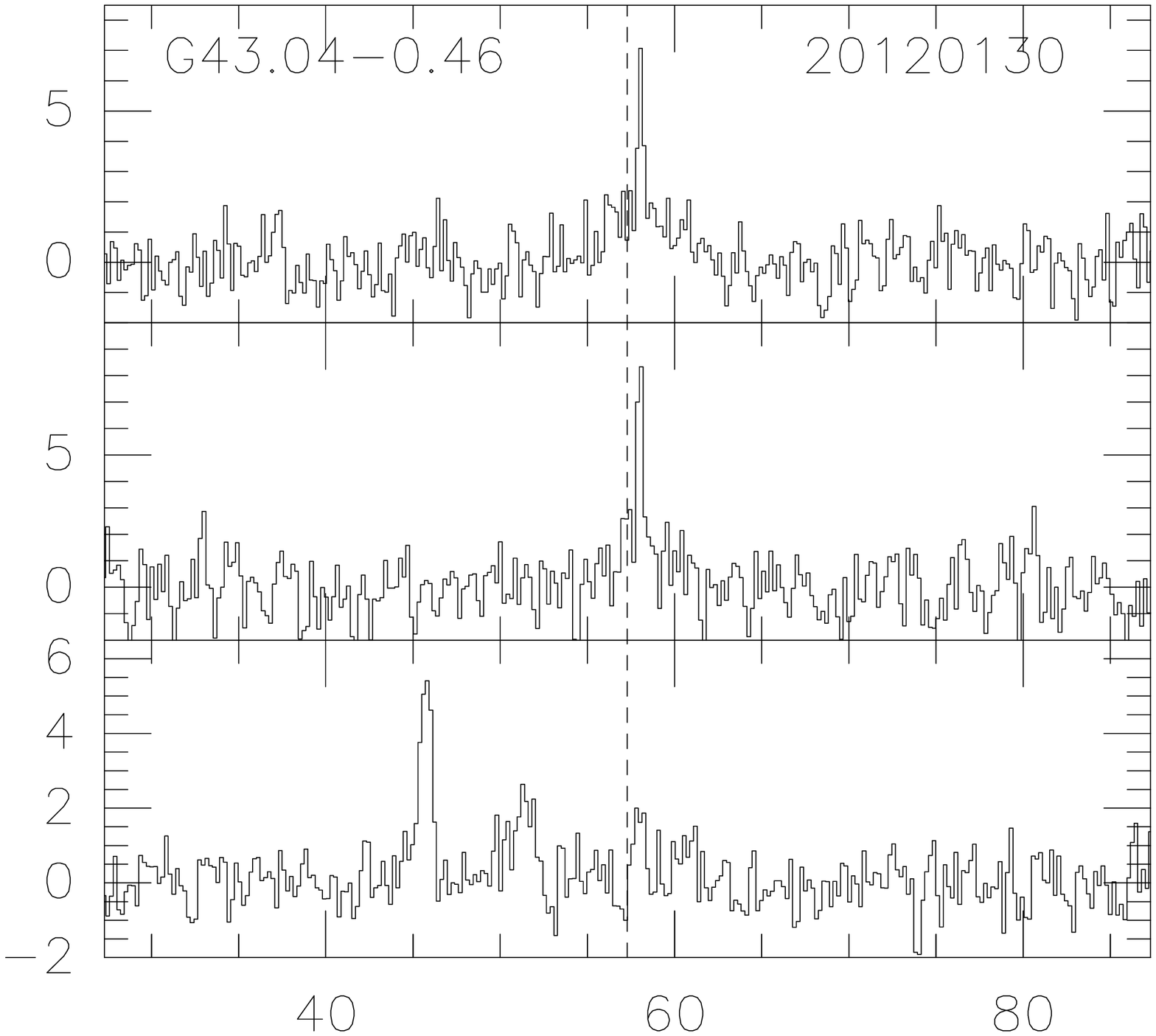} 
&
& 
\includegraphics[width=61mm]{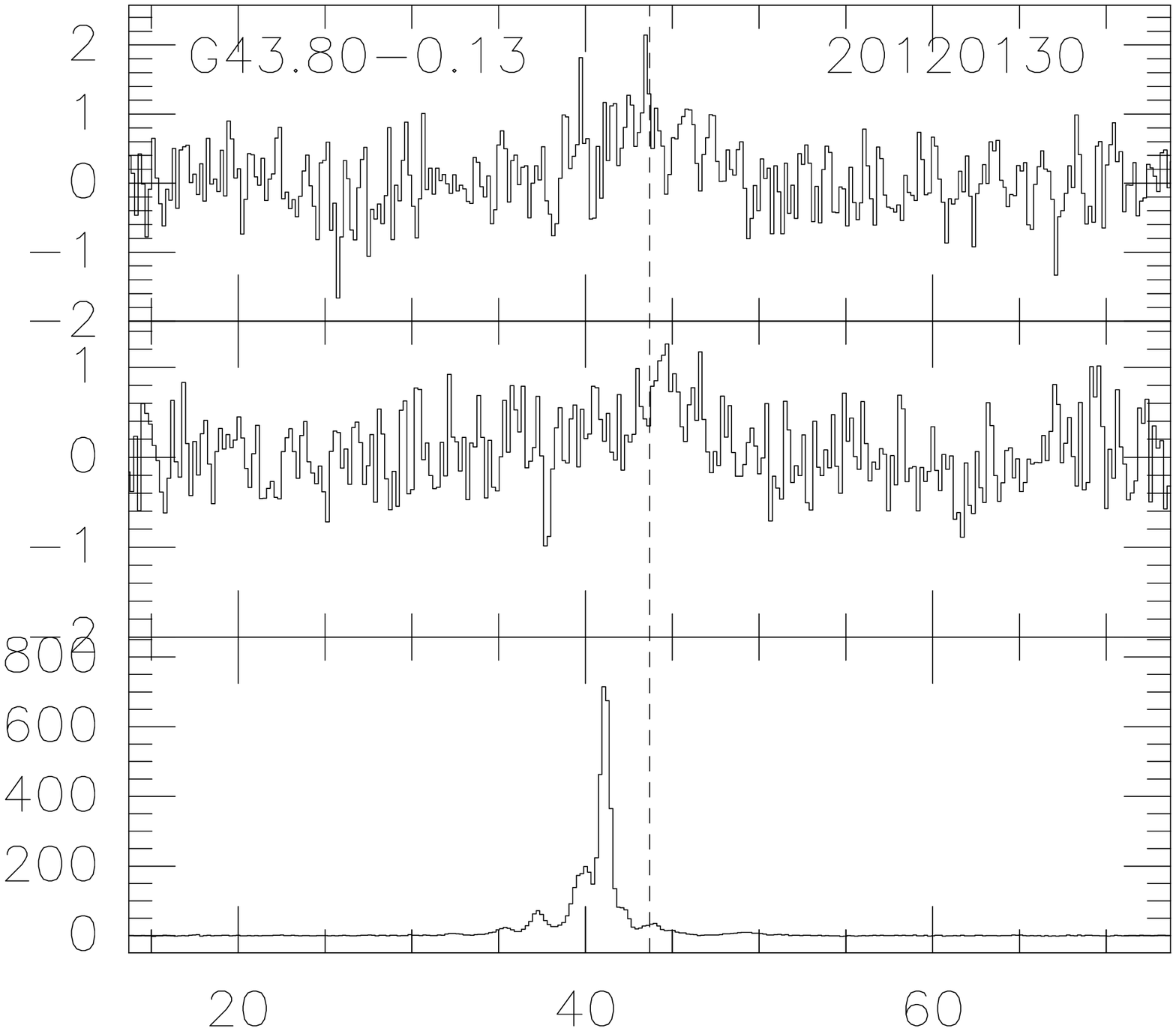} 
\\
&
\includegraphics[width=60mm]{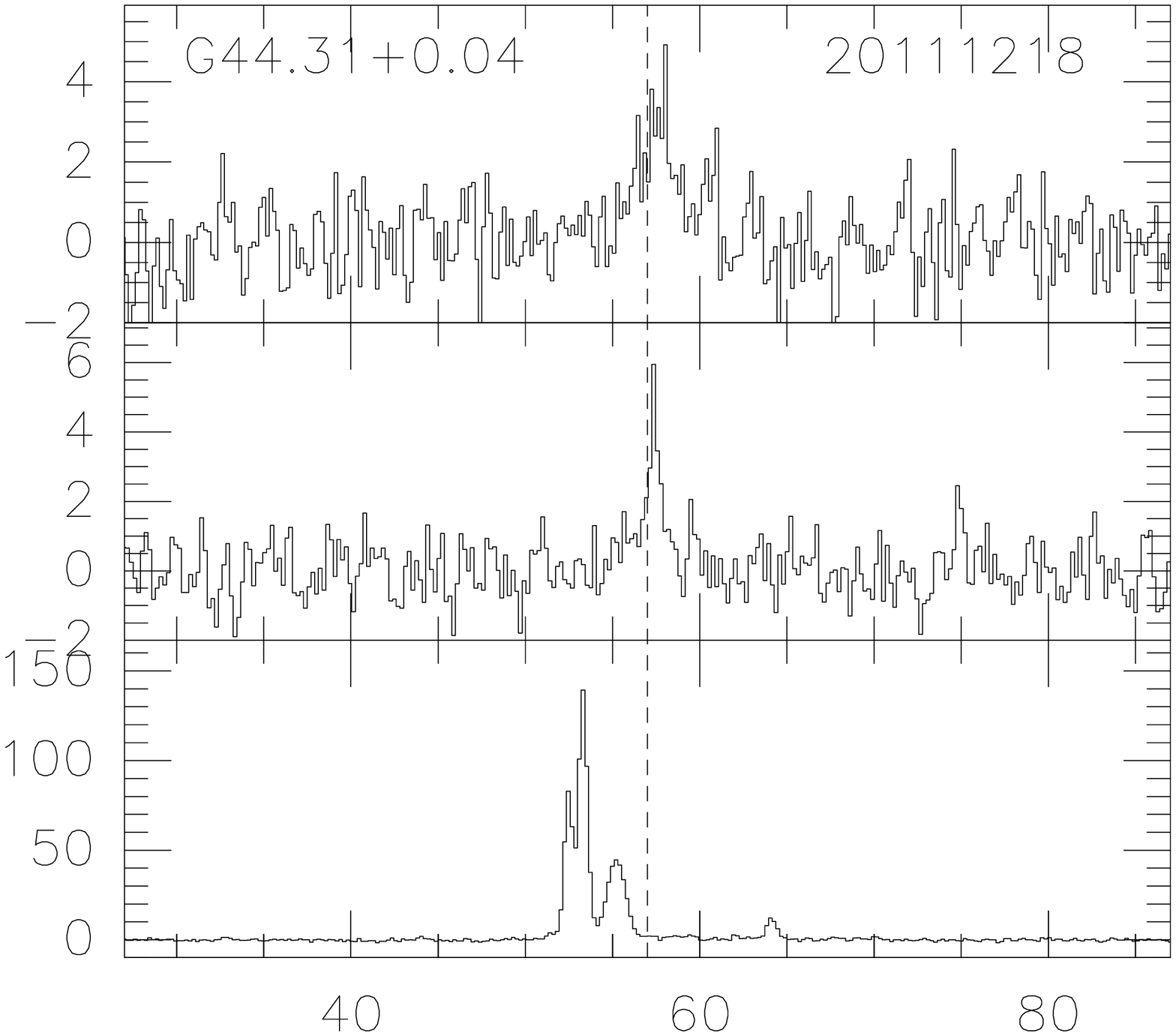} 
&
&  
\includegraphics[width=60mm]{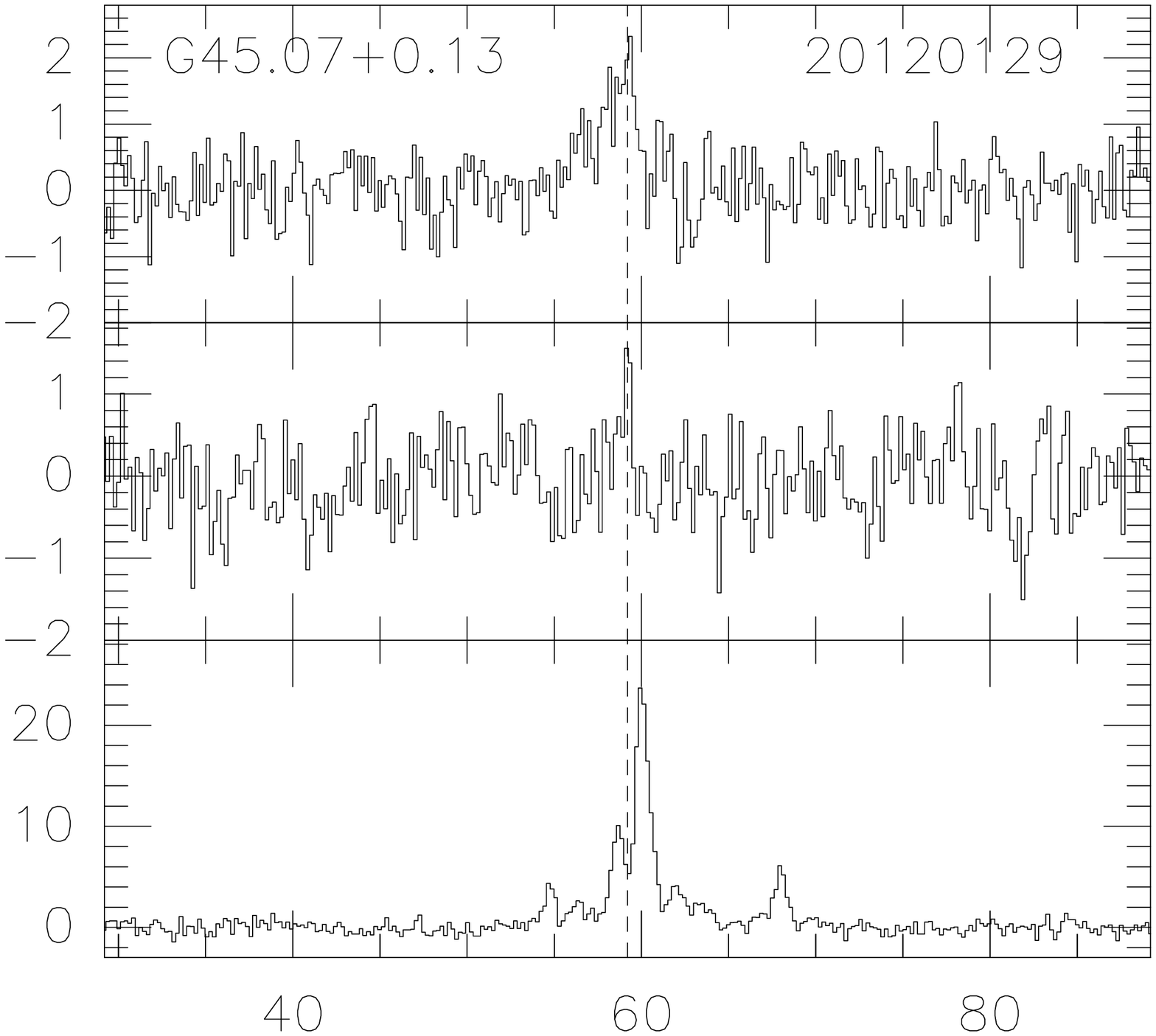} 
\\
&\multicolumn{3}{c}{V$_{lsr}$ (\kms)} \\
\end{tabular}
\caption{continued}
\end{figure}
\clearpage
\begin{figure}
\ContinuedFloat
\begin{tabular}{cccc}
\multirow{3}*[4ex]{\rotatebox{90}{S (Jy)}}
&
\includegraphics[width=60mm]{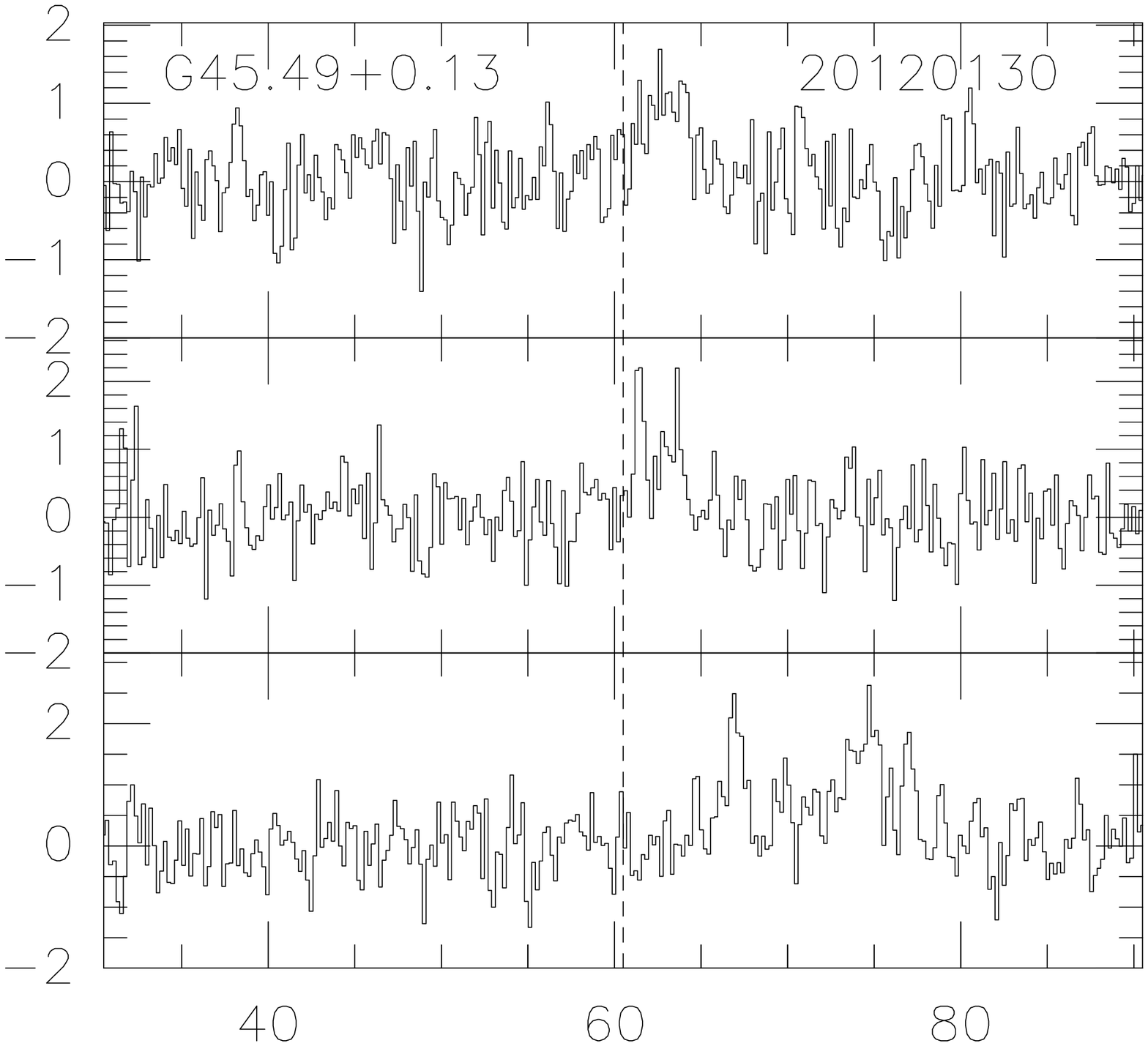} 
& 
&
\includegraphics[width=60mm]{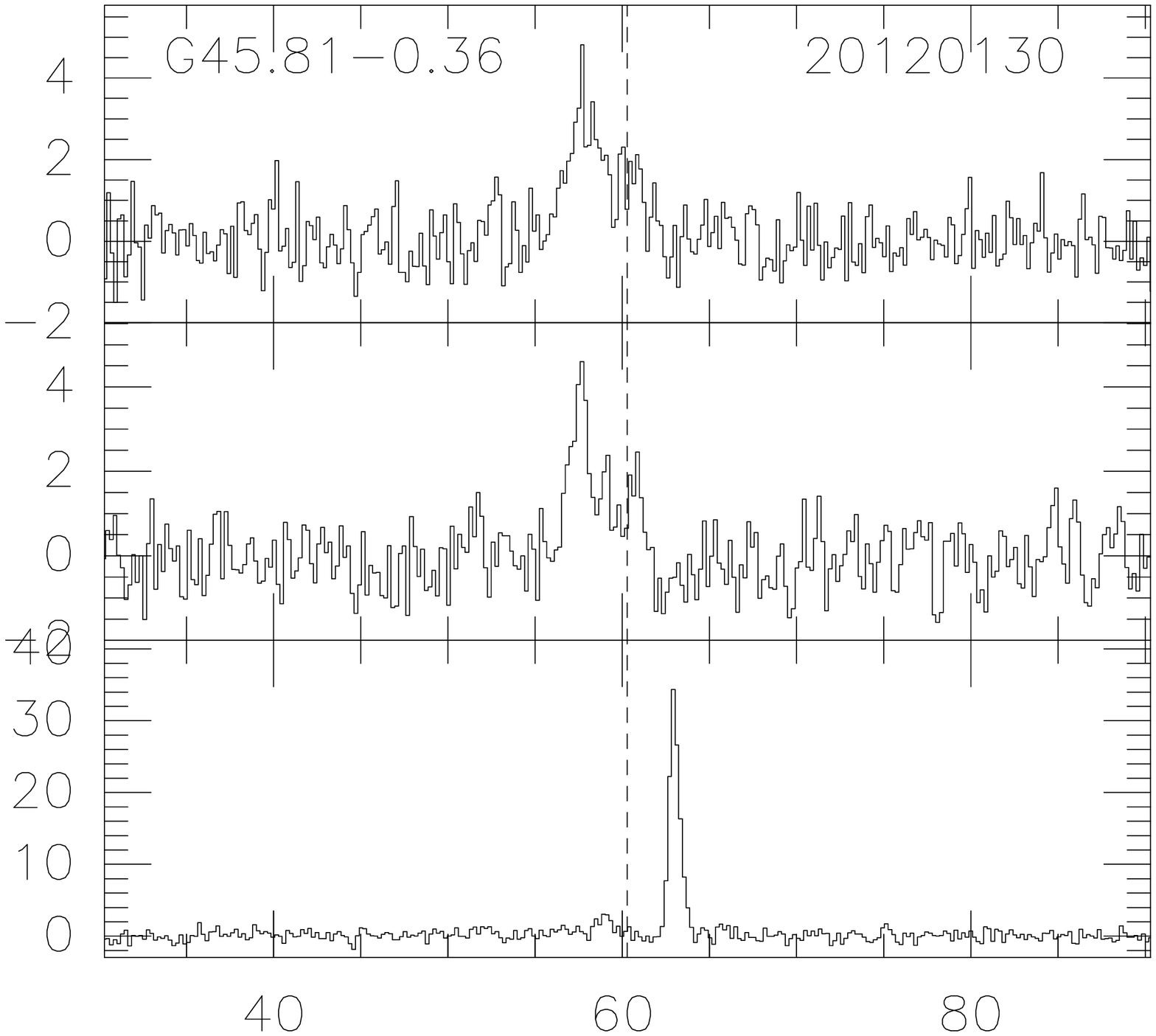} 
\\
&
\includegraphics[width=60mm]{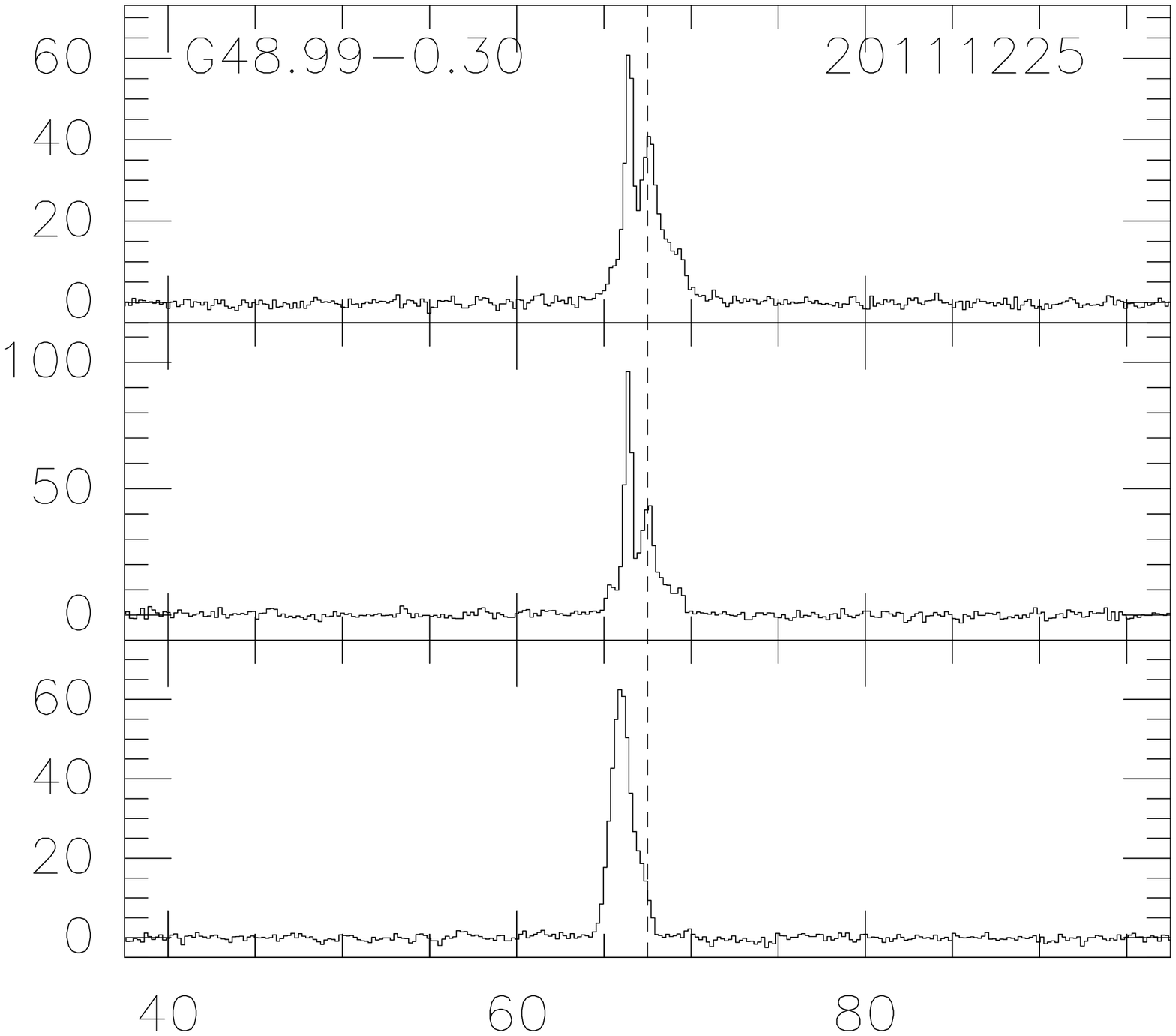} 
&
& 
\includegraphics[width=60mm]{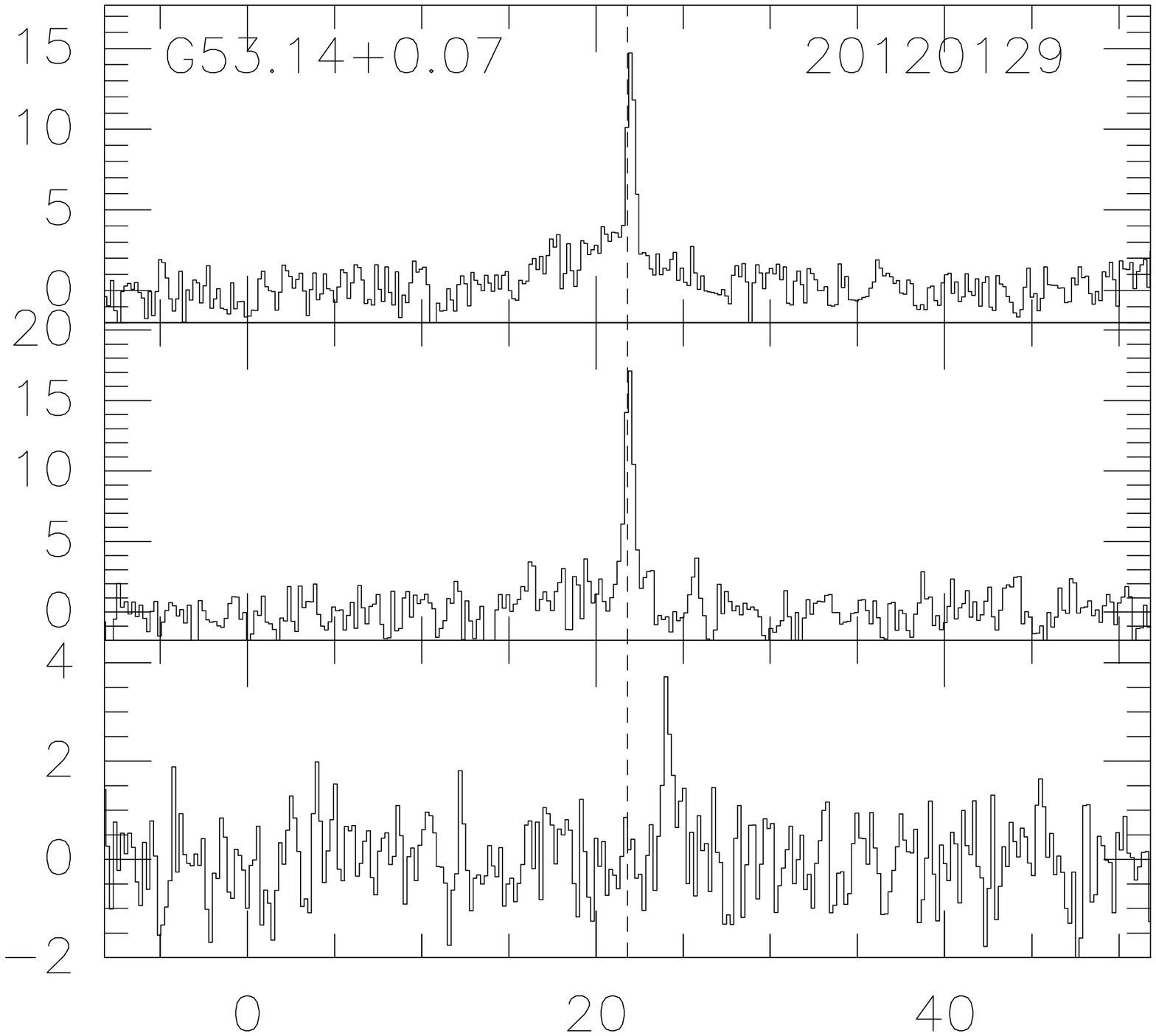} 
\\
&\multicolumn{3}{c}{V$_{lsr}$ (\kms)} \\
\end{tabular}
\caption{continued}
\end{figure}
\clearpage
\begin{figure}
\begin{center}
\begin{tabular}{cc}
& 
\includegraphics[width=60mm]{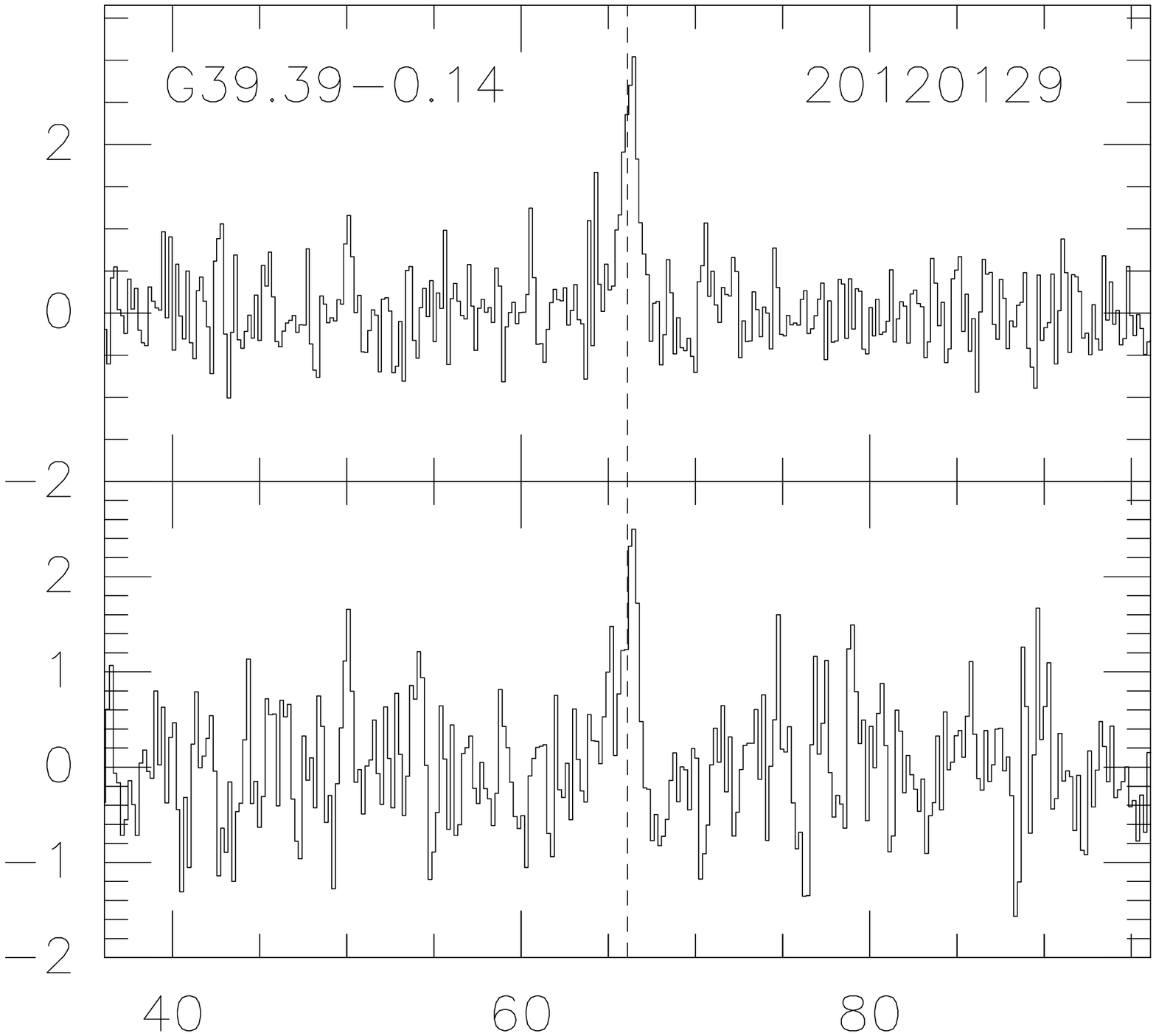} \\
{\rotatebox{90}{\qquad\qquad\qquad S (Jy)}}
&
\includegraphics[width=60mm]{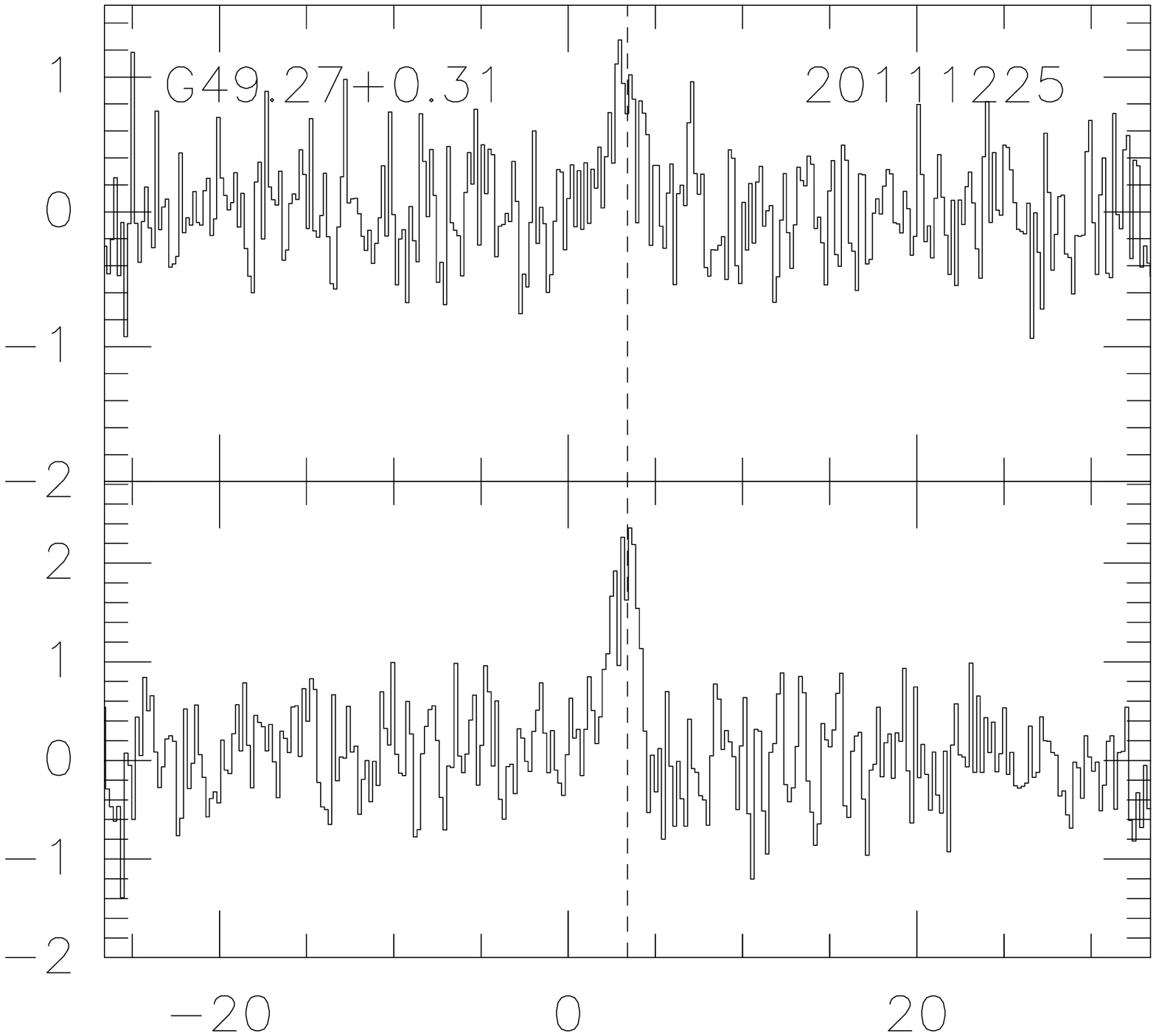} 
\\
& 
\includegraphics[width=60mm]{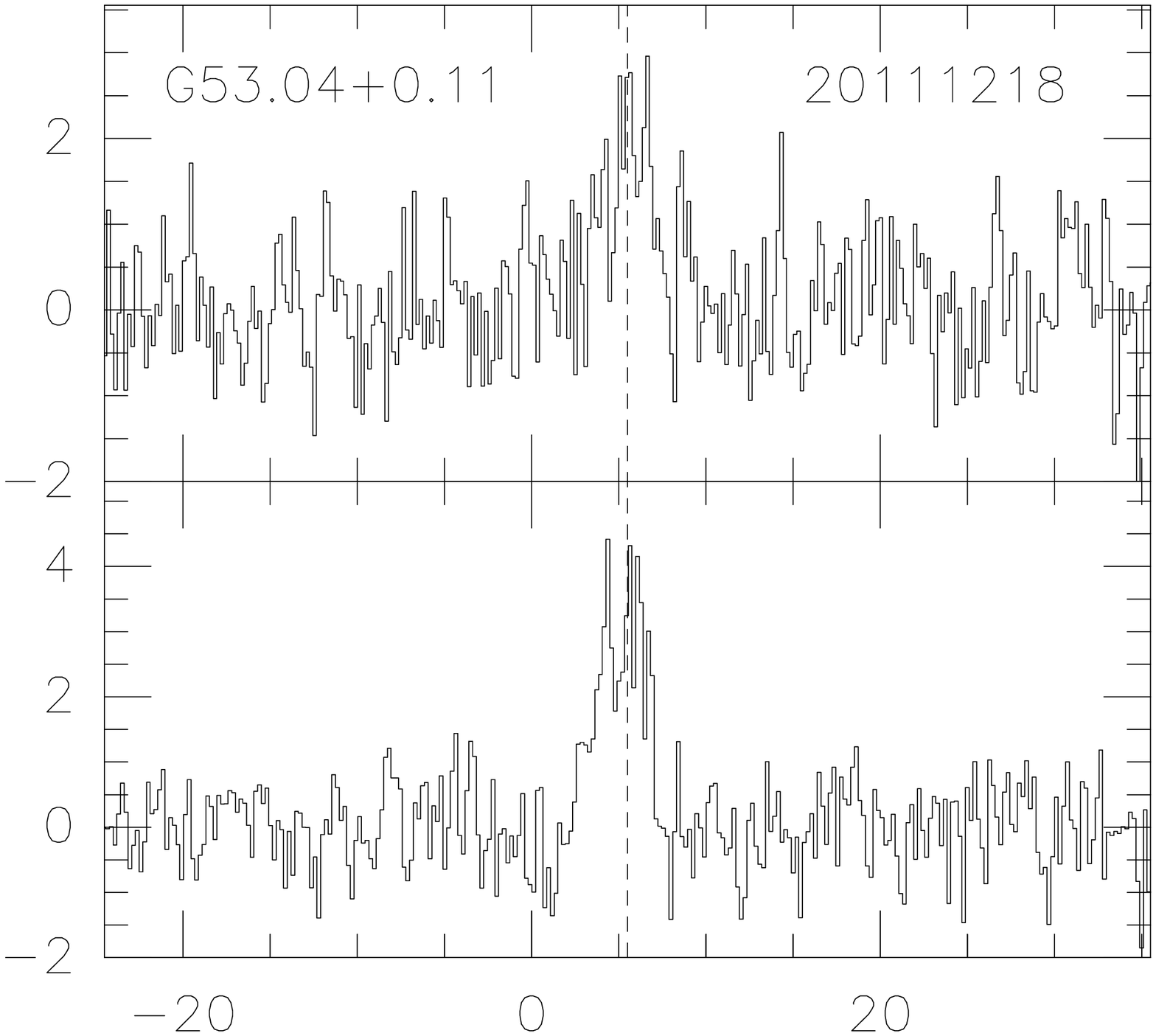} \\
&{V$_{lsr}$ (\kms)} \\
\end{tabular}
\end{center}
\caption{Same as in Figure 2 except for sources detected both in the 44\ghz~(lower) and 95\ghz~(upper) class I methanol masers.\label{fig:4495}}
\end{figure}
\clearpage
\begin{figure}
\begin{tabular}{cccc}
\multirow{2}*[20ex]{\rotatebox{90}{S (Jy)}}
&
\includegraphics[width=60mm]{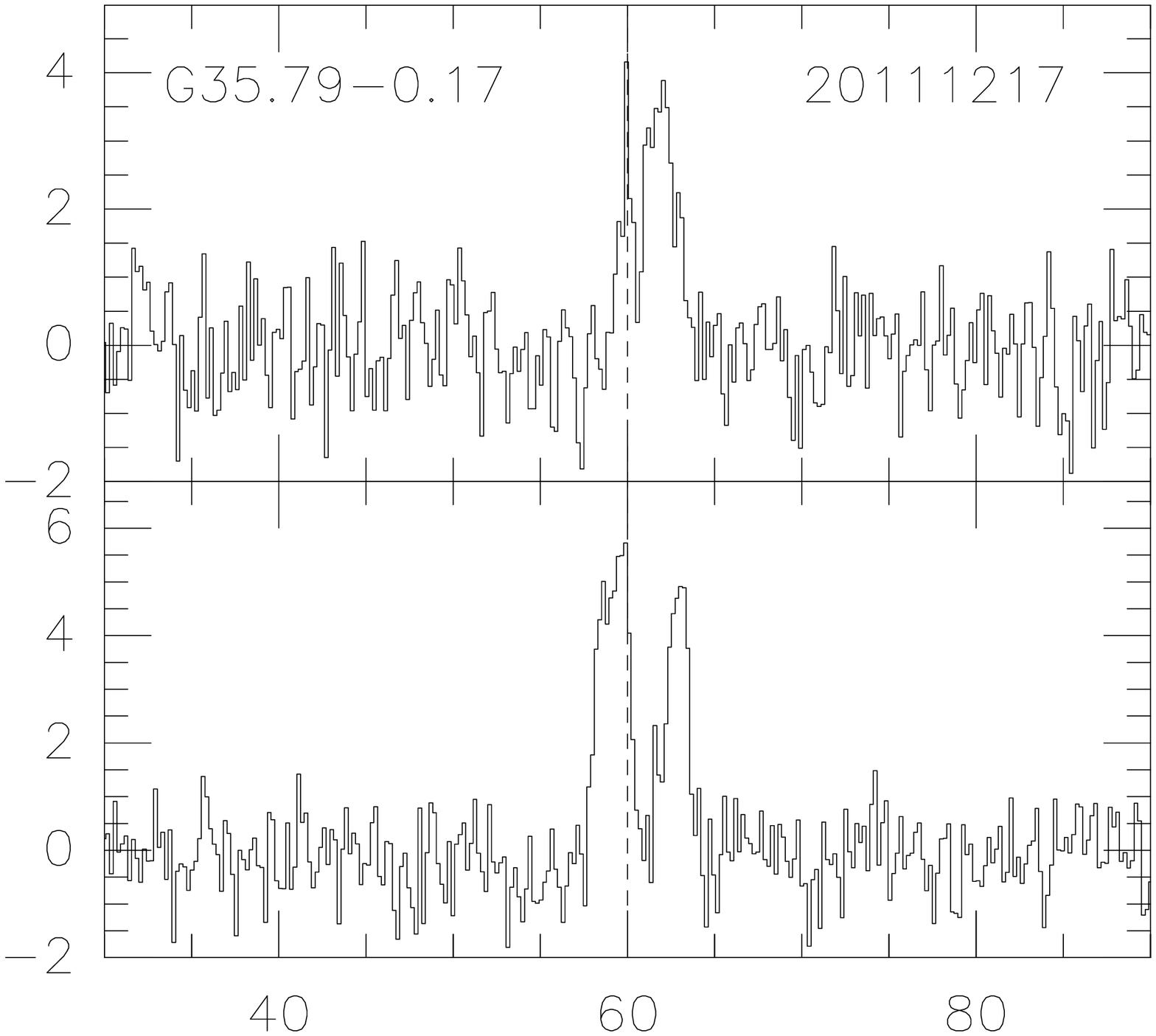} 
&
&
\includegraphics[width=60mm]{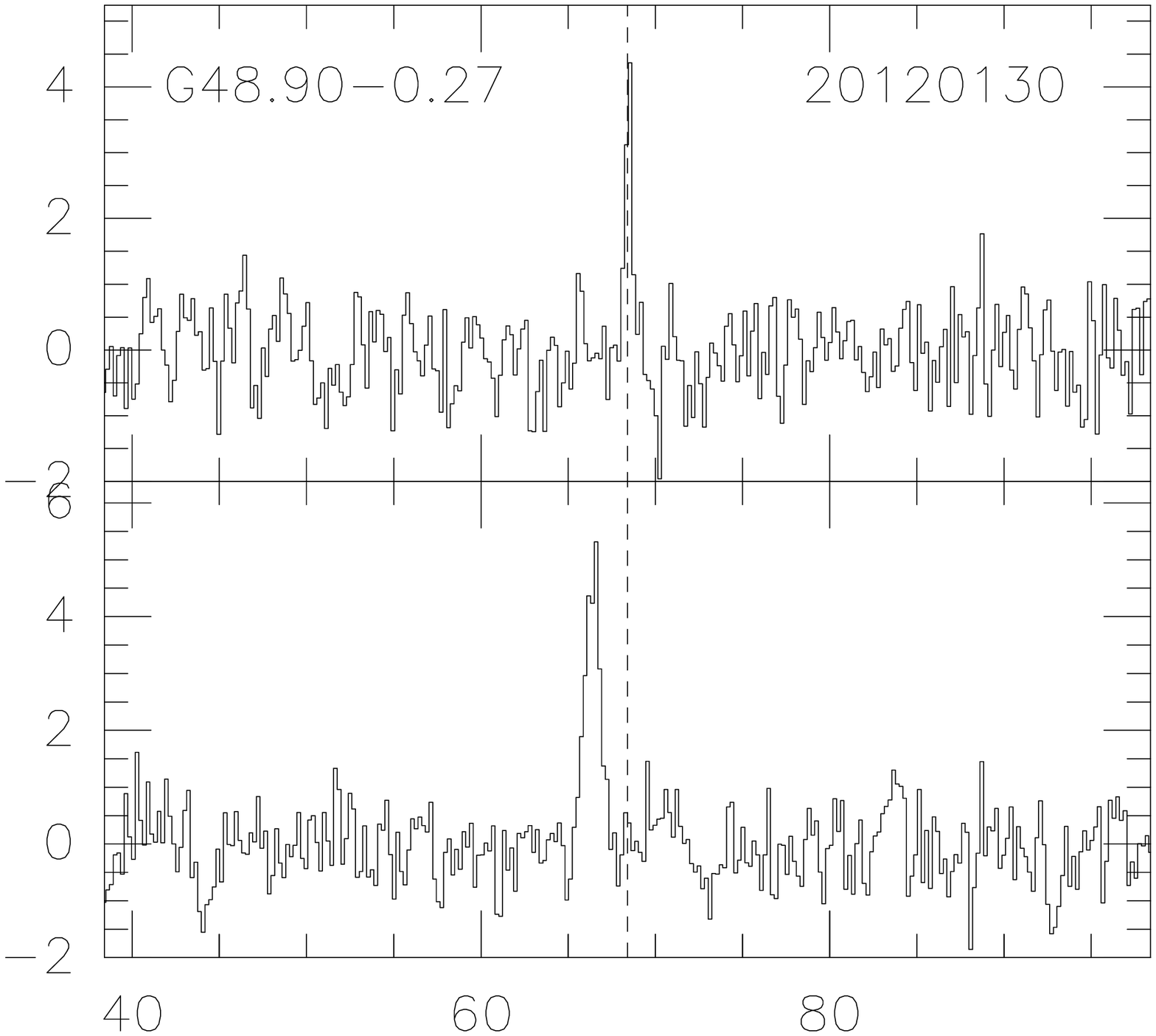} 
\\
&\multicolumn{3}{c}{V$_{lsr}$ (\kms)} \\
\end{tabular}
\caption{Same as in Figure 2 except for sources detected both in the 22\ghz\ water (lower) and the 44\ghz\ class I methanol maser (upper).\label{fig:2244}}
\end{figure}
\clearpage
\begin{figure}
\begin{tabular}{cccc}
\multirow{3}{*}{\rotatebox{90}{T$_{A^{*}}$ (K)}}
&
\includegraphics[width=60mm]{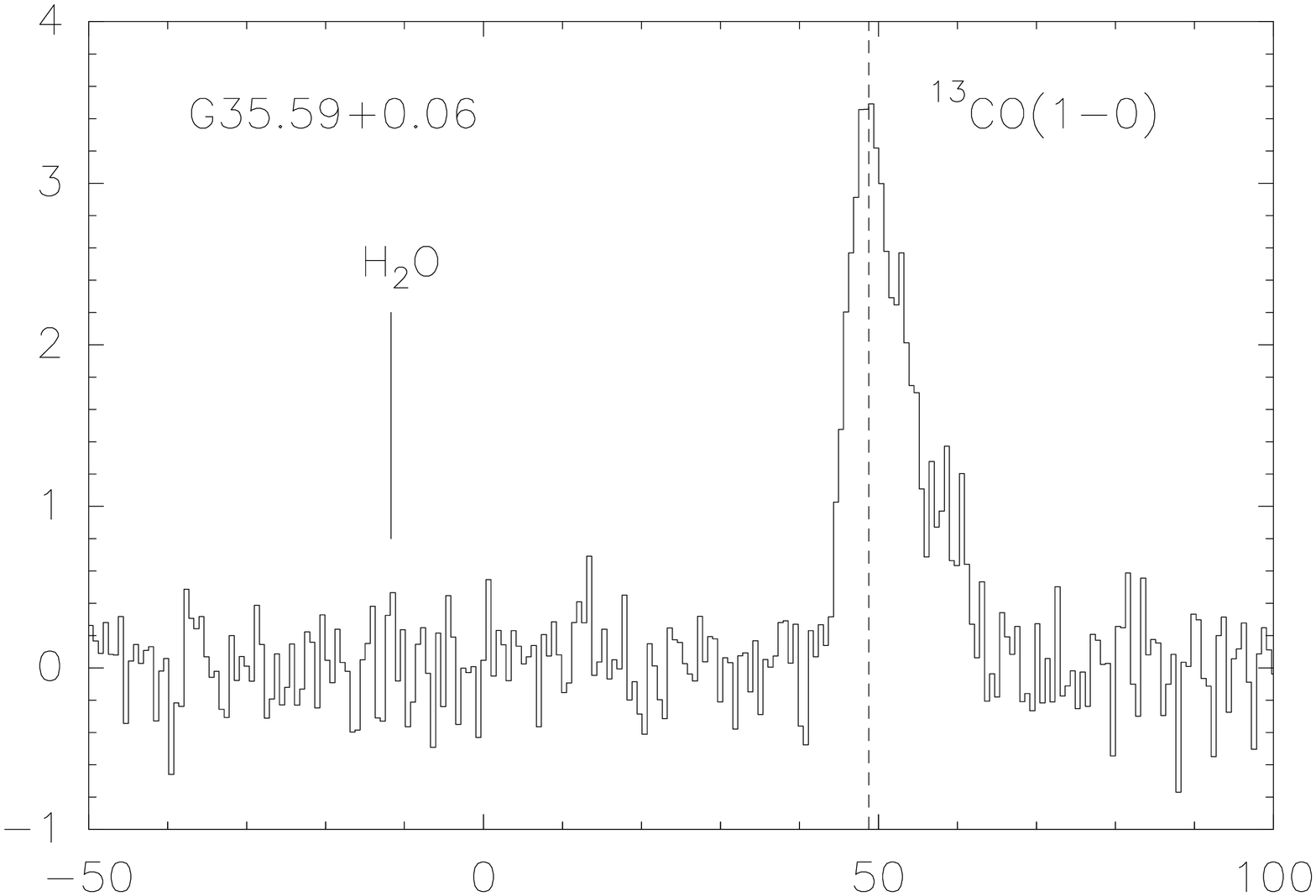} 
& 
&
\includegraphics[width=60mm]{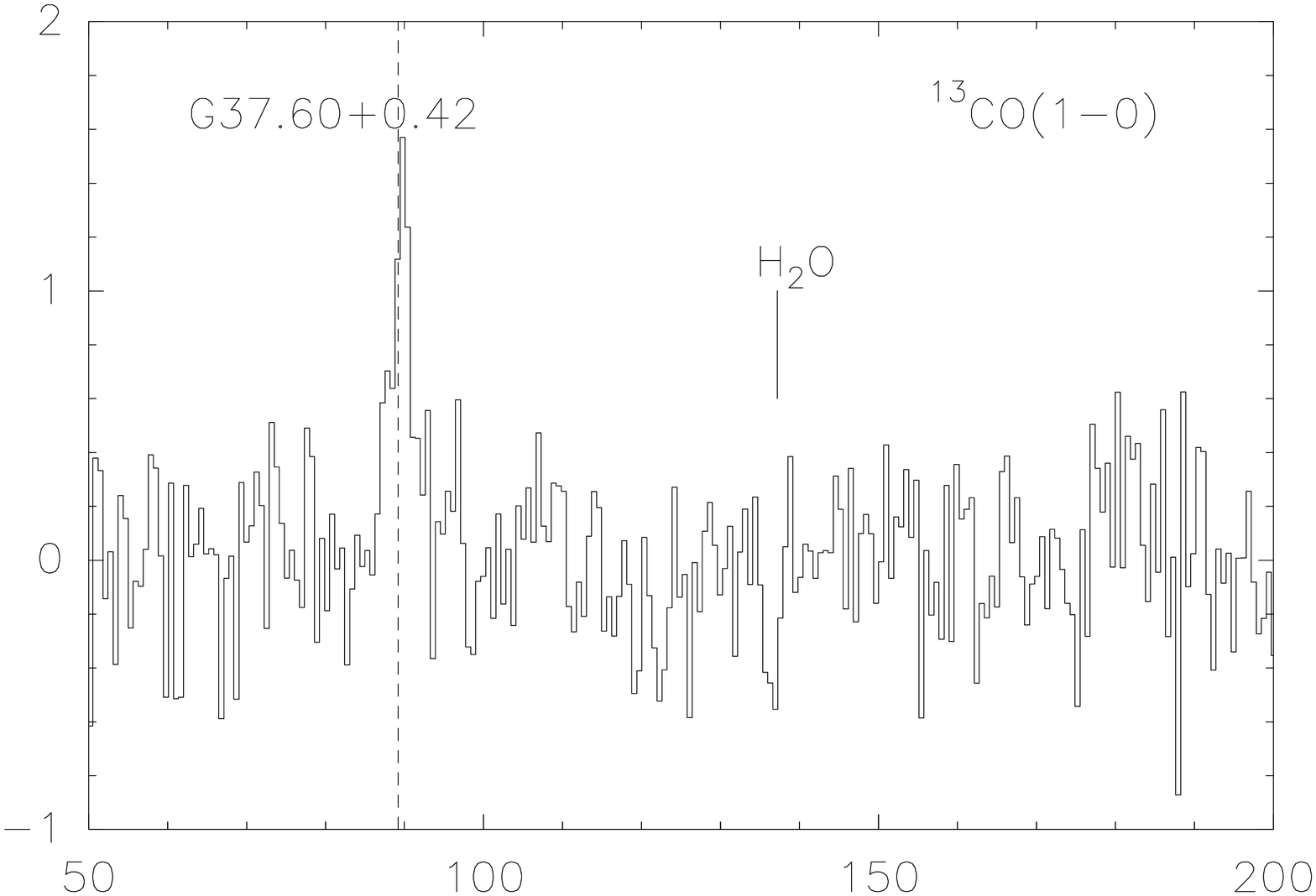} 
\\
&
\includegraphics[width=60mm]{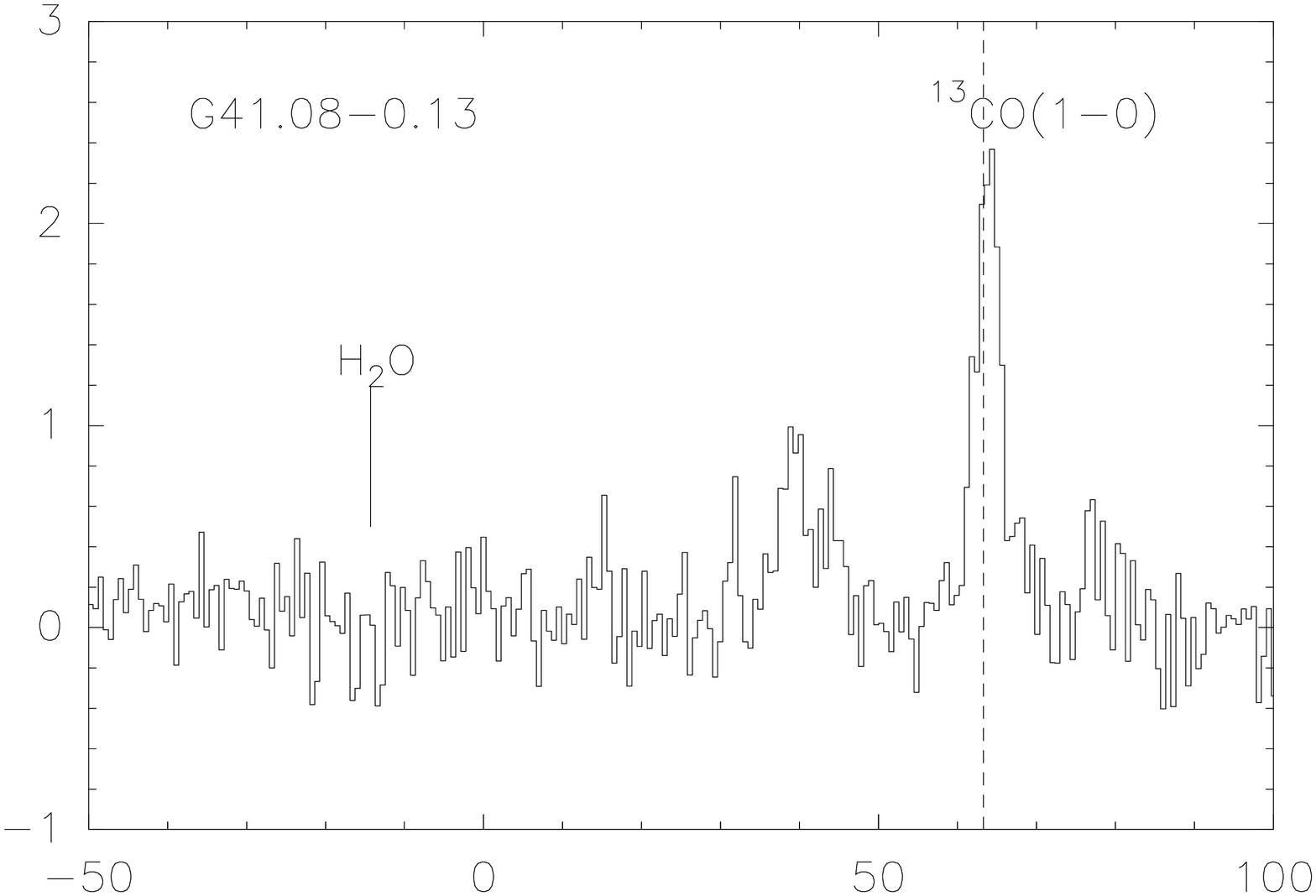} 
&
&
\includegraphics[width=59mm]{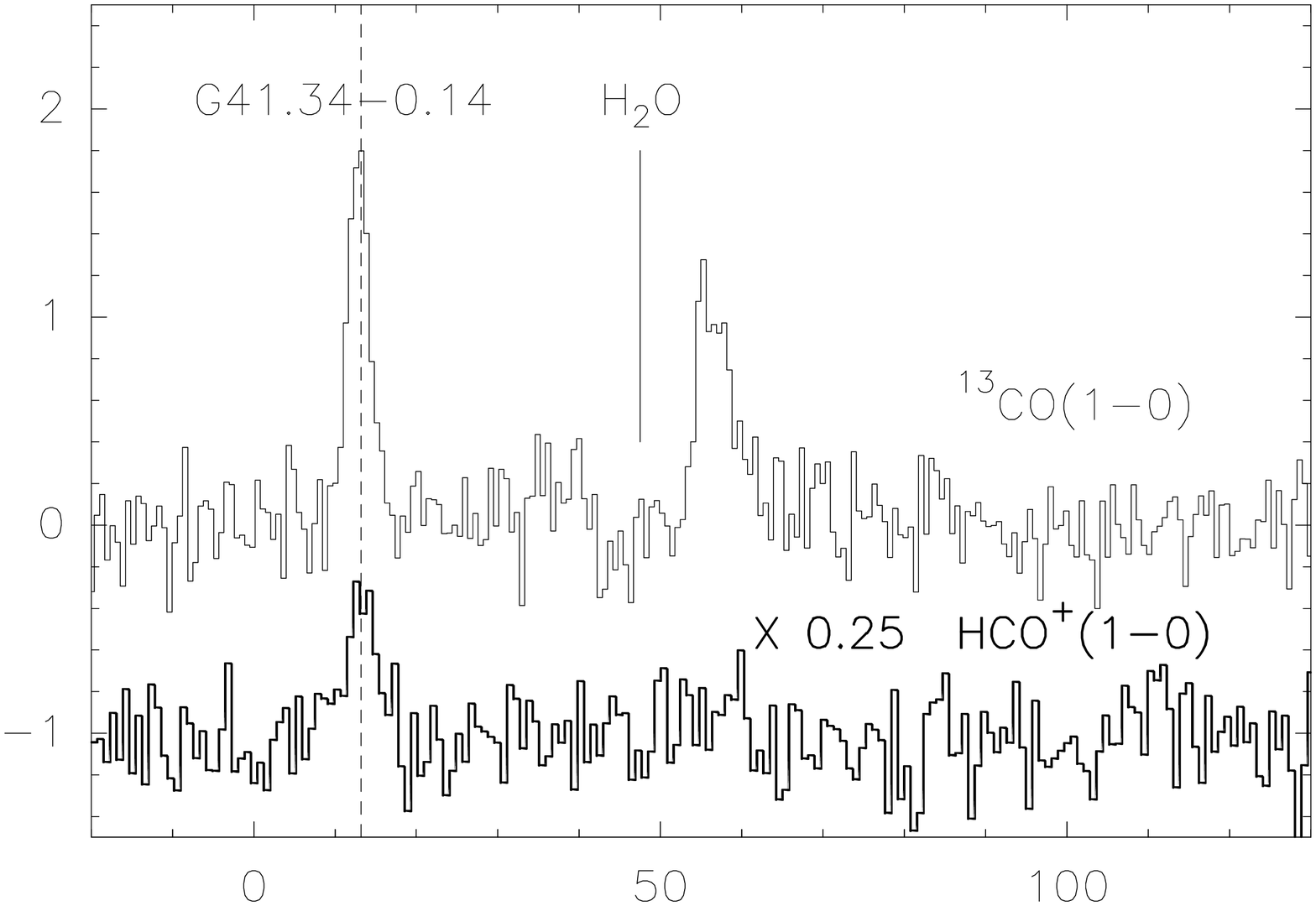} 
\\ 
&\multicolumn{3}{c}{V$_{lsr}$ (\kms)} \\
\end{tabular}
\caption{$^{13}$CO~(J=1$-$0) and HCO$^+$~(J=1$-$0) line spectra of the four water maser sources with large ($>$30~\kms) velocity offsets. The vertical dashed lines show the systemic velocities, and solid lines show the positions of high velocity figures of water maser. The HCO$^+$ spectrum of G41.34$-$0.14 is magnified by a factor of 4.\label{fig:hvtrao}}
\end{figure}
\clearpage
\begin{figure}
\begin{center}
\begin{tabular}{cc}
\includegraphics[width=60mm,angle=90]{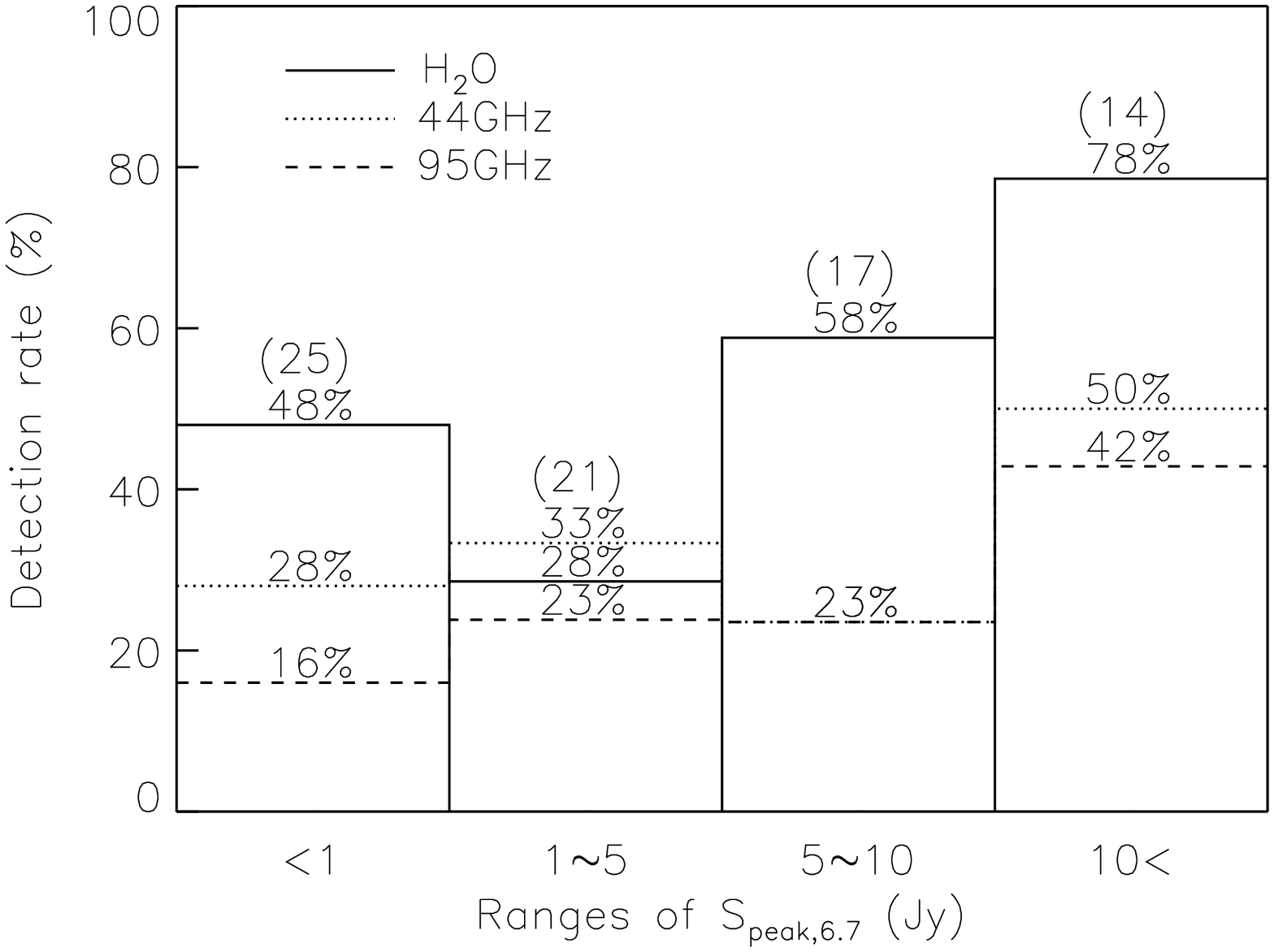} 
&
\includegraphics[width=60mm,angle=90]{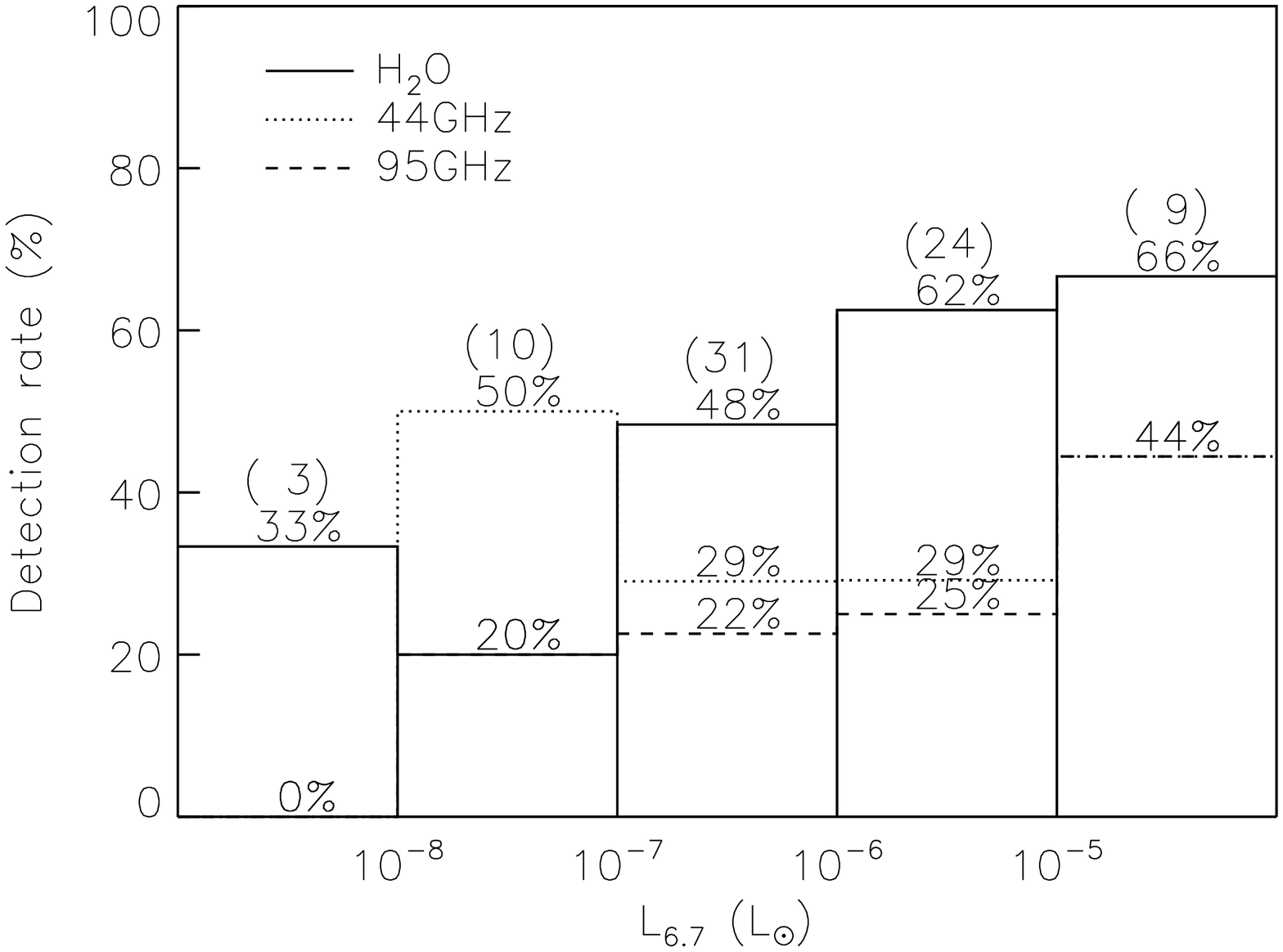} 
\\
\end{tabular}
\caption{Detection rates of 22\ghz\ water~(solid line), 44\ghz~(dotted line), and 95\ghz~(dashed line) methanol masers as a function of the peak flux of 6.7\ghz~class II methanol maser (left) and the luminosity of 6.7\ghz~(right). The numbers of 6.7\ghz~maser sources are presented in parentheses.\label{fig:rate_cutoff}}
\end{center}
\end{figure}

\clearpage
\begin{figure}
\begin{center}
\begin{tabular}{c}
\includegraphics[width=80mm,angle=90]{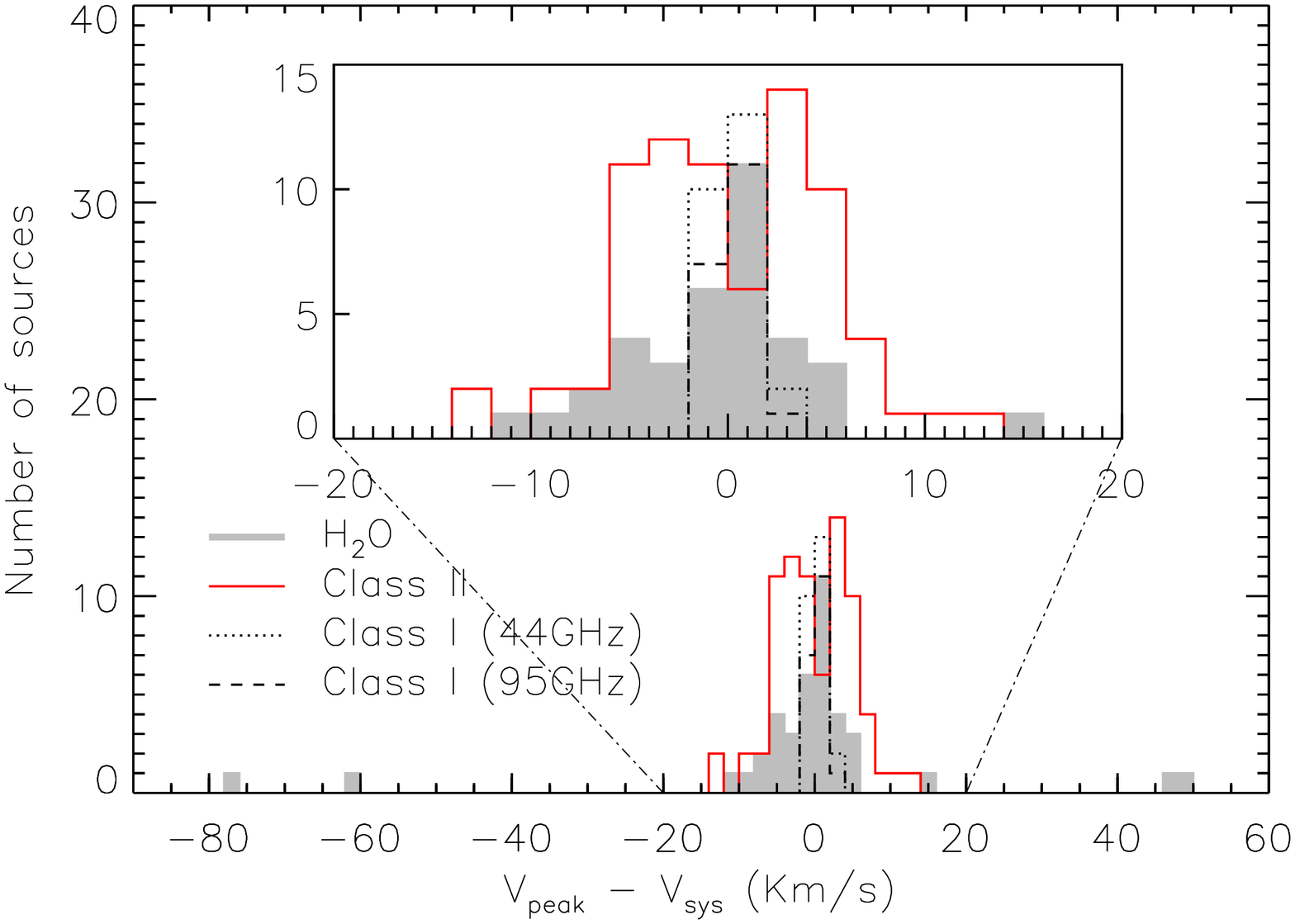} 
\\
\end{tabular}
\caption{Velocity offset distributions of the four maser transitions.
\label{fig:offset}}
\end{center}
\end{figure}

\clearpage
\begin{figure}
\begin{center}
\begin{tabular}{c}
\includegraphics[width=80mm,angle=90]{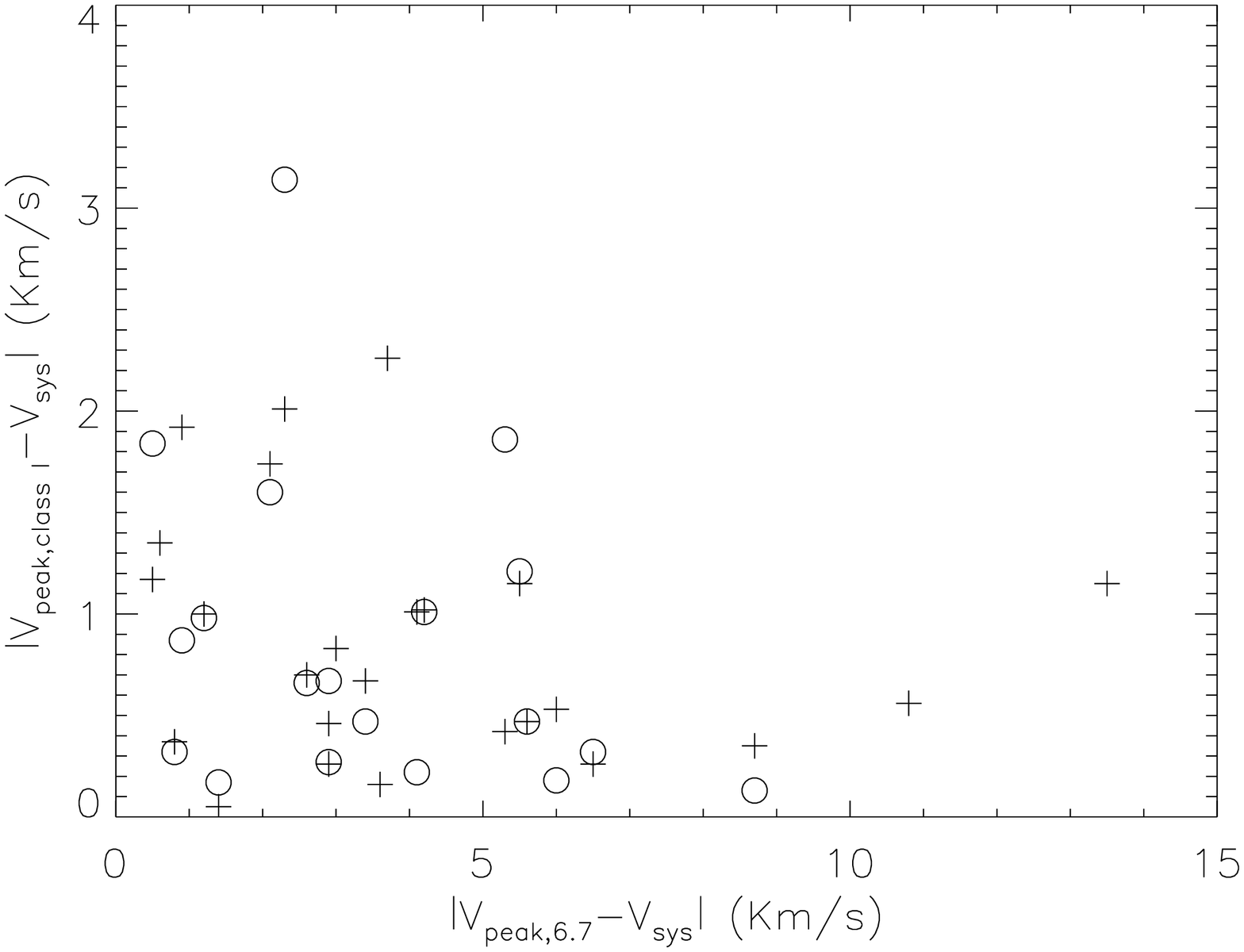}
\\
\end{tabular}
\caption{Plot of the absolute value of velocity offset of class~I methanol maser against that of 6.7\ghz~maser. Plus signs and open circles are the data points of 44\ghz\ and 95\ghz\ masers, respectively.\label{fig:offset6744}}
\end{center}
\end{figure}
\clearpage

\clearpage
\begin{figure}
\begin{center}
\begin{tabular}{c}
\includegraphics[width=80mm,angle=90]{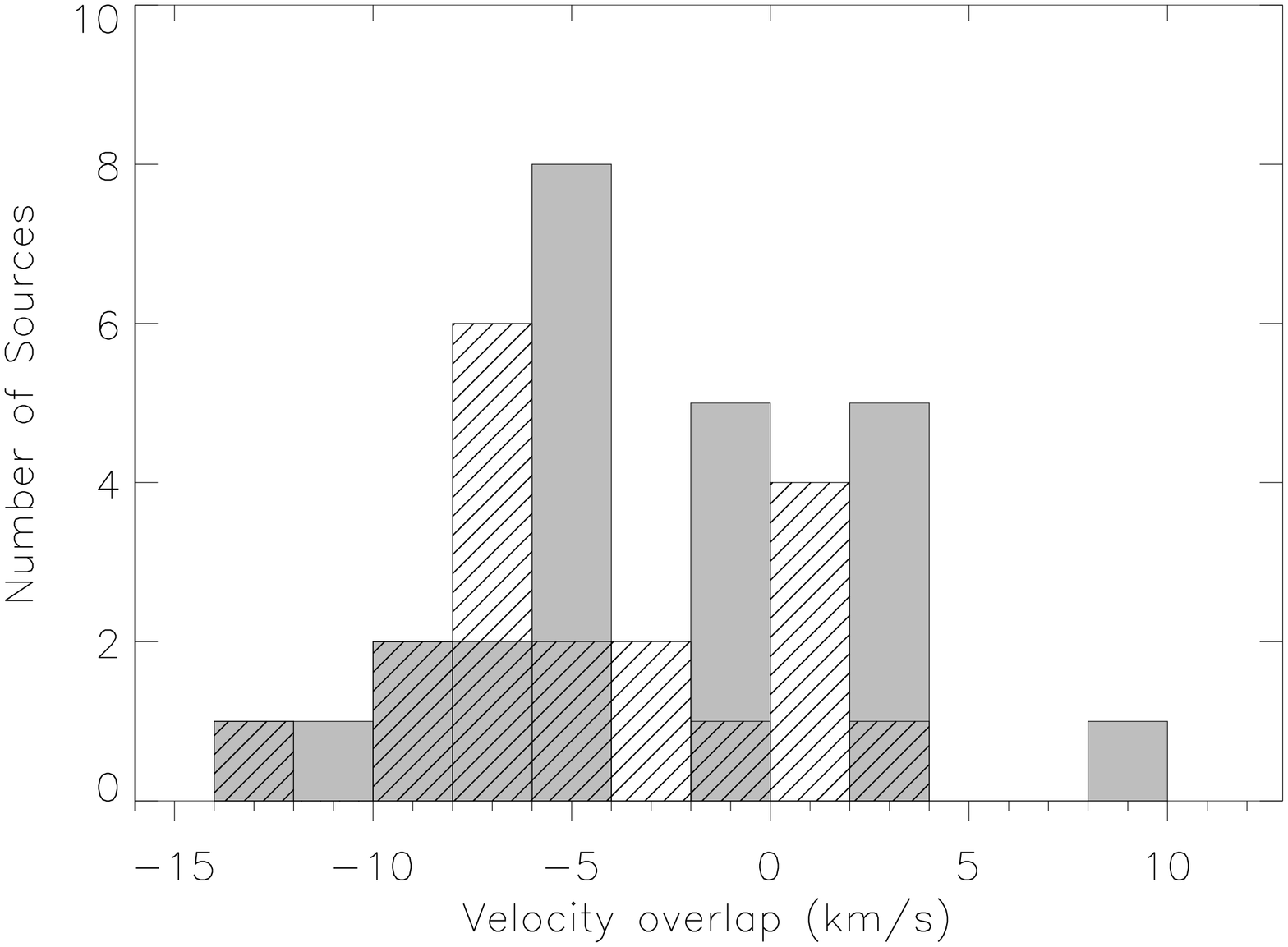}
\\
\end{tabular}
\caption{Histogram of the velocity overlap between class~I and 6.7\ghz\ class II methanol masers. Filled and slant bars indicate 44\ghz\ and 95\ghz\ masers, respectively. A negative value means that the velocity ranges of the two masers overlap (See the text for details).\label{fig:overlap6744}}
\end{center}
\end{figure}
\clearpage

\begin{figure}
\begin{center}
\begin{tabular}{c}
\includegraphics[width=80mm,angle=90]{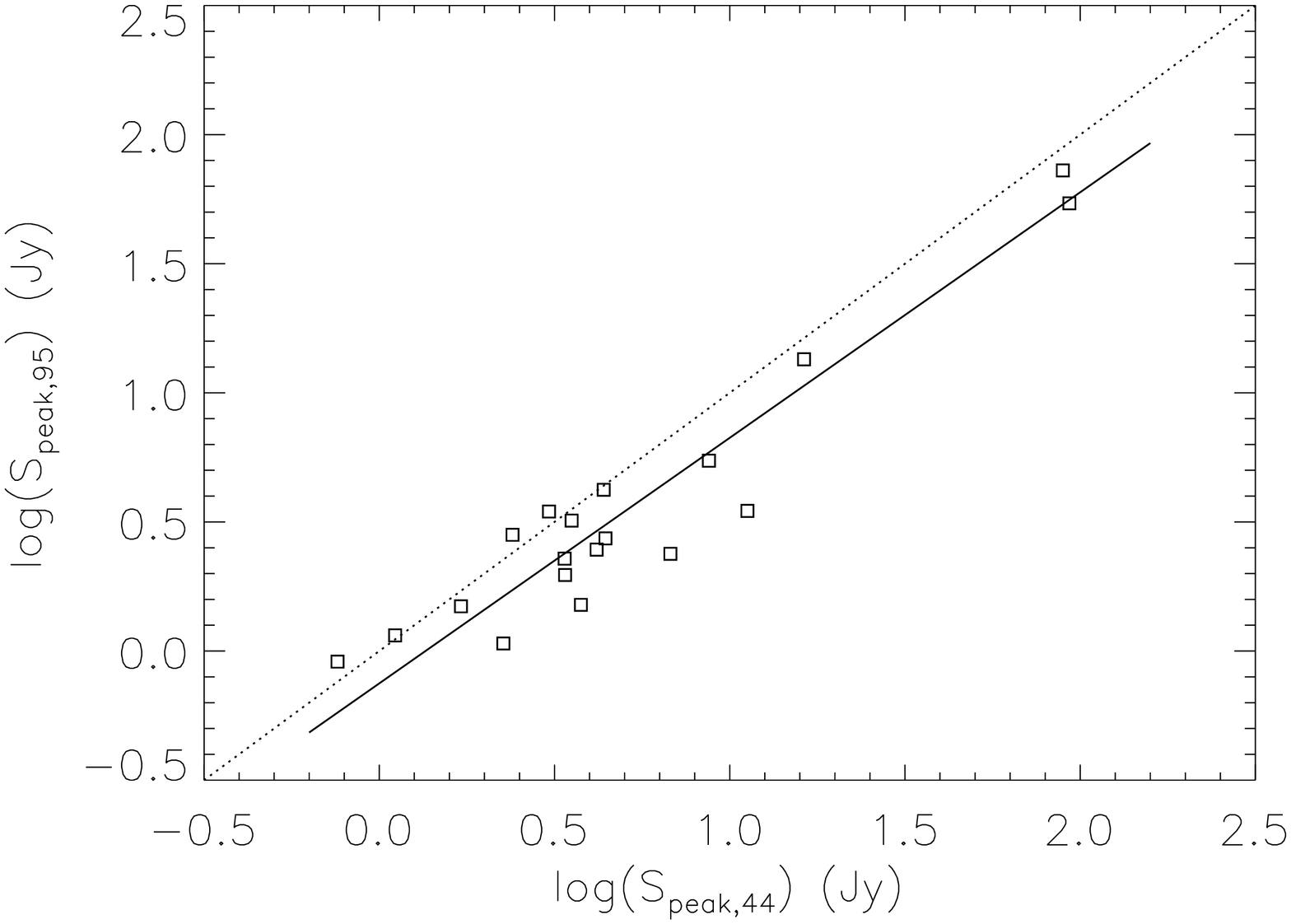} 
\\
\end{tabular}
\caption{Comparison between the peak flux densities of 44\ghz\ and 95\ghz\ class~I methanol masers. A strong correlation exists between the two parameters. The solid line is the least-squares fitting result with a correlation coefficient of 0.95.
\label{fig:flux4495}}
\end{center}
\end{figure}
\clearpage

\begin{figure}
\begin{center}
\begin{tabular}{c}
\includegraphics[width=80mm,angle=90]{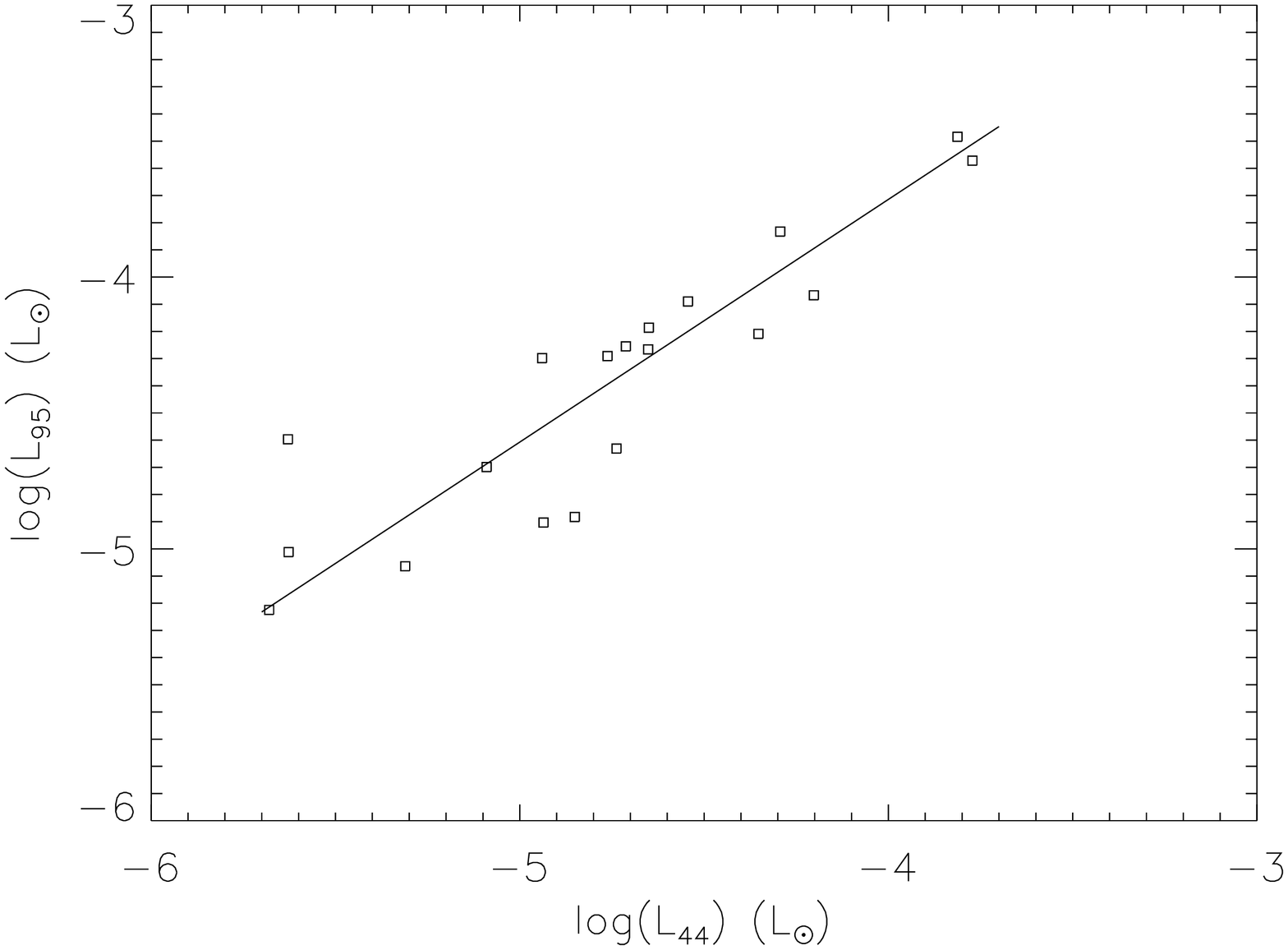} 
\end{tabular}
\caption{Comparison between the isotropic luminosities of 44\ghz\ and 95\ghz\ class~I methanol masers. The solid line is the result of a least-squares fit with a correlation coefficient of $0.89$.
\label{fig:lmaser}}
\end{center}
\end{figure}

\clearpage
\begin{figure}
\begin{center}
\begin{tabular}{c}
\includegraphics[width=80mm,angle=90]{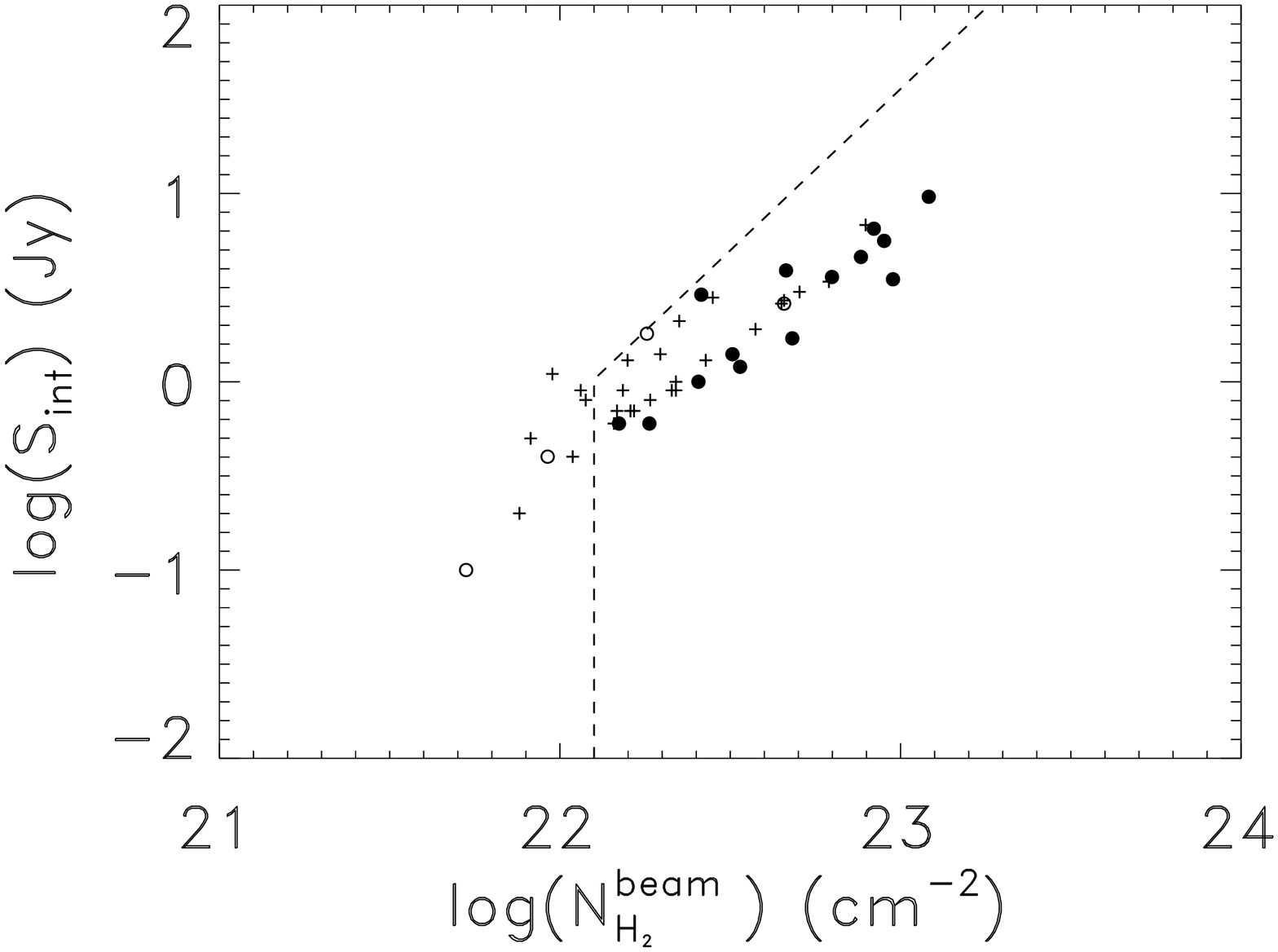} 
\\
\end{tabular}
\caption{Distribution of the associated BGPS sources in a plane of the beam-averaged H$_{2}$ column density and the 1.1~mm integrated continuum flux density. Filled circles, open circles, and plus signs represent the sources with both 44\ghz\ and 95\ghz\ masers,  the ones with only 44\ghz\ maser, and the ones without any of the two, respectively. The dashed lines show the criteria of \cite{chen12}.
\label{fig:bgps_int_n}}
\end{center}
\end{figure}

\clearpage
\begin{figure}
\begin{center}
\begin{tabular}{cc}
\includegraphics[width=60mm,angle=90]{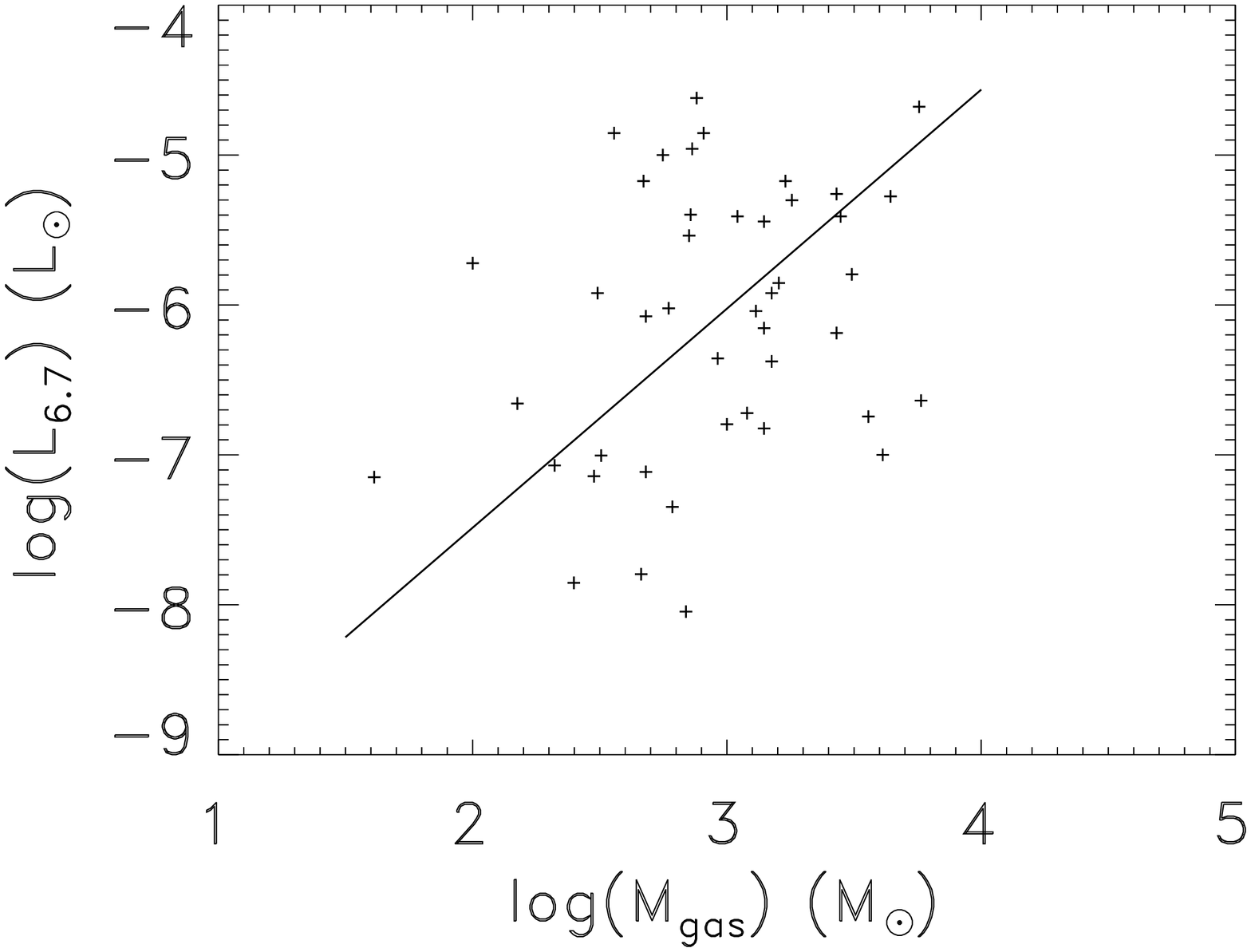} 
&
\includegraphics[width=60mm,angle=90]{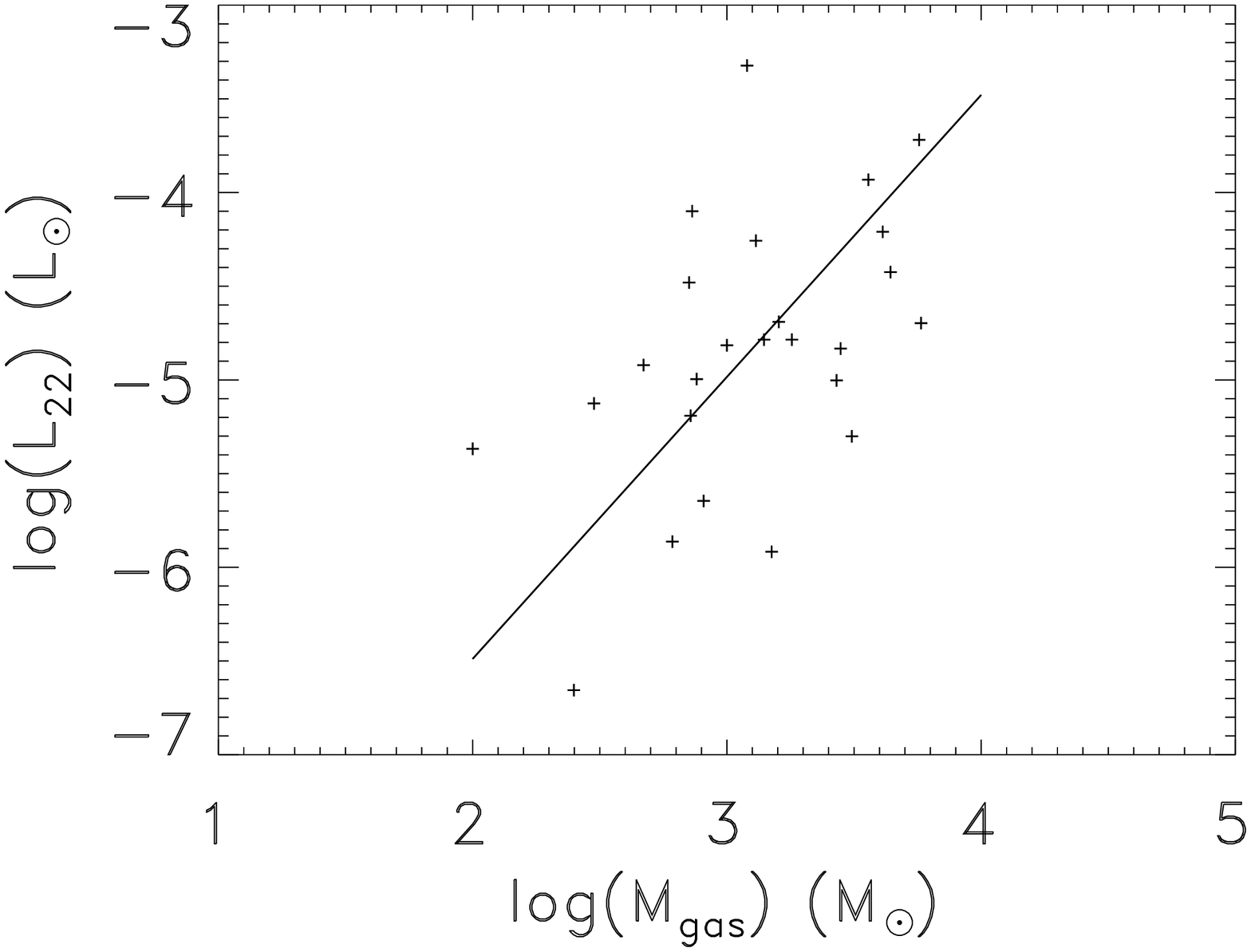} 
\\
\includegraphics[width=60mm,angle=90]{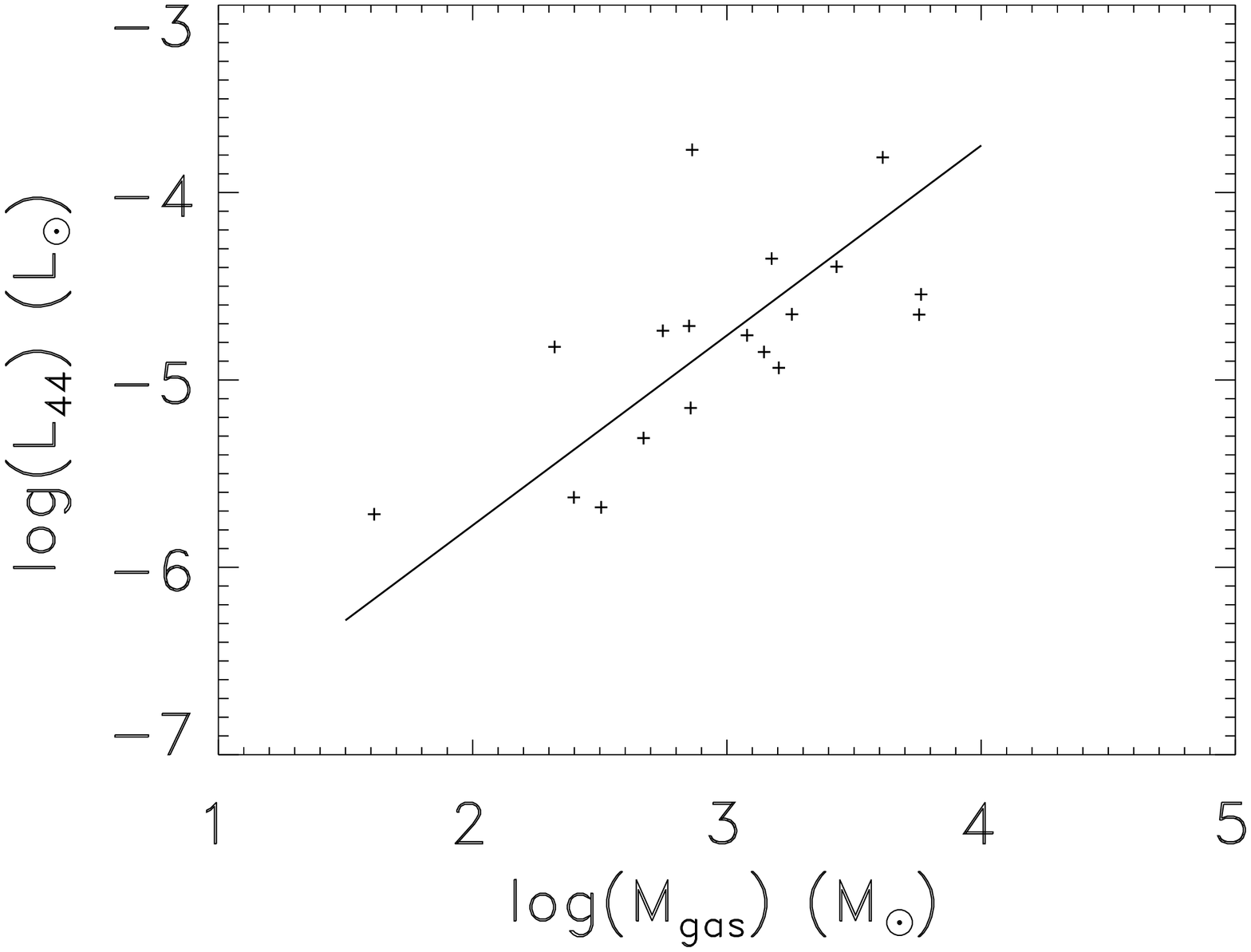} 
&
\includegraphics[width=60mm,angle=90]{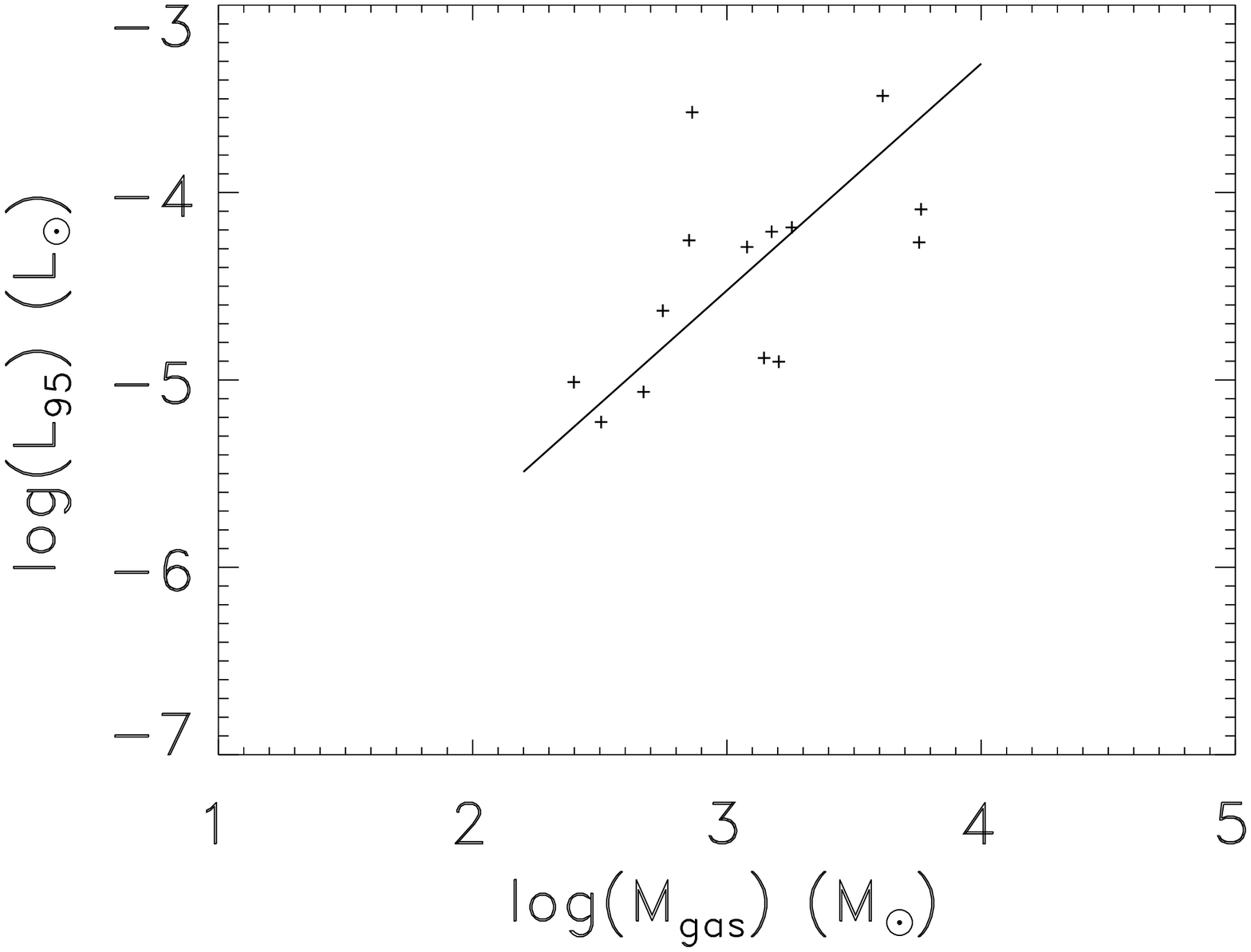} 
\\
\end{tabular}
\end{center}
\caption{Relationship between the isotropic maser luminosity and the associated BGPS source mass for 6.7\ghz\ ({upper left}), 22\ghz\ ({upper right}), 44\ghz\ ({bottom left}), and 95\ghz\ ({bottom right}) masers. The solid lines represent the least-sqaures fitting results. The correlation coefficients are 0.29, 0.50, 0.67, and {0.59} for the 6.7\ghz,
22\ghz, 44\ghz, and 95\ghz~transitions, respectively.
\label{fig:BGPSml}}
\end{figure}

\begin{thebibliography}{}
\bibitem[Bachiller et al.(1998)]{bachiller98} Bachiller, R., Codella, C., Colomer, F., Liechti, S., \& Walmsley, C.~M.\ 1998, \aap, 335, 266 
\bibitem[Bae et al.(2011)]{bae11} Bae, J.-H., Kim, K.-T., Youn, S.-Y., et al.\ 2011, \apjs, 196, 21 
\bibitem[Battersby et al.(2010)]{battersby10} Battersby, C., Bally, J., Jackson, J.~M., et al.\ 2010, \apj, 721, 222 
\bibitem[Bayandina et al.(2012)]{bayandina12}
	Bayandina, O.~S., Val'tts, I.~E., \& Larionov, G.~M. 2012, \azh, 89, 611
\bibitem[Beuther et al.(2002)]{beuther02} Beuther, H., Walsh, A., Schilke, P., et al.\ 2002, \aap, 390, 289 
\bibitem[Breen et al.(2010)]{breen10} Breen, S.~L., Caswell, J.~L., Ellingsen, S.~P., \& Phillips, C.~J.\ 2010, \mnras, 406, 1487 
\bibitem[Breen et al.(2013)]{breen13} Breen, S.~L., Ellingsen, S.~P., Contreras, Y., et al.\ 2013, \mnras, 435, 524
\bibitem[Byun et al.(2012)]{byun12} Byun, D.-Y., Kim, K.-T. \& Bae, J.-H. 2012, in IAU Symp. 287, Cosmic Masers-from OH to H0 (Cambridge: Cambridge Univ. Press), 284
\bibitem[Caswell \& Phillips(2008)]{caswell08} Caswell, J.~L., \& Phillips, C.~J.\ 2008, \mnras, 386, 1521
\bibitem[Caswell et al.(2010)]{caswell10} Caswell, J.~L., Fuller, G.~A., Green, J.~A., et al.\ 2010, \mnras, 404, 1029 
\bibitem[Chen et al.(2011)]{chen11} Chen, X., Ellingsen, S.~P., Shen, Z.-Q., Titmarsh, A., \& Gan, C.-G.\ 2011, \apjs, 196, 9
\bibitem[Chen et al.(2012)]{chen12} Chen, X., Ellingsen, S.~P., He, J.-H., et al.\ 2012, \apjs, 200, 5 
\bibitem[Choi et al.(2012)]{choi12} Choi, M., Kang, M., Byun, D.-Y., \& Lee, J.-E.\ 2012, \apj, 759, 136 
\bibitem[Codella et al.(1994)]{codella94} Codella, C., Felli, M., Natale, V., Palagi, F., \& Palla, F.\ 1994, \aap, 291, 261 
\bibitem[Codella et al.(2004)]{codella04} Codella, C., Lorenzani, A., Gallego, A.~T., Cesaroni, R., \& Moscadelli, L.\ 2004, \aap, 417, 615 
\bibitem[Cragg et al.(1992)]{cragg92} Cragg, D.~M., Johns, K.~P., Godfrey, P.~D., \& Brown, R.~D.\ 1992, \mnras, 259, 203
\bibitem[Cyganowski et al.(2009)]{cyg09} Cyganowski, C.~J., Brogan, C.~L., Hunter, T.~R., \& Churchwell, E.\ 2009, \apj, 702, 1615
\bibitem[Dodson et al.(2004)]{dodson04} Dodson, R., Ojha, R., \& Ellingsen, S.~P.\ 2004, \mnras, 351, 779
\bibitem[Ellingsen(2005)]{ellingsen05} Ellingsen, S.~P.\ 2005, \mnras, 359, 1498 
\bibitem[Elitzur et al.(1989)]{elitzur89} Elitzur, M., Hollenbach, D.~J., \& McKee, C.~F.\ 1989, \apj, 346, 983
\bibitem[Fontani et al.(2010)]{fontani10} Fontani, F., Cesaroni, R., \& Furuya, R.~S.\ 2010, \aap, 517, A56
\bibitem[Forster \& Caswell(1999)]{forster99} Forster, J.~R., \& Caswell, J.~L.\ 1999, \aaps, 137, 43 
\bibitem[Furuya et al.(2003)]{furuya03} Furuya, R.~S., Kitamura, Y., Wootten, A., Claussen, M.~J., \& Kawabe, R.\ 2003, \apjs, 144, 71 
\bibitem[Garay et al.(2002)]{garay02} Garay, G., Mardones, D., Rodr{\'{\i}}guez, L.~F., Caselli, P., \& Bourke, T.~L.\ 2002, \apj, 567, 980
\bibitem[Genzel \& Downes(1977)]{genzel77} Genzel, R., \& Downes, D.\ 1977, \aaps, 30, 145
\bibitem[Green et al.(2010)]{green10} Green, J.~A., Caswell, J.~L., Fuller, G.~A., et al.\ 2010, \mnras, 409, 913
\bibitem[Gwinn et al.(1992)]{gwinn92} Gwinn, C.~R., Moran, J.~M., \& Reid, M.~J.\ 1992, \apj, 393, 149
\bibitem[Gwinn(1994)]{gwinn94} Gwinn, C.~R.\ 1994, \apj, 429, 253
\bibitem[Han et al.(1998)]{han98} Han, F., Mao, R.~Q., Lu, J., et al.\ 1998, \aaps, 127, 181 
\bibitem[Han et al.(2008)]{han08} Han, S.-T., Lee, J.-W., Kang, J., et al. \ 2008, Int. J. Infrared Millim. Waves, 29, 69
\bibitem[Isobe et al.(1990)]{isobe90} Isobe, T., Feigelson, E.~D., Akritas, M.~G., \& Babu, G.~J.\ 1990, \apj, 364, 104
\bibitem[Jordan et al.(2015)]{jordan15} Jordan, C.~H., Walsh, A.~J., Lowe, V., et al.\ 2015, \mnras, 448, 2344 
\bibitem[Kalenskii et al.(2010)]{kalenskii10} Kalenskii, S.~V., Johansson, L.~E.~B., Bergman, P., et al.\ 2010, \mnras, 405, 613
\bibitem[Kim et al.(2011)]{kim11} Kim, K.-T., Byun, D.-Y., Je, D.-H., et al.\ 2011, Journal of Korean Astronomical Society, 44, 81 
\bibitem[Kurtz et al.(2004)]{kurtz04} Kurtz, S., Hofner, P., \& {\'A}lvarez, C.~V.\ 2004, \apjs, 155, 149
\bibitem[Lee et al.(2011)]{lee11} Lee, S.-S., Byun, D.-Y., Oh, C.~S., et al.\ 2011, \pasp, 123, 1398 
\bibitem[Menten et al.(1988)]{menten88} Menten, K.~M., Walmsley, C.~M., Henkel, C., \& Wilson, T.~L.\ 1988, \aap, 198, 253 
\bibitem[Menten(1991)]{menten91} Menten, K.\ 1991, Atoms, Ions and Molecules: New Results in Spectral Line Astrophysics, 16, 119
\bibitem[Minier et al.(2000)]{minier00} Minier, V., Booth, R.~S., \& Conway, J.~E.\ 2000, \aap, 362, 1093 
\bibitem[Minier et al.(2003)]{minier03} Minier, V., Ellingsen, S.~P., Norris, R.~P., \& Booth, R.~S.\ 2003, \aap, 403, 1095 
\bibitem[Moscadelli et al.(2006)]{mosca06} Moscadelli, L., Testi, L., Furuya, R.~S., et al.\ 2006, \aap, 446, 985
\bibitem[Norris et al.(1993)]{nor93} Norris, R.~P., Whiteoak, J.~B., Caswell, J.~L., Wieringa, M.~H., \& Gough, R.~G.\ 1993, \apj, 412, 222 
\bibitem[Ochsenbein et al.(2000)]{ochsenbein00} Ochsenbein, F., Bauer, P., \& Marcout, J.\ 2000, \aaps, 143, 23 
\bibitem[Palagi et al.(1993)]{palagi93} Palagi, F., Cesaroni, R., Comoretto, G., Felli, M., \& Natale, V.\ 1993, \aaps, 101, 153 
\bibitem[Pandian et al.(2007)]{pand07} Pandian, J.~D., Goldsmith, P.~F., \& Deshpande, A.~A.\ 2007, \apj, 656, 255 
\bibitem[Pandian et al.(2009)]{pand09} Pandian, J.~D., Menten, K.~M., \& Goldsmith, P.~F.\ 2009, \apj, 706, 1609 
\bibitem[Pandian et al.(2010)]{pand10} Pandian, J.~D., Momjian, E., Xu, Y., Menten, K.~M., \& Goldsmith, P.~F.\ 2010, \aap, 522, A8 
\bibitem[Pandian et al.(2011)]{pand11} Pandian, J.~D., Momjian, E., Xu, Y., Menten, K.~M., \& Goldsmith, P.~F.\ 2011, \apj, 730, 55 
\bibitem[Pandian et al.(2012)]{pand12} Pandian, J.~D., Wyrowski, F., \& Menten, K.~M.\ 2012, \apj, 753, 50 
\bibitem[Pestalozzi et al.(2004)]{pes04} Pestalozzi, M.~R., Elitzur, M., Conway, J.~E., \& Booth, R.~S.\ 2004, \apjl, 603, L113 
\bibitem[Pestalozzi et al.(2005)]{pes05} Pestalozzi, M.~R., Minier, V., \& Booth, R.~S.\ 2005, \aap, 432, 737 
\bibitem[Plambeck \& Wright(1988)]{plambeck88} Plambeck, R.~L., \& Wright, M.~C.~H.\ 1988, \apjl, 330, L61 
\bibitem[Plambeck \& Menten(1990)]{plambeck90} Plambeck, R.~L., \& Menten, K.~M.\ 1990, \apj, 364, 555 
\bibitem[Pratap et al.(2008)]{pratap08} Pratap, P., Shute, P.~A., Keane, T.~C., Battersby, C., \& Sterling, S.\ 2008, \aj, 135, 1718 
\bibitem[Purcell et al.(2006)]{purcell06} Purcell, C.~R., Balasubramanyam, R., Burton, M.~G., et al.\ 2006, \mnras, 367, 553 
\bibitem[Rodriguez et al.(1980)]{rodri80} Rodriguez, L.~F., Moran, J.~M., Ho, P.~T.~P., \& Gottlieb, E.~W.\ 1980, \apj, 235, 845
\bibitem[Rosolowsky et al.(2010)]{rosolowsky10} Rosolowsky, E., Dunham, M.~K., Ginsburg, A., et al.\ 2010, \apjs, 188, 123 
\bibitem[Sch{\"o}ier et al.(2005)]{schoier05} Sch{\"o}ier, F.~L., van der Tak, F.~F.~S., van Dishoeck, E.~F., \& Black, J.~H.\ 2005, \aap, 432, 369 
\bibitem[Slysh et al.(1994)]{slysh94} Slysh, V.~I., Kalenskii, S.~V., Valtts, I.~E., \& Otrupcek, R.\ 1994, \mnras, 268, 464 
\bibitem[Szymczak et al.(2005)]{szymczak05} Szymczak, M., Pillai, T., \& Menten, K.~M.\ 2005, \aap, 434, 613 
\bibitem[Torrelles et al.(1997)]{torrelles97} Torrelles, J.~M., G{\'o}mez, J.~F., Rodr{\'{\i}}guez, L.~F., et al.\ 1997, \apj, 489, 744 
\bibitem[Urquhart et al.(2011)]{urquhart11} Urquhart, J.~S., Morgan, L.~K., Figura, C.~C., et al.\ 2011, \mnras, 418, 1689 
\bibitem[Valdettaro et al.(2001)]{valdettaro01} Valdettaro, R., Palla, F., Brand, J., et al.\ 2001, \aap, 368, 845 
\bibitem[Val'tts et al.(2000)]{valtts00} Val'tts, I.~E., Ellingsen, S.~P., Slysh, V.~I., et al.\ 2000, \mnras, 317, 315 
\bibitem[van der Tak et al.(2007)]{van07} van der Tak, F.~F.~S., Black, J.~H., Sch{\"o}ier, F.~L., Jansen, D.~J., \& van Dishoeck, E.~F.\ 2007, \aap, 468, 627
\bibitem[Voronkov et al.(2006)]{voronkov06} Voronkov, M.~A., Brooks, K.~J., Sobolev, A.~M., et al.\ 2006, \mnras, 373, 411 
\bibitem[Voronkov et al.(2014)]{voronkov14} Voronkov, M.~A., Caswell, J.~L., Ellingsen, S.~P., Green, J.~A., \& Breen, S.~L.\ 2014, \mnras, 439, 2584
\bibitem[Walsh et al.(1998)]{walsh98} Walsh, A.~J., Burton, M.~G., Hyland, A.~R., \& Robinson, G.\ 1998, \mnras, 301, 640
\bibitem[Xu et al.(2008)]{xu08} Xu, Y., Li, J.~J., Hachisuka, K., et al.\ 2008, \aap, 485, 729 
\bibitem[Xu et al.(2009)]{xu09} Xu, Y., Voronkov, M.~A., Pandian, J.~D., et al.\ 2009, \aap, 507, 1117 
\end{thebibliography}
\end{document}